\documentclass[prb, twocolumn, aps, amsmath,amssymb,amsfonts]{revtex4}
\usepackage{graphicx}
\usepackage{graphics}
\usepackage{dcolumn}
\usepackage{bm}
\usepackage{amssymb}
\usepackage{amsmath}
\usepackage{amsfonts}
\usepackage{epsfig}
\newcommand {\bra} [1] {\langle #1 |}
\newcommand {\ket} [1] {| #1 \rangle}
\newcommand {\bkt} [1] {\langle #1 \rangle}

\newcommand {\pd} [2] {\frac{\partial #1}{\partial #2}}
\newcommand {\td} [2] {\frac{d #1}{d #2}}

\begin{document}
\title{Transport in three-dimensional topological insulators: theory and experiment}
\author{Dimitrie Culcer} 
\affiliation{ICQD, Hefei National Laboratory for Physical Sciences at the Microscale, University of Science and Technology of China, Hefei 230026, Anhui, China}
\begin{abstract}
This article reviews recent theoretical and experimental work on transport due to the surface states of three-dimensional topological insulators. The theoretical focus is on longitudinal transport in the presence of an electric field, including Boltzmann transport, quantum corrections and weak localization, as well as longitudinal and Hall transport in the presence of both electric and magnetic fields and/or magnetizations. Special attention is paid to transport at finite doping, to the $\pi$-Berry phase, which leads to the absence of backscattering, Klein tunneling and half-quantized Hall response. Signatures of surface states in ordinary transport and magnetotransport are clearly identified. The review also covers transport experiments of the past years, reviewing the initial obscuring of surface transport by bulk transport, and the way transport due to the surface states has increasingly been identified experimentally. Current and likely future experimental challenges are given prominence and the current status of the field is assessed.  
\end{abstract}
\date{\today}
\maketitle

\section{Introduction}

The conventional band theory picture divides solids into metals, semiconductors and insulators based on the size of their band gaps and occupation of the conduction band. Within this picture, an insulator is understood to be a material with a band gap exceeding room temperature by orders of magnitude. In nature, other types of insulators exist, for example induced by a large disorder concentration, as in the Anderson insulator, or by strong Coulomb repulsion between electrons, as in the Mott insulator. In this context, the discovery of topological insulators (TI), first posited in the groundbreaking research of Kane and Mele, \cite{KaneMele_QSHE_PRL05} has been a milestone galvanizing physics research. \cite{KaneMele_QSHE_PRL05, Hasan_TI_RMP10, Qi_TI_RMP_10, Qi_TFT_PRB08} These novel materials are band insulators in the bulk while conducting along the surfaces, possessing gapless surface states with a well-defined spin texture protected by time-reversal symmetry. The landmark discovery of topological insulators has revolutionized the understanding of insulating behavior. In addition to a variety of exotic effects, TI-based structures may help solve long standing puzzles in physics as well as contribute to the development of quantum computation. Specifically, the interface between a topological insulator and a superconductor was predicted to support Majorana fermions, \cite{Fu_Proximity_Majorana_PRL08} that is, a species of Bogolyubov quasiparticles represented by real fermions that are their own antiparticles. Topological insulators are being explored with a view towards applications, as a potential platform for quantum computation, \cite{SDS_TQC_RMP08} and as rich physical systems in their own right.

\subsection{Topological order}

On a more profound level, topological insulators are a manifestation of topological order. Unlike other examples of topological order, such as the fractional quantum Hall effect or chiral $p$-wave superconductors, the emergence of topologically insulating behavior is a one-particle phenomenon, related to the existence of strong spin-orbit coupling. A topologically insulating state in 2D, the quantum spin Hall insulator, was predicted in HgCdTe quantum wells. The topological behavior in this case is related to the existence of edge states. \cite{Bernevig_QSHE_Science06} The key concept is band inversion consider: two materials, of which one has an inverted band structure. In the case demonstrated so far this was a CdTe/HgTe quantum well. Each state can be labeled by a $Z_2$ topological invariant $\nu$ (topological quantum number.) As the well width is increased the band gap is inverted and the $Z_2$ invariant of the ground state changes. From a lay perspective, in the context of the quantum spin Hall effect the $Z_2$ invariant can be thought of as a topological invariant that counts the number of edge states. The band inversion transition is not characterized by symmetry breaking, \cite{Hasan_TI_RMP10, Qi_TI_RMP_10} it is rather a topological phase transition: the fundamental symmetry of the lattice is the same on either side of the transition, but there is a change in the topological invariant $\nu$ of the ground state. Kane and Mele \cite{KaneMele_QSHE_PRL05} demonstrated that in two dimensions the invariant $\nu$ can be zero or one, the former representing ordinary insulators (equivalent to the vacuum) and the latter topological insulators. The two band structures corresponding to $\nu = 0$ and $\nu = 1$ cannot be deformed into one another. The material with an inverted band structure possesses a pair of crossing edge states which realize a quantum spin Hall insulator, that is, they give rise to a quantum spin Hall effect (in fact it possesses two pairs, by the fermion doubling theorem.) In the quantum spin Hall effect the charge current is carried by edge states with different spin polarizations. Because of spin-momentum locking due to the spin-orbit interaction electrons traveling along the edges in different directions have opposite spin orientations, and the chiral band structure associated with the surface states prohibits backscattering. The quantum spin Hall insulating state was observed experimentally shortly after its prediction \cite{Koenig_HgTe_QSHE_Science07}. For an excellent review of this effect, which will not be considered here, the reader is referred to the work of Koenig \textit{et al.} \cite{Koenig_QSHE_JPSJ08}.

Topological insulators were also predicted to exist in 3D, \cite{Fu_3DTI_PRL07, Roy_Z2_PRB09, Moore_TRI_TI_Invariants_PRB07} where analogous sets of crossing surface states are present. Three-dimensional topological insulators will be the focus of this review. In 3D there are four $Z_2$ topological invariants characterizing the band structure, \cite{Moore_TRI_TI_Invariants_PRB07} which are related to the time-reversal invariant momenta and \textit{time-reversal polarization} at eight special points in the Brillouin zone. The four $Z_2$ topological invariants in 3D are customarily labeled $\nu_0$, $\nu_1$, $\nu_2$ and $\nu_3$. They can be calculated easily if the system has inversion symmetry. \cite{Fu_TI_InvSym_PRB07} The $Z_2$ topological invariant $\nu_0$ distinguishes between strong and weak topological insulators: weak TI have $\nu_0 = 0$ whereas strong TI have $\nu_0 = 1$. In a weak TI the surface states are not protected by time reversal and can be localized by strong disorder. A weak TI can also be interpreted as a 3D stack of 2D quantum spin-Hall insulators. On the other hand, a strong TI is a new class of material and cannot be directly related to the quantum spin-Hall insulator. Interestingly, it was found that the crossover from a 2D to a 3D TI occurs in an oscillatory manner, \cite{Liu_TI_Crossover_PRB10} with the material alternating between topologically trivial and nontrivial phases as the layer thickness is increased. 

In an alternative picture of TI that does not rely on $Z_2$ invariants, Murakami studied the physics of gap closing in 2D and 3D, \cite{Murakami_TI_GapClose_10} focusing on whether the gap closes as certain parameters are tuned. Murakami also found that 3D inversion-asymmetric systems possess a finite crossover region of gapless phase between the ordinary insulator and topological insulator regimes. 

The surface states of 3D TI are well described by a Rashba spin-orbit Hamiltonian. \cite{BychkovRashba_JETP84, KaneMele_QSHE_PRL05, Hasan_TI_RMP10, Qi_TI_RMP_10} A detailed discussion of the model Hamiltonian for TI has been given in Ref.~ \onlinecite{Liu_TI_ModelHam_PRB10}, and a reduced Hamiltonian including surface and bulk states has been presented in Ref.~\onlinecite{Shen_3DTI_EffMod_NJP10}. These states are chiral, possessing a well-defined spin texture, referred to as spin-momentum locking, and have an energy spectrum akin to a Dirac cone. Chiral Rashba surface states have a nontrivial band topology and are characterized by a $\pi$ Berry phase, which is associated with Klein tunneling, and provides protection against coherent back scattering and therefore weak localization and Anderson localization. 

Several materials were predicted to be topological insulators in three dimensions. The first was the alloy Bi$_{1-x}$Sb$_x$,\cite{Teo_BiSb_SfcStt_PRB08, Zhang_BiSb_SfcStt_PRB09} followed by the tetradymite semiconductors Bi$_2$Se$_3$, Bi$_2$Te$_3$ and Sb$_2$Te$_3$. \cite{Zhang_TI_BandStr_NP09} These materials have a rhombohedral structure composed of quintuple layers oriented perpendicular to the trigonal $c$ axis. The covalent bonding within each quintuple layer is much stronger than weak van der Waals forces bonding neighboring layers. The semiconducting gap is approximately 0.3 eV, and the TI states exists along the (111) direction. In particular Bi$_2$Se$_3$ and Bi$_2$Te$_3$ have long been known from thermoelectric transport as displaying sizable Peltier and Seebeck effects, and their high quality has ensured their place at the forefront of experimental attention. \cite{Hasan_TI_RMP10} Initial predictions of the existence of chiral surface states were confirmed by first principles studies of Bi$_2$Se$_3$, Bi$_2$Te$_3$, and Sb$_2$Te$_3$. \cite{Zhang_TI_1stPrinc_NJP10} Heusler alloys were recently predicted to have topological surface states, \cite{Xiao_Heusler_PRL10} as well as chalcopyrites. \cite{Feng_TI_Chalcop_PRL11}

Unlike graphene, the Hamiltonian of topological insulators is a function of the real spin, rather than a sublattice pseudospin degree of freedom. This implies that spin dynamics will be qualitatively very different from graphene. Moreover, the twofold valley degeneracy of graphene is not present in topological insulators. Despite the apparent similarities, the study of topological insulators is thus not a simple matter of translating results known from graphene. Due to the dominant spin-orbit interaction it is also very different from ordinary spin-orbit coupled semiconductors.

Strictly speaking, the first instance of topologically insulating behavior discovered in nature is the quantum Hall effect (QHE). There is no difference in symmetry between the two sides of the quantum Hall transition, involving localized versus extended states. Rather, the Landau levels in the QHE are identified by their Chern number (TKNN invariant),\cite{TKNN} which is the integral of the geometrical curvature over the Brillouin zone. From a mathematical perspective, the Chern number is a relative of the genus and winding number. An analogy can be made between the Gaussian curvature and genus on the one hand and the Berry curvature and Chern number on the other. In the QHE the Chern number can be thought of as the topological invariant that counts the number of Landau levels, being therefore equivalent to the filling factor. Nevertheless, quantum Hall systems naturally break time-reversal due to the presence of a magnetic field. Haldane \cite{Haldane_ParAnom_PRL88} devised a model of the QHE without an overall magnetic field and therefore without Landau levels, in which local magnetic fields were generated by circulating current loops but averaged to zero over the whole sample. This model realizes the parity anomaly, but to this day remains a theoretical construct.

In summary, in order to determine whether a certain material is a topological insulator it is necessary to calculate a series of topological invariants, for which one requires the band structure of the material. The existence of chiral surface states is protected by time reversal symmetry, and is thus robust against smooth perturbations invariant under time-reversal, such as non-magnetic disorder and electron-electron interactions. In plain parlance, in the same manner that a doughnut \textit{can} be deformed into a coffee cup, the band structure of a material \textit{cannot} be deformed so as to remove the crossing of a pair of surface states (see the excellent figure in Ref.~\onlinecite{Hasan_TI_RMP10}.)


\subsection{Experimental discovery of 3D TI}

Thanks to surface-sensitive probes such as scanning tunneling microscopes (STM) and angle-resolved photoemission (ARPES), the existence of gapless chiral surface states in topological insulators has been established beyond any doubt. Experimental work originally reported unconventional behavior in Bi$_2$Se$_3$ and Bi$_2$Te$_3$, \cite{Urazhdin_PRB04} which was later understood to arise from their TI surface states. Several years later, experimental progress on TI skyrocketed. Hsieh \textit{et al}, \cite{Hsieh_BiSb_QSHI_Nature08} were the first to observe the Dirac cone-like surface states of Bi$_{1-x}$Sb$_x$, with $x\approx 0.1$, using ARPES. Following that, the same group used spin-resolved ARPES to measure the spin polarization of the surface states, demonstrating the correlation between spin and momentum.\cite{Hsieh_BiSb_QmSpinTxtr_Science09} Shortly afterwards experiments identified the Dirac cones in the tetradymite semiconductors Bi$_2$Se$_3$,\cite{Xia_Bi2Se3_LargeGap_NP09, Hsieh_Bi2Se3_Tunable_Nat09, Zhang_Bi2Se3_Film_Epitaxy_APL09} Bi$_2$Te$_3$,\cite{Chen_Bi2Te3_Science09, Hsieh_Bi2Te3_Sb2Te3_PRL09} and Sb$_2$Te$_3$.\cite{Hsieh_Bi2Te3_Sb2Te3_PRL09} Reflection high-energy electron diffraction (RHEED) was used to monitor MBE sample growth dynamics of Bi$_2$Te$_3$, \cite{Li_Bi2Se3_GrowthDyn_AM10} and the electronic structure has been shown to be controllable in thin films. \cite{Wang_Bi2Te3_Ctrl_AM11} TI growth has been covered in recent reviews. \cite{Hasan_TI_RMP10, Qi_TI_RMP_10} On a related note, recently Bi$_2$Se$_3$ nanowires and nanoribbons have also been grown, \cite{Kong_TI_WireRbn_NanoLett10} and a topological phase transition was reported in BiTl(S$_{1-d}$Se$_d$)2, \cite{Xu_TI_e/2_Science11} which may lead to the observation of charge $e/2$.

Zhu \textit{et al.} \cite{Zhu_Bi2Se3_RashbaCtrl_11} used ARPES to study the surface states of Bi$_2$Se$_3$. These researchers employed in-situ potassium deposition to tackle the instability of electronic properties, which otherwise evolve continuously in time. In their samples, as potassium is deposited, new spin-polarized quantum-well states emerge (with no $k_z$ dispersion), originating in the parabolic bulk bands and distinct from the Rashba-Dirac cone. These new states also follow a Rashba-like dispersion, and their spin splitting is tunable as well as reversible. The potassium-induced potential gradient $\partial V/\partial z$ enhances the spin splitting. These states may affect transport data even from pristine surfaces. 

Considerable efforts have been devoted to scanning tunneling microscopy (STM) and spectroscopy (STS), which enable the study of quasiparticle scattering. Scattering off surface defects, in which the initial state interferes with the final scattered state, results in a standing-wave interference pattern with a spatial modulation determined by the momentum transfer during scattering. These manifest themselves as oscillations of the local density of states in real space, which have been seen in several materials with topologically protected surface states. 

Roushan \textit{et al.} \cite{Roushan_BiSb_STM_NoBxct_Nat09} imaged the surface states of Bi$_{0.92}$Sb$_{0.08}$ using STM and spin-resolved ARPES demonstrating that scattering between states of opposite momenta is strongly suppressed in the presence of time-reversal invariant disorder. Gomes \textit{et al.} \cite{Gomes_Sb_STM_NoBxct_09} investigated the (111) surface of Sb, which displays a topological metal phase, and found a similar suppression of backscattering. Zhang \textit{et al.} \cite{Zhang_Bi2Te3_STM_NoBxct_PRL09} and Alpichshev \textit{et al.}\cite{Alp_Bi2Te3_STM_NoBxct_PRL10} observed the suppression of backscattering in Bi$_2$Te$_3$, and used STM to image surface bound states. \cite{Alpichshev_Bi2Se3_BoundStt_PRB11} The absence of backscattering has been studied theoretically in the context of quasiparticle interference seen in STM experiments. \cite{Guo_TI_QptIntfrnc_PRB10, Lee_TI_Qpt_Intfrnc_PRB09} Park \textit{et al.} \cite{Park_Bi2Se3_QPtSct_PRB10} performed ARPES accompanied by theoretical modelling on Bi$_2$Se$_3$. The observed quasiparticle lifetime of the surface states is attributed to quasiparticle decay predominantly into bulk electronic states through electron-electron interaction and defect scattering. Studies on aged surfaces show the surface states to be unaffected by adsorbed atoms or molecules on the surface, indicating protection against weak perturbations.

Experimental studies have provided evidence of the existence of chiral surface states and of their protection by time-reversal symmetry. Experimental research has begun to investigate the fascinating question concerning the transition between the three- and two-dimensional phases in this material. \cite{Zhang_Bi2Se3_Crossover_NP10} Zhang \textit{et al.} \cite{Zhang_Bi2Se3_Crossover_NP10} used ARPES to study a thin 3D Bi$_2$Se$_3$ slab grown by MBE, in which tunnelling between opposite surfaces opens a small, thickness-dependent gap. The energy gap opening is clearly seen when the thickness is below six quintuple layers. The gapped surface states also exhibit sizeable Rashba spin-orbit splitting due to the potential difference \textit{between} the two surfaces induced by the substrate. In other words, one of the most important findings of this experiment is that thin films need to be thicker than approximately six quintuple layers, or 6nm. Bianchi \textit{et al.} \cite{Bianchi_Bi2Se3_TopStt_2DEG_NatComm10} showed that the band bending near the Bi$_2$Se$_3$ surface leads to the formation of a two-dimensional electron gas (2DEG), which coexists with the topological surface state in Bi$_2$Se$_3$. In the setup used by this experimental group, a topological and a non-topological metallic state are confined to the same region of space. 

Several efforts have focused on the role of magnetic impurities on the surface states of 3D TI. The work of Hor \textit{et al.} demonstrated that doping Bi$_2$Te$_3$ with Mn results in the onset of ferromagnetism. \cite{Hor_DopedTI_FM_PRB10} Chen \textit{et al.} \cite{Chen_MTI_MassDirac_Sci10} introduced magnetic dopants into Bi$_2$Se$_3$ to break the time reversal symmetry and further position the Fermi energy inside the gaps by simultaneous magnetic and charge doping. The resulting insulating massive Dirac fermion state was observed by ARPES. More recently, Wray \textit{et al.} \cite{Wray_Bi2Se3_StrongCoul_NP11} studied the topological surface states of Bi$_2$Se$_3$ under Coulomb and magnetic perturbations. The authors deposited Fe on the surface of Bi$_2$Se$_3$, using the fact that Fe has a large Coulomb charge and sizable magnetic moment to modify the spin structure of the Bi$_2$Se$_3$ surface states. Interestingly, it was found that this perturbation leads to the creation of odd multiples of Dirac fermions. In other words, new pairs of Dirac cones appear as more Fe is deposited, also described by Rashba interactions, and the interaction grows in strength as Fe is added. Numerical simulations indicate that the new Dirac cones are due to electrons localized deeper inside the material. At the same time a gap is opened at the original Dirac point, as one expects.

\subsection{The importance of transport}

The key to the eventual success of topological insulators in becoming technological materials is inherently linked to their transport properties. Potential applications of topological surface states necessarily rely on the realization of an edge metal (semimetal), allowing continuous tuning of the Fermi energy through the Dirac point, the presence of a minimum conductivity (maximum resistivity) at zero carrier density, and ambipolar transport. At this point in time experimental and theoretical studies of equilibrium TI abound. Transport has been lagging somewhat but is now picking up in both theory and experiment at a brisk rate. 

Despite the success of photoemission and scanning tunneling spectroscopy in identifying chiral surface states, signatures of the surface Dirac cone have not yet been observed in transport. Simply put, the band structure of topological insulators can be visualized as a band insulator with a Dirac cone within the bulk gap, and to access this cone one needs to ensure the chemical potential lies below the bottom of the bulk conduction band. Given that the static dielectric constants of materials under investigation are extremely large, approximately 100 in Bi$_2$Se$_3$ and 200 in Bi$_2$Te$_3$, gating in order to bring the chemical potential down is challenging. Therefore, in all materials studied to date, residual conduction from the bulk exists due to unintentional doping. The above presentation makes it plain that, despite the presence of the Dirac cone, currently no experimental group has produced a true topological insulator. Consequently, all existing topological insulator systems are in practice either heavily bulk-doped materials or thin films of a few monolayers (thus, by definition, not 3D.) Following the pattern set by the imaging of the surface Dirac cone, the Landau levels observed in STM studies have not been seen in transport.\cite{Cheng_TI_STM_LL_PRL10, Hanaguri_TI_STM_LL_PRB10} These last works illustrate both the enormous potential of the field and the challenges to be overcome experimentally, on which more below.

In this work I will provide a theoretical framework for understanding TI transport, review existing TI transport theories, and survey the impressive experimental breakthroughs in transport registered in the past years. The review will focus on phenomena in electric fields as well as electric and magnetic fields, but not magnetic fields alone. As it stands, fascinating aspects such as the search for Majorana fermions also lie outside the scope of this review, since no transport work exists on these topics. The transport work covered consists of two main parts, which in turn can be subdivided into five smaller parts: ordinary (non-magnetic) transport, comprising Boltzmann transport and weak localization, and magnetotransport, comprising the quantum Hall, ordinary Hall and anomalous Hall effects. Theoretical work in these fields will be covered first, followed by experiment, and I will close with a section on experimental challenges, followed by general conclusions and an outline of future directions.

\section{Effective Hamiltonian} 
\label{sec:Ham}

The effective Hamiltonian describing the surface states of non-magnetic topological insulators can be written in the form of a Rashba spin-orbit interaction \cite{Zhang_TI_BandStr_NP09, BychkovRashba_JETP84}
\begin{equation}\label{Ham}
H_{0{\bm k}} = A \, {\bm \sigma} \cdot {\bm k} \times \hat{\bm z} \equiv - Ak \, {\bm \sigma} \cdot \hat{\bm \theta}. 
\end{equation}
It is understood that ${\bm k} = (k_x, k_y)$, and $\hat{\bm \theta}$ is the tangential unit vector corresponding to ${\bm k}$. The constant A = 4.1eV$\AA$ for Bi$_2$Se$_3$. \cite{Zhang_TI_BandStr_NP09} The eigenenergies are denoted by $\varepsilon_\pm = \pm Ak$, so that the spectrum possesses particle-hole symmetry. The Hamiltonian is similar to that of graphene, with the exception that the vector of Pauli spin matrices ${\bm \sigma}$ represents the true electron spin. A scalar term $Dk^2$ also exists in principle. The scalar and spin-dependent terms in the Hamiltonian are of the same magnitude when $D \, k_F = A$, corresponding to $k_F \approx 10^9$ m$^{-1}$ in Bi$_2$Se$_3$, that is, a density of $10^{14}$cm$^{-2}$, which is higher than the realistic densities in transport experiments. Consequently, in this review scalar terms will not be considered. I will also not take into account small anisotropy terms in the scalar part of the Hamiltonian, such as those discussed in Ref. \onlinecite{Zhang_TI_BandStr_NP09}. Terms cubic in ${\bm k}$ in the spin-orbit interaction are in principle also present, in turn making the energy dispersion anisotropic, but they are much smaller than the ${\bm k}$-linear terms in Eq.\ (\ref{Ham}) and are not expected to contribute significantly to quantities discussed in this review. All theories described here require $\varepsilon_F \tau/\hbar \gg 1$ to be applicable, with $\tau$ the momentum relaxation time. 

A topological insulator can doped with magnetic impurities, which give a net magnetic moment as well as spin-dependent scattering. Magnetic impurities are described by an additional interaction Hamiltonian
\begin{equation}
{H}_\mathrm{mag} ({\bm r}) = {\bm \sigma} \cdot \sum_I \mathcal{V}({\bm r} - {\bm R}_I)\, {\bm s}_I,
\end{equation} 
where the sum runs over the positions ${\bm R}_I$ of the magnetic ions with ${\bm s}_I$. To capture the physics discussed in this review it is sufficient to approximate the potential $\mathcal{V}({\bm r} - {\bm R}_I) = J \, \delta({\bm r} - {\bm R}_I)$, with $J$ the exchange constant between the localized moments and itinerant carriers. The magnetic ions are assumed spin-polarized so that ${\bm s}_I = s \, \hat{\bm z}$. Fourier transforming to the crystal momentum representation $\{ \ket{{\bm k}} \}$, the ${\bm k}$-diagonal term, $H_\mathrm{mag}^{{\bm k} = {\bm k}'} = n_\mathrm{mag}\, J \, s \sigma_z \equiv M \, \sigma_z$, gives the magnetization, with $n_\mathrm{mag}$ the density of magnetic ions. The ${\bm k}$-off-diagonal term causes spin-dependent scattering
\begin{equation}\label{Hmagod}
\begin{array}{rl}
\displaystyle {H}_\mathrm{mag}^{{\bm k} \ne {\bm k}'} = & \displaystyle \frac{Js}{V} \, \sigma_z \sum_I e^{i({\bm k} - {\bm k}')\cdot{\bm R}_I},
\end{array}
\end{equation}
with $V$ the volume. The effective band Hamiltonian incorporating a mean-field magnetization takes the form
\begin{equation}\label{magHam}
H_{m{\bm k}} = - Ak \, {\bm \sigma} \cdot \hat{\bm \theta} + {\bm \sigma}\cdot{\bm M} \equiv \frac{\hbar}{2} \, {\bm \sigma} \cdot {\bm \Omega}_{\bm k}.
\end{equation}
Retaining the same notation as in the non-magnetic case, the eigenenergies are $\varepsilon_\pm = \pm \sqrt{A^2k^2 + M^2}$. The bulk of the work on magnetic TI has concentrated on out-of-pane magnetizations and magnetic fields. An in-plane magnetization or magnetic field does not alter the Rashba-Dirac cone, though coherence between layers can induce a quantum phase transition. \cite{Zyuzin_TI_B_parallel_QPT_PRB11}

Interaction with a static, uniform electric field ${\bm E}$ is quite generally contained in $H_{E, {\bm k}{\bm k}'} = H_{E, {\bm k}{\bm k}'}^{sc} + H_{E, {\bm k}{\bm k}'}^{sj}$, the scalar part arising from the ordinary position operator $H_{E, {\bm k}{\bm k}'}^{sc} = (e{\bm E}\cdot\hat{\bm r})_{{\bm k}{\bm k}'} \openone$, with $\openone$ the identity matrix in spin space, and the side-jump part $H_{E, {\bm k}{\bm k}'}^{sj} = e\lambda \, {\bm \sigma}\cdot({\bm k}\times{\bm E}) \, \delta_{{\bm k}{\bm k}'}$ arising from the spin-orbit modification to the position operator, $\hat{\bm r} \rightarrow \hat{\bm r} + \lambda {\bm \sigma} \times \hat{\bm k}$, \cite{Sinitsyn_AHE_Review_JPCM08} with $\lambda$ a material-specific constant. 

Elastic scattering off charged impurities, static defects and magnetic ions, but not phonons or other electrons, is contained in the disorder scattering potential $U_{{\bm k}{\bm k}'} = \bar{U}_{{\bm k}{\bm k}'} \sum_{J} e^{i({\bm k} - {\bm k}')\cdot{\bm R}_J}$, with ${\bm R}_J$ the random locations of the impurities. The potential $U_{{\bm k}{\bm k}'}$ incorporates the magnetic scattering term ${H}_\mathrm{mag}^{{\bm k} \ne {\bm k}'}$ of Eq. (\ref{Hmagod}). The potential of a single impurity $\bar{U}_{{\bm k}{\bm k}'} = (1 - i \, \lambda \, {\bm \sigma} \cdot {\bm k} \times {\bm k}')\, \mathcal{U}_{{\bm k}{\bm k}'} - Js \, \sigma_z $, with $\mathcal{U}_{{\bm k}{\bm k}'}$ the matrix element of a Coulomb potential between plane waves, and $\lambda k_F^2 \ll 1$. Note that $\bar{U}_{{\bm k}{\bm k}'}$ has the dimensions of energy times volume.

Charged impurities are described by a screened Coulomb potential. The screening function can be evaluated in the random phase approximation (RPA.) In this approximation the polarization function is obtained by summing the lowest bubble diagram, and takes the form \cite{Hwang_Gfn_Screening_PRB07} ($\lambda, \lambda' = \pm$)
\begin{equation}
\Pi (q, \omega) = -\frac{1}{A} \sum_{{\bm k}\lambda\lambda'} \frac{f_{0{\bm k}\lambda} - f_{0{\bm k}'\lambda'}}{\omega + \varepsilon_{{\bm k}\lambda} - \varepsilon_{{\bm k}'\lambda'} + i\eta} \, \bigg(\frac{1 + \lambda\lambda' \cos\gamma}{2}\bigg), 
\end{equation}
where $f_{0{\bm k}\lambda} \equiv f_0(\varepsilon_{{\bm k}\lambda})$ is the equilibrium Fermi distribution function. The static dielectric function, which is of relevance to the physics discussed in this review, can be written as $\epsilon(q) = 1 + v(q) \,\Pi(q)$, where $v(q) = e^2/(2\epsilon_0\epsilon_r q)$, where $\epsilon_r$ is the relative permittivity. To determine $\Pi (q)$, we assume $T=0$ and use the Dirac cone approximation as in Ref.\ \onlinecite{Hwang_Gfn_Screening_PRB07}. This approximation is justified in the regime $T/T_F \ll 1$, with $T_F$ the Fermi temperature. At $T=0$ for charged impurity scattering the long-wavelength limit of the dielectric function is \cite{Hwang_Gfn_Screening_PRB07}
\begin{equation}
\epsilon(q) = 1 + \frac{e^2}{4\pi \epsilon_0 \epsilon_r A} \, \bigg(\frac{k_F}{q}\bigg),
\end{equation}
yielding the Thomas-Fermi wave vector as $k_{TF} = e^2  k_F/(4\pi \epsilon_0 \epsilon_r A)$. As a result, in topological insulators the matrix element $\bar{U}_{{\bm k}{\bm k}'}$ of a screened Coulomb potential between plane waves is given by
\begin{equation}\label{eq:W}
\arraycolsep 0.3ex
\begin{array}{rl}
\displaystyle \bar{U}_{{\bm k}{\bm k}'} = & \displaystyle \frac{Ze^2}{2\epsilon_0\epsilon_r}\, \frac{1}{ |{\bm k} - {\bm k}'| + k_{TF}},
\end{array}
\end{equation}
where $Z$ is the ionic charge (which I will assume for simplicity to be $Z = 1$) and $k_{TF}$ is the Thomas-Fermi wave vector. The polarization function was also calculated in Ref.\ \onlinecite{Raghu_HlcLqd_PRL10}. The Wigner-Seitz radius $r_s$, which parametrizes the relative strength of the kinetic energy and electron-electron interactions, is a constant for the Rashba-Dirac Hamiltonian, and is given by $r_s = e^2/(2\pi \epsilon_0\epsilon_r A)$.

The full Hamiltonian $H_{{\bm k}} = H_{0{\bm k}} + H_{E{\bm k}{\bm k}'} + U_{{\bm k}{\bm k}'} + {H}_\mathrm{mag}^{{\bm k} = {\bm k}'}$. The current operator ${\bm j}$ has contributions from the band Hamiltonian,
\begin{equation}\label{j}
{\bm j} = \frac{eA}{\hbar} \,{\bm \sigma} \times \hat{\bm z},
\end{equation}
as well as from the applied electric field ${\bm j}_E = \frac{2e^2 \lambda}{\hbar} \, {\bm \sigma} \times {\bm E}$, and from the disorder potential ${\bm j}_U =  \frac{2ie\lambda}{\hbar} \, {\bm \sigma}\times ({\bm k} - {\bm k}') \, \mathcal{U}_{{\bm k }{\bm k}'}$. These latter two cancel on physical grounds, as they represent the force acting on the system. 

Transport experiments on topological insulators seek to distinguish the conduction due to the surface states from that due to the bulk states. As mentioned in the introduction, it has proven challenging to lower the chemical potential beneath the bottom of the conduction band, so that the materials studied at present are not strictly speaking insulators. It is necessary for $\varepsilon_F$ to be below the bulk conduction band, so that one can be certain that there is only surface conduction, and at the same time I recall the requirement that $\varepsilon_F \tau/\hbar \gg 1$ for the kinetic equation formalism to be applicable. These assumptions will be made throughout the theoretical presentation outlined in what follows.

\section{Transport theory for topological insulators}

As stated above, work on transport in topological insulators has begun to expand energetically. The focus of experimental research in particular has shifted to the separation of bulk conduction from surface conduction, and, ideally, strategies for the elimination of the former. Hence it is necessary to know what contributions to the conductivity are expected from the chiral surface states.

Transport problems are customarily approached within the framework of the linear response formalism. The resulting Kubo formula is frequently expressed in terms of the Green's functions of the system. This can be done in the Keldysh formulation, which has been covered in countless textbooks, one of the clearest and most comprehensive being Ref.~ \onlinecite{Vasko}. Most generally, the finite-frequency conductivity tensor is given by
\begin{widetext}
\begin{equation}\label{Kubo}
\sigma_{\alpha\beta} (\omega) = \frac{ine^2}{m\omega} \, \delta_{\alpha\beta} + \frac{e^2}{2\pi\omega V}\, \bigg\langle {\rm Tr} \int d\varepsilon \, \rho_{0, \varepsilon}[ \hat{v}_\alpha\hat{G}^R_{\varepsilon + \hbar \omega}\hat{v}_\beta(\hat{G}^A_\varepsilon - G^R_\varepsilon) + \hat{v}_\alpha (\hat{G}^A_\varepsilon - G^R_\varepsilon) \hat{v}_\beta \hat{G}^A_{\varepsilon - \hbar \omega}] \bigg\rangle
\end{equation}
\end{widetext}
In the above formula, $n$ stands for the carrier number density, Tr is the operator trace, $\bkt{}$ denotes the average over impurity configurations, $\rho_0$ is the equilibrium density matrix, $\hat{\bm v}$ is the velocity operator, and $G^{R, A}$ are the usual retarded and advanced Green's functions. In topological insulators the indices $\alpha, \beta = x, y$. One important feature of the Kubo formula is that the conductivity is effectively given by the impurity average of the product of two Green's functions, which can be iterated in the strength of the disorder potential, or alternatively in powers of the quantity $\hbar/(\varepsilon_F\tau)$, which as we recall is required to be small for the transport theory outlined here to be valid. This iteration is encompassed within the Bethe-Salpeter equation, and can be expressed diagrammatically. Depending on the approximation required, several sets of diagrams must be summed. In a non-magnetic system, the contribution to leading-order in $\hbar/(\varepsilon_F\tau)$ recovers the Boltzmann conductivity, and corresponds to the sum of ladder diagrams. Since the sum of ladder diagrams can be interpreted classically as representing diffusion of carriers, this terms also bears the name of diffuson. In the next order, which yields a term independent of $\tau$, the weak localization contribution to the conductivity is obtained as the sum of maximally-crossed diagrams, or the Cooperon (plus some minor corrections). Weak localization is usually unimportant for $\varepsilon_F \tau/\hbar \gg 1$ unless a magnetic field is applied to measure the magnetoresistance. However when the Hamiltonian is spin-dependent/chiral it is possible to have terms independent of $\tau$ even in the weak momentum scattering limit. Finally, it must be borne in mind that, in topological insulators, identifying and explaining all the relevant contributions to transport necessitates a matrix formulation, thus in Eq.\ (\ref{Kubo}) the equilibrium density matrix, the velocity operators and Green's functions are matrices.

The Kubo formula is extremely well known in all its forms, and there appears to be little motivation for this review to focus on conventional linear response. An alternative matrix formulation, which contains the same physics and is potentially more transparent, relies on the quantum Liouville equation to derive a kinetic equation for the density matrix. This theory was first discussed for graphene monolayers \cite{Culcer_Gfn_Transp_PRB08} and bilayers, \cite{Culcer_Bil_PRB09} and was recently  extended to topological insulators including the full scattering term to linear order in the impurity density. \cite{Culcer_TI_Kineq_PRB10} Peculiarities of topological insulators, such as the absence of backscattering, which reflects the $\pi$ Berry phase and leads to Klein tunneling, are built into this theory in a transparent fashion. 

The derivation of the kinetic equation for a system driven by an electric field in the presence of random, uncorrelated impurities  begins with the quantum Liouville equation for the density operator $\hat \rho$,\cite{Culcer_TI_Kineq_PRB10}
\begin{equation}
\td{\hat\rho}{t} + \frac{i}{\hbar} \, [\hat{H}, \hat \rho] = 0,
\end{equation}
where the full Hamiltonian $\hat{H} = \hat{H}_0 + \hat{H}_E + \hat U$, the band Hamiltonian $\hat{H}_0$ is defined in Eq.\ (\ref{Ham}), $\hat{H}_E = e{\bm E}\cdot\hat{\bm r}$ represents the interaction with external fields, $\hat{\bm r}$ is the position operator, and $\hat{U}$ is the impurity potential. We consider a set of time-independent states $\{ \ket{{\bm k}s} \}$, where ${\bm k}$ indexes the wave vector and $s$ the spin. The matrix elements of $\hat \rho$ are $\rho_{{\bm k}{\bm k}'} \equiv \rho^{ss'}_{{\bm k}{\bm k}'} = \bra{{\bm k}s} \hat\rho \ket{{\bm k}'s'}$, with the understanding that $\rho_{{\bm k}{\bm k}'}$ is a matrix in spin space. The terms $\hat{H}_0$ and $\hat{H}_E$ are diagonal in wave vector but off-diagonal in spin, while for elastic scattering in the first Born approximation $U_{{\bm k}{\bm k}'}^{ss'} = U_{{\bm k}{\bm k}'}\delta_{ss'}$. The impurities are assumed uncorrelated and the average of $\bra{{\bm k}s}\hat U\ket{{\bm k}'s'}\bra{{\bm k}'s'}\hat U\ket{{\bm k}s}$ over impurity configurations is $(n_i |\bar{U}_{{\bm k}{\bm k}'}|^2 \delta_{ss'})/V$, where $n_i$ is the impurity density, $V$ the crystal volume and $\bar{U}_{{\bm k}{\bm k}'}$ the matrix element of the potential of a single impurity.

The density matrix $\rho_{{\bm k}{\bm k}'}$ is written as $\rho_{{\bm k}{\bm k}'} = f_{{\bm k}} \, \delta_{{\bm k}{\bm k}'} + g_{{\bm k}{\bm k}'}$, where $f_{\bm k}$ is diagonal in wave vector (i.e. $f_{{\bm k}{\bm k}'} \propto \delta_{{\bm k}{\bm k}'}$) while $g_{{\bm k}{\bm k}'}$ is off-diagonal in wave vector (i.e. ${\bm k} \ne {\bm k}'$ always in $g_{{\bm k}{\bm k}'}$.) The quantity of interest in determining the charge current is $f_{\bm k}$, since the current operator is diagonal in wave vector. We therefore derive an effective equation for this quantity by first breaking down the quantum Liouville equation into
\begin{subequations}
\begin{eqnarray}
\label{eq:f} \td{f_{{\bm k}}}{t} + \frac{i}{\hbar} \, [H_{0{\bm k}}, f_{{\bm k}}] & = & - \frac{i}{\hbar} \, [H^E_{\bm k}, f_{{\bm k}}] - \frac{i}{\hbar} \, [\hat U, \hat g]_{{\bm k}{\bm k}}, \\ [1ex]
\label{eq:g} \td{g_{{\bm k}{\bm k}'}}{t} + \frac{i}{\hbar} \, [\hat{H}, \hat g]_{{\bm k}{\bm k}'} & = & - \frac{i}{\hbar} \, [\hat U, \hat f + \hat g]_{{\bm k}{\bm k}'}.
\end{eqnarray}
\end{subequations}
The term $\displaystyle - \frac{i}{\hbar} \, [\hat U, \hat g]_{{\bm k}{\bm k}}$ on the RHS of Eq.\ (\ref{eq:f}) will become the scattering term. Solving Eq.\ (\ref{eq:g}) to first order in the scattering potential $\hat{U}$
\begin{equation}\label{eq:gsol}
g_{{\bm k}{\bm k}'} = - \frac{i}{\hbar} \, \int_0^\infty dt'\, e^{- i \hat H t'/\hbar} \left[\hat U, \hat f (t - t') \right] e^{i \hat H t'/\hbar}\bigg|_{{\bm k}{\bm k}'}.
\end{equation}
We are interested in variations which are slow on the scale of the momentum relaxation time, consequently we do not take into account memory effects and $\hat f(t - t') \approx \hat{f} (t)$. Time integrals such as the one appearing in Eq.\ (\ref{eq:gsol}) 
are performed by inserting a regularizing factor $e^{- \eta t'}$ and letting $\eta \rightarrow 0$ in the end. It is necessary to carry out an average over impurity configurations, keeping the terms to linear order in the impurity density $n_i$. 

The solution of Eq.\ (\ref{eq:g}) to first order in $\hat{U}$ represents the first Born approximation, which is sufficient for ordinary Boltzmann transport, with or without magnetic interactions. In order to recover phenomena such as skew scattering, elaborated upon below, one needs to work to second order in $\hat{U}$. The total scattering term $\hat{J} (f_{\bm k}) = \hat{J}^{Born}(f_{\bm k}) + \hat{J}^{3rd}(f_{\bm k})$, with
\begin{equation}
\label{JBorn}
\hat{J}^{Born}(f_{\bm k}) = \bigg\langle\int_0^{\infty} \frac{dt'}{\hbar^2} \, [\hat U, e^{- \frac{i \hat H t'}{\hbar}}[\hat U, \hat f]\, e^{ \frac{i \hat H t'}{\hbar}}]\bigg\rangle_{{\bm k}{\bm k}},
\end{equation}
while $\bkt{}$ represents averaging over impurity configurations. In the absence of scalar terms in the Hamiltonian the scalar and spin-dependent parts of the density matrix, $n_{\bm k}$ and $S_{\bm k}$, are decoupled in the first Born approximation. In the second Born approximation, which is needed in magnetic systems, one obtains the additional scattering term
\begin{widetext}
\begin{equation}
\label{Jss3rd}
\hat{J}^{3rd}(f_{\bm k}) = - \frac{i}{\hbar^3} \bigg\langle\int_0^{\infty} dt' \int_0^{\infty} dt'' [\hat U, e^{- \frac{i \hat H t'}{\hbar}}[\hat U, e^{- \frac{i \hat H t''}{\hbar}}[\hat U, \hat f]\, e^{\frac{i \hat H t''}{\hbar}} ]\, e^{\frac{i \hat H t'}{\hbar}}]\bigg\rangle_{{\bm k}{\bm k}}.
\end{equation}
\end{widetext}
The former can be further broken down into $\hat{J}^{Born}(f_{\bm k}) = \hat{J}_0(f_{\bm k}) + \hat{J}^{Born}_{ss}(f_{\bm k}) + \hat{J}^{Born}_{sj}(f_{\bm k})$. The first term, $\hat J_0(f_{{\bm k}})$, in which $\lambda = {\bm E} = 0$ in the time evolution operator, represents elastic, spin-independent, pure momentum scattering. In $\hat{J}^{Born}_{ss}(f_{\bm k})$ we allow $\lambda$ to be finite but ${\bm E} = 0$. In $\hat{J}^{Born}_{sj}(f_{\bm k})$ both $\lambda$ and ${\bm E}$ are finite, thus $\hat{J}^{Born}_{sj}(f_{\bm k})$ acts on the equilibrium density matrix $f_{0{\bm k}}$. Beyond the Born approximation we retain the leading term $\hat{J}^{3rd}(f_{\bm k}) \equiv \hat{J}^{3rd}_{ss}(f_{\bm k})$, with $\lambda$ finite but ${\bm E} = 0$, and which is customarily responsible for skew scattering \cite{Sinitsyn_AHE_Review_JPCM08}. The contribution due to magnetic impurities is also $\propto \sigma_z$ and is contained in $\hat{J}^{Born}_{ss}(f_{\bm k})$.

First consider the case in which the magnetization ${\bm M} = 0$. In this case all terms $\propto \lambda$ above are zero, and it is safe to take $\lambda = 0$. To first order in ${\bm E}$, the diagonal part $f_{\bm k}$ satisfies \cite{Culcer_TI_Kineq_PRB10}
\begin{equation}
\label{f} 
\td{f_{{\bm k}}}{t} + \frac{i}{\hbar} \, [H_{0{\bm k}}, f_{{\bm k}}] + \hat{J}(f_{\bm k}) = \mathcal{D}_{\bm k}. 
\end{equation}
The driving term $\mathcal{D}_{\bm k} = - \frac{i}{\hbar} \, [H^E_{\bm k}, f_{0{\bm k}}]$, and $f_{0{\bm k}}$ is the equilibrium density matrix, which is diagonal in ${\bm k}$. The ${\bm k}$-diagonal part of the density matrix $f_{\bm k}$ is a $2 \times 2$ Hermitian matrix, which is decomposed into a scalar part and a spin-dependent part. One writes $f_{\bm k} = n_{\bm k}\openone + S_{\bm k}$, where $S_{\bm k}$ is a $2 \times 2$ Hermitian matrix, which represents the spin-dependent part of the density matrix and is written purely in terms of the Pauli $\sigma$ matrices. Every matrix in this problem can be written in terms of a scalar part, labeled by the subscript $n$, and ${\bm \sigma}$. Rather than choosing Cartesian coordinates to express the latter, it is more natural to identify two orthogonal directions in reciprocal space, denoted by $\hat{\bm \Omega}_{\bm k} = -\hat{\bm \theta}$ and $\hat{\bm k}$. One projects ${\bm \sigma}$ as $\sigma_{{\bm k}, \parallel} = {\bm \sigma} \cdot \hat{\bm \Omega}_{\bm k}$ and $\sigma_{{\bm k}, \perp} = {\bm \sigma} \cdot \hat{\bm k}$. Note that $\sigma_{{\bm k}, \parallel}$ commutes with $H_{0{\bm k}}$, while $\sigma_{{\bm k}, \perp}$ does not. One projects $S_{\bm k}$ onto the two directions in ${\bm k}$-space, writing $S_{\bm k} = S_{{\bm k}, \parallel} + S_{{\bm k}, \perp}$, and defines $S_{{\bm k}, \parallel} = (1/2) \, s_{{\bm k}, \parallel} \, \sigma_{{\bm k}, \parallel}$ and $S_{{\bm k}, \perp} = (1/2) \, s_{{\bm k}, \perp} \, \sigma_{{\bm k}, \perp}$. There is no coupling of the scalar and spin distributions because of the particle-hole symmetry inherent in the Rashba-Dirac Hamiltonian.

Introducing projection operators $P_\parallel$ and $P_\perp$ onto the scalar part, $\sigma_{{\bm k}, \parallel}$ and $\sigma_{{\bm k}, \perp}$ respectively, equation~(\ref{f}) can be written as
\begin{subequations}\label{eq:Spp}
\begin{eqnarray}
\td{S_{{\bm k}, \parallel}}{t} + P_\| \hat J (f_{{\bm k}}) & = & \mathcal{D}_{{\bm k},\parallel}, \\ [0.5ex]
\label{Sperp}
\td{S_{{\bm k}, \perp}}{t} + \frac{i}{\hbar} \, [H_{{\bm k}}, S_{{\bm k}\perp}] + P_\perp \hat J (f_{{\bm k}}) & = & \mathcal{D}_{{\bm k},\perp}.
\end{eqnarray}
\end{subequations}
The projector $P_\parallel$ acts on a matrix $\mathcal{M}$ as ${\rm tr} \, (\mathcal{M}\sigma_{{\bm k}, \parallel})$, with tr the matrix trace, while $P_\perp$ singles out the part orthogonal to $H_{0{\bm k}}$. 

Let $\gamma$ represent the angle between $\hat{\bm k}$ and $\hat{\bm k}'$. The following projections of the scattering term acting on the spin-dependent part of the density matrix are needed
\begin{equation}
\arraycolsep 0.3ex
\begin{array}{rl}
\displaystyle P_\parallel \hat J(S_{{\bm k}\parallel}) = & \displaystyle \frac{kn_i\, \sigma_{{\bm k}\parallel}}{8\hbar \pi A} \int d\theta' \, |\bar{U}_{{\bm k}{\bm k}'}|^2 \, (s_{{\bm k}\parallel} - s_{{\bm k}'\parallel})(1 + \cos\gamma) \\ [3ex]
\displaystyle P_\perp \hat J(S_{{\bm k}\parallel}) = & \displaystyle \frac{kn_i\, \sigma_{{\bm k} \perp} }{8\hbar \pi A} \int\! d\theta' \, |\bar{U}_{{\bm k}{\bm k}'}|^2\, (s_{{\bm k}\parallel} - s_{{\bm k}'\parallel}) \sin\gamma \\ [3ex]
\displaystyle P_\parallel \hat J(S_{{\bm k}\perp}) = & \displaystyle \frac{kn_i \, \sigma_{{\bm k}\parallel}}{8\hbar \pi A} \int\! d\theta' \, |\bar{U}_{{\bm k}{\bm k}'}|^2\, \big(s_{{\bm k}\perp} + s_{{\bm k}'\perp}\big) \sin\gamma.
\end{array}
\end{equation} 
Notice the factors of $(1 + \cos\gamma)$ and $\sin\gamma$ which prohibit backscattering and give rise to Klein tunneling. In fact, the scattering integrals originally contain energy $\delta$-functions $\delta(\varepsilon'_\pm - \varepsilon_\pm) = (1/A) \, \delta(k' - k)$, which have already been integrated over above. The $\delta$-functions of $\varepsilon_-$ are needed in the scattering term in order to ensure agreement with Boltzmann transport. This fact is already seen for a Dirac cone dispersion in graphene where the expression for the conductivity found in Ref.\ \onlinecite{Culcer_Gfn_Transp_PRB08} using the density-matrix formalism agrees with the Boltzmann transport formula of Ref.\ \onlinecite{SDS_Gfn_RMP11} (the definition of $\tau$ differs by a factor of two in these two references.) The necessity of keeping the $\varepsilon_-$ terms is a result of Zitterbewegung: the two branches $\varepsilon_\pm$ are mixed, thus scattering of an electron requires conservation of $\varepsilon_+$ as well as $\varepsilon_-$. 

For ${\bm M} = 0$ the equilibrium density matrix $\rho_{0{\bm k}} = (1/2) \, (f_{0{\bm k}+} + f_{0{\bm k}-}) \, \openone - (1/2)\, (f_{0{\bm k}+} - f_{0{\bm k}-}) \, {\bm\sigma}\cdot\hat{\bm \theta}$. One decomposes the driving term into a scalar part $\mathcal{D}_n$ (not given) and a spin-dependent part $\mathcal{D}_s$. The latter is further decomposed into a part parallel to the Hamiltonian $\mathcal{D}_\parallel$ and a part perpendicular to it $\mathcal{D}_\perp$
\begin{equation}
\arraycolsep 0.3ex
\begin{array}{rl}
\displaystyle \mathcal{D}_{\parallel} = & \displaystyle \frac{1}{2} \, \frac{e{\bm E}\cdot\hat{\bm k}}{\hbar} \, (\pd{f_{0{\bm k}+}}{k} - \pd{f_{0{\bm k}-}}{k}) \, \sigma_{{\bm k}\parallel} = \frac{1}{2} \, d_{{\bm k}\parallel} \, \sigma_{{\bm k}\parallel} \\ [3ex] 
\displaystyle \mathcal{D}_{\perp} = & \displaystyle \frac{1}{2} \, \frac{e{\bm E}\cdot \hat{\bm \theta}}{\hbar k} \, (f_{0{\bm k}+} - f_{0{\bm k}-}) \, \sigma_{{\bm k}\perp} = \frac{1}{2} \, d_{{\bm k}\perp} \, \sigma_{{\bm k}\perp} .
\end{array}
\end{equation}
Equation\ (\ref{eq:Spp}) must be solved perturbatively in the small parameter $\hbar/(\varepsilon_F\tau)$ (which is proportional to the impurity density $n_i$) as described in Ref. \onlinecite{Culcer_TI_Kineq_PRB10}. In transport this expansion starts at order $[\hbar/(\varepsilon_F\tau)]^{-1}$ (i.e. $n_i^{-1}$), reflecting the competition between the driving electric field and impurity scattering resulting in a shift of the Fermi surface. The leading terms in $[\hbar/(\varepsilon_F\tau)]^{-1}$ appear in $n_{\bm k}$ and $S_{{\bm k}, \parallel}$. \cite{Culcer_TI_Kineq_PRB10} To leading order in $\hbar/(\varepsilon_F\tau)$ one finds
\begin{equation}\label{Spar}
\arraycolsep 0.3ex
\begin{array}{rl}
\displaystyle S_{E{\bm k}, \|} = & \displaystyle \frac{\tau\, e \,{\bm E}\cdot\hat{\bm k}}{4\hbar} \, \pd{f_{0+}}{k} \, \sigma_{{\bm k} \|} \\ [2ex]
\displaystyle \frac{1}{\tau} = & \displaystyle \frac{kn_i}{4\hbar A}\int \frac{d\gamma}{2\pi} \, |\bar{U}_{{\bm k}{\bm k}'}|^2\sin^2\gamma.
\end{array} 
\end{equation}
The $\sin^2\gamma$ represents the product $(1 + \cos\gamma)(1 - \cos\gamma)$. The first term in this product is characteristic of TI and ensures backscattering is suppressed, while the second term is characteristic of transport, eliminating the effect of small-angle scattering. The conductivity linear in $\tau$ arises from the parallel part of the density matrix and is given by
\begin{equation}\label{eq:ord}
\sigma^\mathrm{Boltz}_{xx} = \frac{e^2}{h} \, \frac{A \, k_F \, \tau}{4\hbar}.
\end{equation}
We emphasize again that this result tends to the result for graphene\cite{SDS_Gfn_RMP11, Culcer_Gfn_Transp_PRB08} when $D \rightarrow 0$ (note that the definition of $\tau$ in Refs.\ \onlinecite{SDS_Gfn_RMP11, Culcer_Gfn_Transp_PRB08} differs by a factor of 2.) The contribution to the conductivity independent of $\tau$ is is zero for finite $\mu$, which is the case considered here. Thus, the leading order term in the conductivity is $\propto \tau$, and in the limit $\varepsilon_F\tau/\hbar \gg 1$ there is no term of order $(\tau)^0$.

The charge current is proportional to the spin operator, as can be seen from Eq.\ (\ref{j}). Therefore a nonzero steady-state surface charge current automatically translates into a nonzero steady-state surface spin density. The spin density response function can easily be found by simply multiplying the charge current by $\hbar^2/(-2eA)$, yielding a spin density of
\begin{equation}
s_y = -\frac{eE_x}{4\pi} \, \frac{A \, k_F \, \tau}{4}.
\end{equation}
For a sample in which the impurities are located on the surface, with $n/n_i = 0.5$ and $E_x = 25000V/m$, the spin density is $\approx 5 \times 10^{14}$ $spins/m^2$ (where spin $\equiv \hbar$), which corresponds to approximately $10^{-4}$ spins per unit cell area. This number, although small, can be detected experimentally using a surface spin probe such as Kerr rotation. It is also a conservative estimate: for very clean samples, having either a smaller impurity density or impurities located further away from the surface, this number can reach much higher values. Current-induced spin polarization is a definitive signature of two-dimensional surface transport in topological insulators since there is no spin polarization from the bulk: bulk spin densities vanish if the bulk has inversion symmetry, which is the case in the materials Bi$_2$Se$_3$ and Bi$_2$Te$_3$. \cite{Zhang_TI_BandStr_NP09}

I will discuss next implications of these results for transport experiments. \cite{Culcer_TI_Kineq_PRB10} Below I will use the terminology \textit{high density} as meaning that the density is high enough that a scalar term in the Hamiltonian, of the form $Dk^2$, has a noticeable effect (meaning $n \approx 10^{13}$cm$^{-2}$ and higher), with \textit{low density} reserved for situations in which this term is negligible. The conductivity is a function of two main parameters accessible experimentally: the carrier number density and the impurity density/scattering time. The dependence of the conductivity on the carrier number density arises through its direct dependence on $k_F$ and through its dependence on $n$ through $\tau$. In terms of the number density the Fermi wave vector is given by $k_F^2 = 4\pi n$. The number-density dependence of $\tau$ depends on the dominant form of scattering and whether the number density is high or low as defined above. For charged impurity scattering the conductivity contains terms $\propto n$ and $n^{3/2}$ (the latter being due to the scalar term not given in detail here.) At the same time, in a two-dimensional system surface roughness gives rise to short-range scattering as discussed in Ref. \onlinecite{Ando_RMP82}. For short range scattering the two terms in the conductivity are a constant and $n^{1/2}$ (the latter again due to the scalar term.) These results are summarized in Table \ref{tab:scattering}, where I have listed only the number density dependence explicitly, replacing the constants of proportionality by generic constants.

The frequently invoked topological protection of TI surface transport represents protection only against localization by back scattering, not protection against impurity or defect scattering, i.e. against resistive scattering in general. The presence of impurities and defects will certainly lead to scattering of the TI surface carriers and the surface 2D conductivity will be strongly affected by such scattering. If such scattering is strong (which will be a non-universal feature of the sample quality), then the actual surface resistivity will be very high and the associated mobility very low. There is no guarantee or protection against low carrier mobility in the TI surface states whatsoever, and unless one has very clean surfaces, there is little hope of studying surface transport in 2D topological states, notwithstanding their observation in beautiful band structure measurements through ARPES or STM experiments. Strong scattering will lead to low mobility, and the mobility is not a topologically protected quantity. To see surface transport one therefore requires very clean samples.


\begin{table}[tbp] 
  \caption{\label{tab:scattering} Carrier density dependence of the longitudinal conductivity $\sigma_{xx}$. The numbers $a_1 - a_4$ represent constants. The table is adapted from Ref.~\onlinecite{Culcer_TI_Kineq_PRB10}. }
  $\arraycolsep 1em
   \begin{array}{c@{\hspace{2em}}cc} \hline\hline
     & {\rm Screened \,\, charges} & {\rm Short-range} \\ \hline
         {\rm Low \,\, density}   & a_1n & a_3 \\ 
         {\rm High \,\, density}  & a_1n + a_2n^{3/2} & a_3 + a_4 n^{1/2} \\ \hline\hline
  \end{array}$
\end{table}

\begin{figure}[tbp]
\includegraphics[width=\columnwidth]{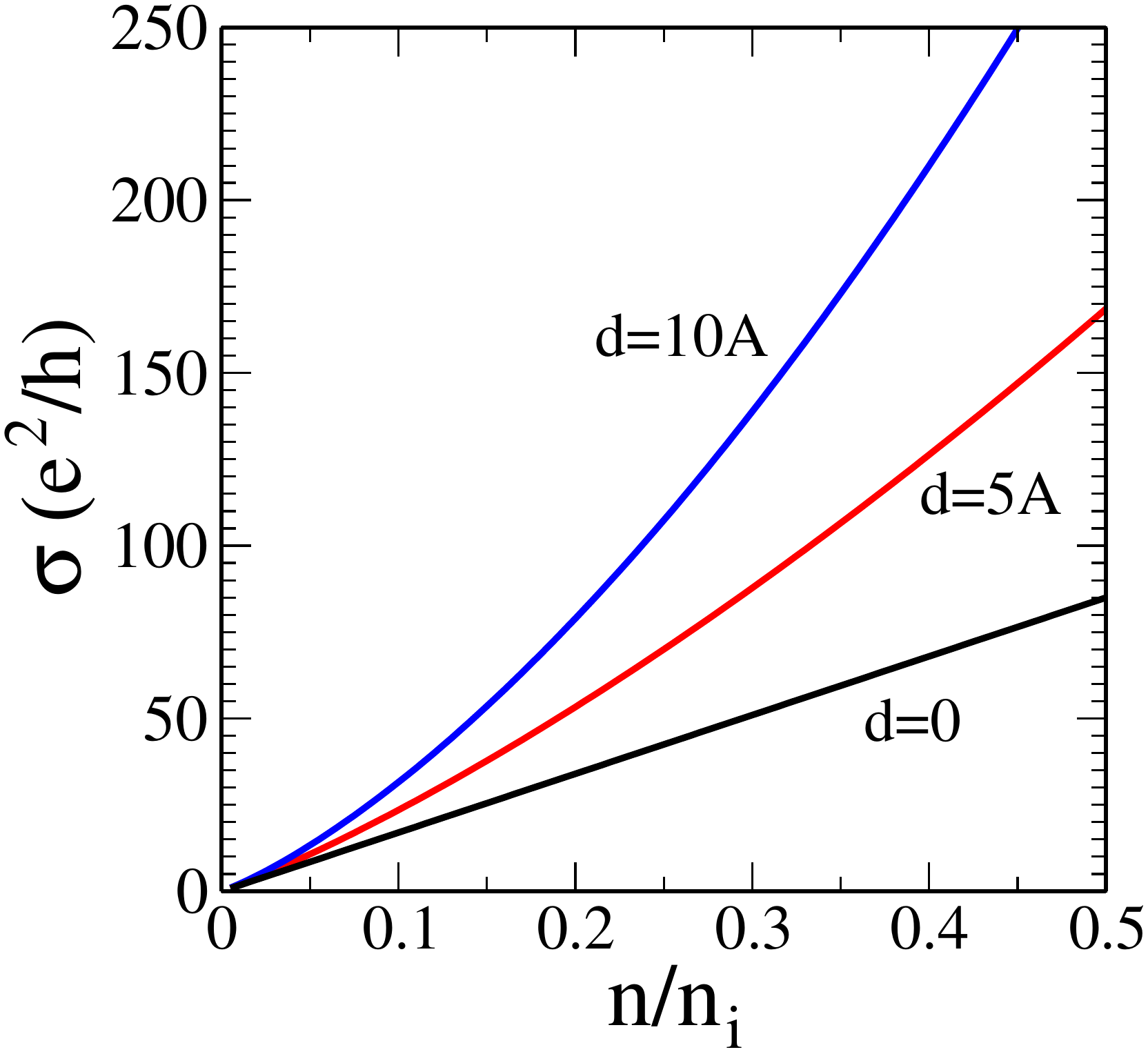}
\caption{\label{sigma}
Calculated conductivity limited by screened charged impurities as
  a function of the surface carrier density for different impurity
  locations $d=0$, 5, 10\AA. In this figure we use the
  following parameters: an impurity density $n_i=10^{13}$ cm$^{-2}$,
  $A=4.1$ eV\AA, which corresponds to the Fermi velocity
  $v_F=6.2\times 10^7$ cm/s, and the static dielectric constant $\epsilon_r=100$. The figure is adapted from Ref.~\onlinecite{Culcer_TI_Kineq_PRB10}.
}
\end{figure}

At carrier densities such that $\varepsilon_F \tau/\hbar \gg 1$ the conductivity is given by $\sigma^{Boltz}_{xx}$ from Eq.\ (\ref{eq:ord}). As the number density is tuned continuously through the Dirac point, $n \rightarrow 0$ and the condition $\varepsilon_F \tau/\hbar \gg 1$ is inevitably violated, and a strong renormalization of the conductivity is expected due to charged impurities.\cite{Adam_Gfn_PNAS07, Rossi_Gfn_RandCrgImp_PRL08, Rossi_Gfn_EffMdm_PRB09} Quite generally, charged impurities give rise to an inhomogeneous Coulomb potential, which is screened by \textit{both} electrons and holes. The net effects of this potential are an inhomogeneity in the carrier density itself and a shift in the Dirac point as a function of position. At high densities the spatial fluctuations in the carrier density are of secondary importance, and do not modify the linear dependence of the conductivity on $n$, yet as the chemical potential approaches the Dirac point, where the average carrier density $\bkt{n} = 0$, these fluctuations play the dominant role in conduction.\cite{Adam_Gfn_PNAS07} At low densities, the carriers cluster into puddles of electrons \textit{and} holes, and a residual density of carriers is always present, making it impossible to reach the Dirac point experimentally. \cite{Adam_Gfn_PNAS07} Consequently the renormalization due to the presence of electron and hole puddles displays a strong sample dependence.\cite{SDS_Gfn_RMP11, Rossi_Gfn_RandCrgImp_PRL08} In addition, exchange and correlation effects make a significant contribution to this minimum conductivity. \cite{Rossi_Gfn_EffMdm_PRB09} An accurate determination of this enhancement for topological insulators requires detailed knowledge of the impurity density distribution, which can only be determined experimentally or modeled numerically by means of e.g. an effective medium theory. \cite{Rossi_Gfn_EffMdm_PRB09} However, a self consistent transport theory provides a physically transparent way to identify the approximate minimum conductivity. The magnitude of voltage fluctuations can be calculated in the random phase approximation, and the result used to determine the residual density and the critical number density at which the transition occurs to the regime of electron and hole puddles, where the carrier density will be highly inhomogeneous. Both the residual density and the critical density are proportional to $n_i$ by a factor of order unity, \cite{Adam_Gfn_PNAS07} and as a first approximation one may take both of them as $\approx n_i$. In view of these observations, the minimum conductivity plateau will be seen approximately at $\sigma_{xx}^{Boltz}$, in which with the carrier density $n$ is replaced by $n_i$. Using the expressions found above for $\sigma_{xx}^{Boltz}$ and $\tau$, this yields for topological insulators
\begin{equation}
\sigma_{xx}^{min} \approx \frac{e^2}{h} \, \bigg(\frac{8}{I_{tc}}\bigg),
\end{equation}
where the dimensionless $I_{tc}$ is determined by $r_s$. \cite{Culcer_TI_Kineq_PRB10} With $r_s$ much smaller than in graphene due to the large dielectric constant, the minimum conductivity may be substantially larger. \cite{Rossi_Gfn_EffMdm_PRB09} (Note that for a Rashba-Dirac cone the definition of $r_s$ contains some arbitrariness.\cite{SDS_Gfn_RMP11})

\begin{figure}[tbp]
\begin{center}
\includegraphics[width=0.45\textwidth]{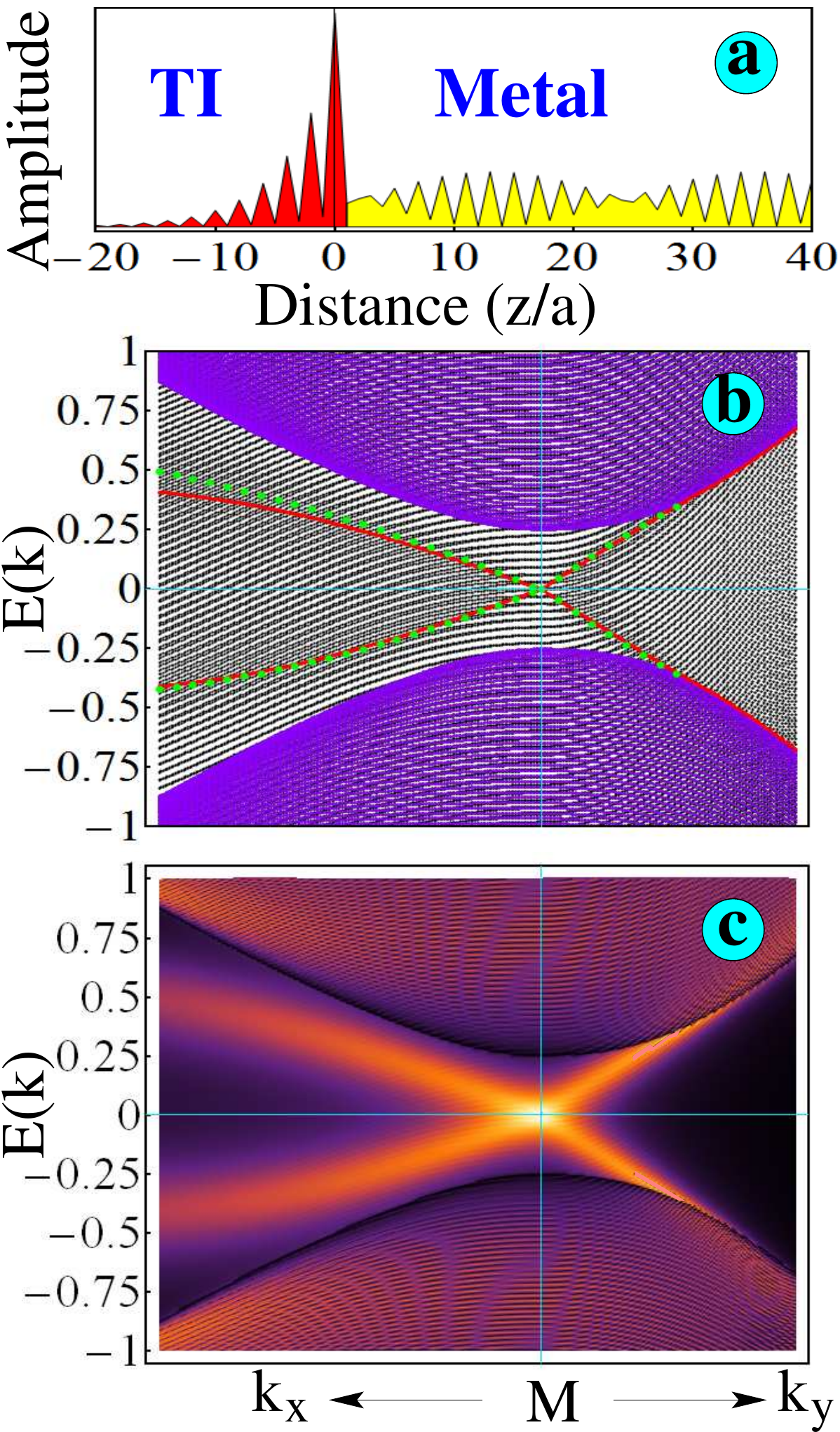}
\vspace{-3mm}
\end{center}
\caption{(a) Amplitude of a metallic state hybridized with a topological insulator surface state as function of distance from the interface (in units of interlayer spacing). (b) Spectrum of a topological insulator in contact with a metal (black). The states of a topological insulator with a free surface are also shown (purple - bulk states, red - surface states). (c) Density plot of the total amplitude within a thin region of topological insulator in the vicinity of the interface (red/dark grey area in panel a). Note that, instead of a well-defined surface mode, one has a diffuse distribution of states with boundary contributions. The dispersion of the maxima of this distribution is represented by green points in panel (b). The figure is adapted from Ref.~\onlinecite{Culcer_TI_Kineq_PRB10}.} 
\label{FigContacts}
\end{figure}

Transport measurements require the addition of metallic contacts on the surface of the topological insulator. Since the properties of the surface states depend crucially on the boundary conditions, the natural question is how these properties are affected by the metallic contacts. To address this question numerical calculations for the minimal tight-binding model studied in Ref. \onlinecite{Stanescu_TI_Proximity_PRB10} were performed, focusing on the case of the case of large area contacts. \cite{Culcer_TI_Kineq_PRB10} When the TI surface is in contact with a metal, the surface states penetrate inside the metal and hybridize with the metallic states. A typical hybridized state is shown in Fig. \ref{FigContacts}a. The amplitude of this state near the boundary of the topological insulator is reduced by a factor $1/L_z^{(m)}$ with respect to the amplitude of a pure surface state with the same energy, where $L_z^{(m)}$ is the width of the metal in the direction perpendicular to the interface. However, the local density of states near the boundary is not reduced, as the number of hybridized states also scales with $L_z^{(m)}$. The spectrum of a topological insulator in contact with a metal is shown in Fig. \ref{FigContacts}b. For each state the total amplitude within a thin layer of topological insulator in the vicinity of the boundary was calculated, shown in Fig. \ref{FigContacts}c. Instead of the sharply defined Dirac cone that characterizes the free surface, one has a diffuse distribution of states with boundary contributions. The properties of the interface amplitude distribution shown in Fig. \ref{FigContacts}c, i.e., its width and the dispersion of its maximum, depend on the strength of the coupling between the metal and the topological insulator. If the distribution is sharp enough and the dispersion does not deviate significantly from the Dirac cone, the transport analysis presented above should be applied using bare parameters (i.e. parameters characterizing the spectrum of a free surface.) Nonetheless, significant deviations from the free surface dispersion are possible. For example, the location of the Dirac point in the model calculation considered here is fixed by symmetry. However, in real systems the energy of the Dirac point can be easily modified by changing the boundary conditions. Consequently, the effective parameters (e.g. $A \rightarrow A_{\rm eff}$) entering transport coefficients may differ significantly from the corresponding parameters extracted from ARPES measurements. 

In topological insulator thin films tunneling may be possible between the top and bottom surfaces. Evidently inclusion of this physics requires one to start with a $4 \times 4$ Hamiltonian, rather than the $2 \times 2$ model discussed so far. Interlayer hybridization induces a gap in the surface states of thin films, as was discussed in Ref. \onlinecite{Lu_TI_Film_PRB10}. Once the gap has been taken into account, the theory can be reduced to an effective $2 \times 2$ Hamiltonian for each surface, each of which contains a $\sigma_z$ term due to interlayer tunneling. This mass term turns the original Dirac cones of the surface states into two massive parabolas. Twofold Kramers degeneracy is preserved since time-reversal symmetry is not broken. Consequently, the parabolas are indexed by $\tau_z = \pm 1$, describing two sets of Dirac fermions (rather than two independent surfaces). As a result of the mass term the Berry curvature of the surface states is finite, as is the orbital magnetic moment arising from the self-rotation of Bloch electron wave packets. The mass term oscillates as a function of film thickness and changes sign at critical thicknesses, where a topological phase transition occurs due to a discontinuous change in Chern number, in the same fashion as in the 2D quantum spin-Hall insulator.

\begin{figure}[tbp]
\bigskip
\includegraphics[width=\columnwidth]{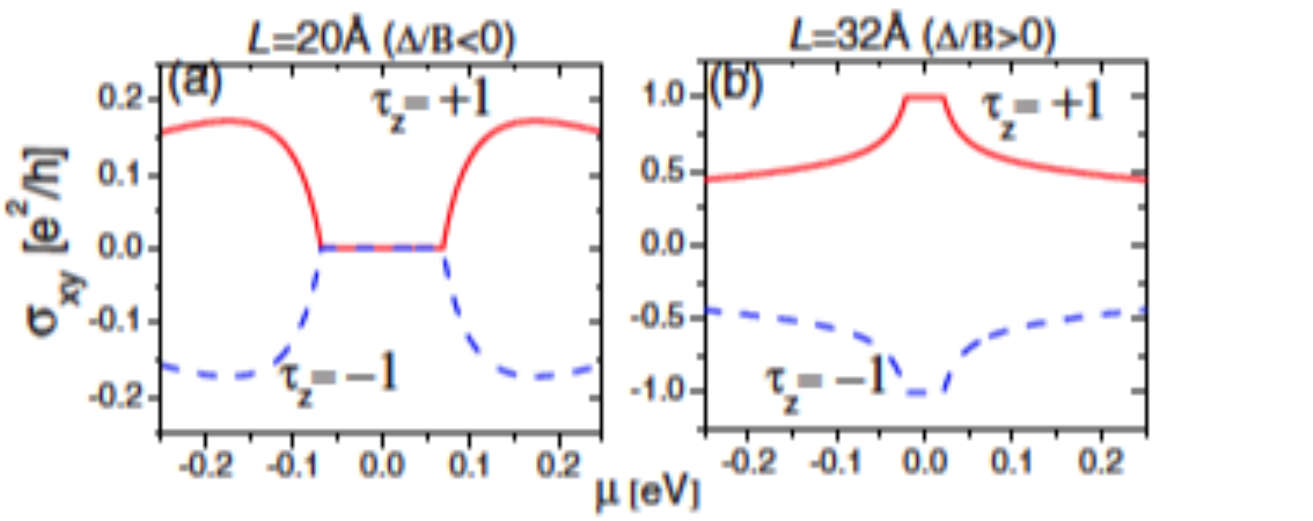}
\caption{\label{Lu_Hall}
Contributions to the Hall conductivity of a TI thin film due to the orbital magnetic moment plotted as a function of Fermi level. The two plots are for film thicknesses of (a) $20\AA$ and (b) $32\AA$ respectively. This effect can be understood as a spin-Hall effect in much the same manner as the 2D quantum spin-Hall insulator, and is accompanied by a topological phase transition as the film thickness is changed. This figure has been adapted from Ref.~\onlinecite{Lu_TI_Film_PRB10}.}
\end{figure}

Li \textit{et al.} \cite{Qiuzi_TI_SDW_AnisoTrans_PRB11} studied transport in topological insulators in the neighborhood of a helical spin density wave. The nontrivial magnetic structure results in a spatially localized breakdown of time reversal symmetry on the TI surface. The helical spin density wave has the same effect as an external one-dimensional periodic magnetic potential (of period $L$) for spins on the surface of the TI, with an additional term in the Hamiltonian 
\begin{equation}
U (x) = U_y \sigma_y \cos\bigg(\frac{2\pi x}{L}\bigg) + U_z \sigma_z \sin\bigg(\frac{2\pi x}{L}\bigg).
\end{equation}
As expected the Dirac cone of the TI surfaces becomes highly anisotropic, and, as a result, transport due to the topological surface states displays a strong anisotropy. The helical spin density wave on the TI surface leads to striking anisotropy of the Dirac cones and group velocity of the surface states. The decrease in group velocity along the direction defined by the one-dimensional periodic magnetic potential is twice as large as that perpendicular to this direction. More importantly, the researchers found that at the Brillouin zone boundaries, the periodic magnetic structure gives rise to new semi-Dirac points that have a linear dispersion along the direction defined by the magnetic potential but a quadratic dispersion perpendicular to this direction. The group velocity of electrons at these new semi-Dirac points is predictably also highly anisotropic. The authors suggest that such a structure could be realized in a topological insulator grown on a multiferroic substrate. It could be the direct manifestation of the elusive 2D surface TI transport even in the presence of considerable bulk conduction, since the bulk transport is isotropic and presumably unaffected by the presence of the helical spin density wave.

\begin{figure}[tbp]
\bigskip
\includegraphics[width=\columnwidth]{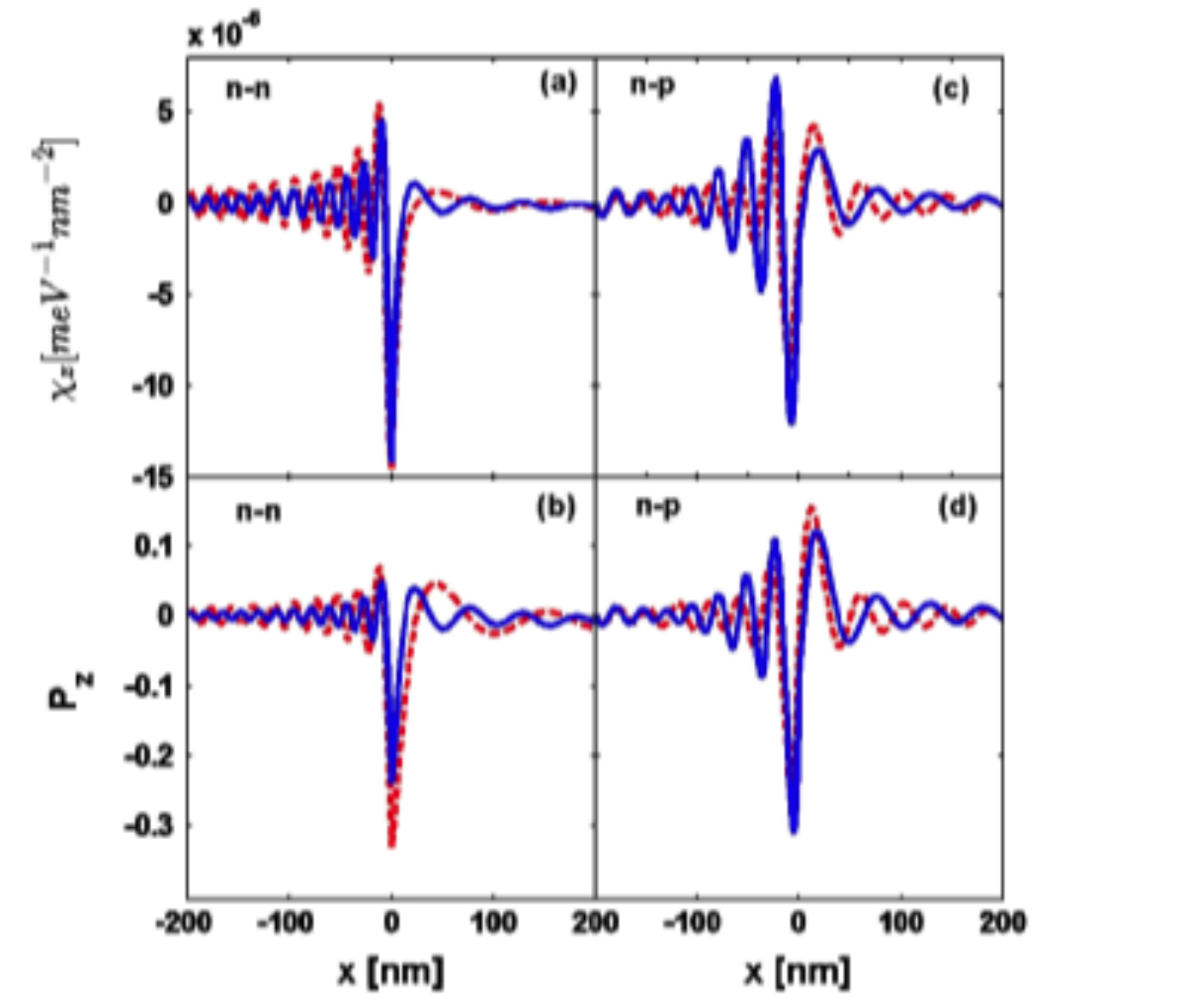}
\caption{\label{Gao_SHE}
Spatial distribution of local spin density $S_z$ for a single Dirac fermion with a Fermi velocity of $5 \times 10^5$ms$^{-1}$. (a) $n-n$ junction, $V_2 = 40$ meV, and the incoming electron energy is 60 meV. The angle of incidence $\theta = 63\deg$ [(blue) solid line] and $\theta = 18\deg$ [(red) dashed line]. (b) $n-p$ junction, $V_2 = 60$ meV and the incoming electron energy is 35 meV. The angle of incidence $\theta = 63\deg$ [(blue) solid line] and $\theta = 36 \deg$ [(red) dashed line]. This figure has been adapted from Ref.~ \onlinecite{Gao_TI_GiantSHE_PRL11}.}
\end{figure}

A fascinating publication recently predicted a giant polarization due to the topological surface states of Bi$_2$Se$_3$. \cite{Gao_TI_GiantSHE_PRL11} The structure proposed by the authors envisages a TI thin film to which a step-function potential is applied by means of a back gate, given by $V(x) = V_2 \Theta(x)$. A voltage $V_y$ across the surface drives a charge current. The charge current in turn generates a spin-Hall current along the $\hat{\bm x}$-direction, which strikes the step-potential boundary at $x = 0$. Near the potential boundary at $x = 0$, a spin-$z$ polarization ($20\%$) exists along the $\hat{\bm x}$-direction, induced by an electric current along the $\hat{\bm y}$-direction in the ballistic (disorder-free) transport regime at zero temperature. Thanks to the Klein paradox, electrons with energies less than $V_2$ are not confined. However, the incident electrons are no longer in eigenstates on the $x > 0$ side of the potential boundary. The spin polarization oscillates across the potential boundary with a period given by $1/k_F$, reflecting the interference of electron and hole states. This is shown in Fig.~\ref{Gao_SHE}. Note that the plot represents the spin polarization of a \textit{single} electron. The local spin density has a strong dependence on the angle of incidence of the electrons. The induced spin polarization is found to be insensitive to the surface $\varepsilon_F$, because the spin polarization is approximately inversely proportional to the Fermi velocity, which is a constant for Dirac electrons. The authors expect the effect to be observable in a spin resolved STM.

Still on the topic of spin transport, Dora and Moessner studied the spin-Hall effect of massive Dirac fermions in topological insulators in \textit{strong} electric fields, near the threshold of electrical breakdown. \cite{Dora_TI_Gfn_Hall_PRB11} It is well known that a single Dirac cone with a mass term gives rise to a quantized spin-Hall effect, as discussed above in the case of Ref.~\onlinecite{Lu_TI_Film_PRB10}. Dora and Moessner consider switching on an electric field, following which a steady-state Hall current is reached via damped oscillations of the transient components. They find that the spin-Hall conductivity remains quantized as long as the electric field does not induce Landau-Zener transitions. However, as the Landau-Zener threshold is approached. the spin-Hall conductance quantization breaks down, and the conductivity decreases as $1/\sqrt{E}$. In other words, electrical breakdown affects not only longitudinal transport, but also transverse transport.

\begin{figure}[tbp]
\bigskip
\includegraphics[width=\columnwidth]{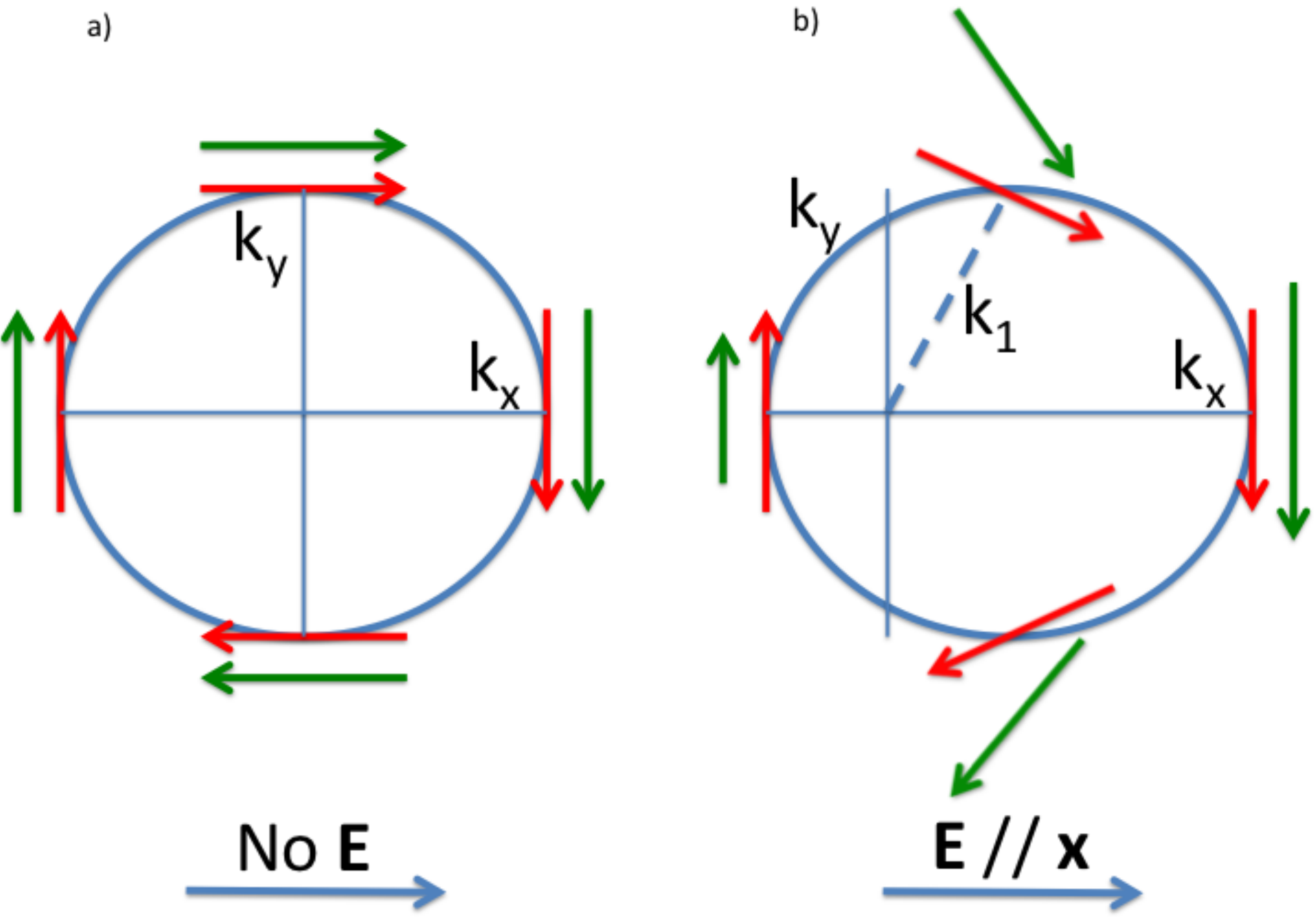}
\caption{\label{ee_SpinPol} Spin polarization in interacting TI. At each ${\bm k}$, red arrows represent the effective spin-orbit field ${\bm \Omega}_{\bm k}$ and green arrows the spin. (a) Equilibrium non-interacting case. The spin follows the local ${\bm \Omega}_{\bm k}$ and the spin polarization averages to zero. Out of equilibrium (not shown) the spin at ${\bm k}$ also follows ${\bm \Omega}_{\bm k}$, resulting in a net spin polarization. (b) Non-equilibrium interacting case. Interactions tilt the spin away from ${\bm \Omega}_{\bm k}$ (here the tilt $ \parallel \hat{\bm y}$) and enhance the non-equilibrium spin polarization. Scattering from ${\bm k} = {\bm k}_1$ to ${\bm k} = k \hat{\bm x}$ further increases the \textit{local} (i.e. ${\bm k}$-dependent) spin polarization, while scattering from ${\bm k} = {\bm k}_1$ to ${\bm k} = - k \hat{\bm x}$ decreases it. In an electric field the asymmetry of the Fermi surface means that the net effect of scattering is to increase the spin polarization. This figure is adapted from Ref.~\onlinecite{Culcer_TI_Int_11}.}
\end{figure}
 
The topological order present in TI is a result of one-particle physics. Interaction effects have been studied in TI with a focus on phenomena in equilibrium and in the quantum Hall regime. A fundamental question is whether basic TI phenomenology survives interactions \textit{out of equilibrium} (i.e. in an electric field), since in transport topology only protects against \textit{back}-scattering. A second fundamental question concerns the non-equilibrium spin polarization due to spin-momentum locking in an electric field. In interacting TI, the mean-field Hamiltonian is spin-dependent, and it is natural to ask whether a Stoner criterion exists for the non-equilibrium spin polarization. A systematic understanding of electron-electron interactions in \textit{non-equilibrium} TI remains to be constructed. It was demonstrated recently \cite{Culcer_TI_Int_11} that many-body interactions enhance the charge conductivity and non-equilibrium spin polarization of TI, and that the non-equilibrium enhancement is intimately linked to Zitterbewegung and not divergent under any circumstances. Although disorder renormalizations give a nontrivial doubling of the enhancement, the equivalent of the Stoner criterion is never fulfilled for TI. These findings can be understood by means of the following physical picture, illustrated in Fig.~\ref{ee_SpinPol}. The enhancement of the charge conductivity and non-equilibrium spin polarization reflects the interplay of spin-momentum locking and many-body correlations. A spin at ${\bm k}$ feels the effect of two competing interactions. The Coulomb interaction between Bloch electrons with ${\bm k}$ and ${\bm k}'$ tends to align a spin at ${\bm k}$ with the spin at ${\bm k}'$, equivalent to a $\hat{\bm z}$-rotation. The total mean-field interaction tends to align the spin at ${\bm k}$ with the sum of all spins at all ${\bm k}'$, and the effective (Hartree-Fock) Hamiltonian describing electron-electron interactions encapsulates the amount by which the spin at ${\bm k}$ is tilted as a result of the mean-field interaction with all other spins on the Fermi surface. The effective field ${\bm \Omega}_{\bm k}$ tends to align the spin with itself. As a result of this latter fact, out of equilibrium, an electrically-induced spin polarization is already found in the non-interacting system, \cite{Culcer_TI_Kineq_PRB10} as discussed extensively above. Interactions tend to align electron spins with the existing polarization. The effective $\hat{\bm z}$-rotation explains the counterintuitive observation that the enhancement is related to Zitterbewegung, originating as it does in $S_{{\bm k}\perp}$. Many-body interactions give an effective ${\bm k}$-dependent magnetic field $\parallel\hat{\bm z}$, such that for ${\bm E} \parallel \hat{\bm x}$ the spins $s_y$ and $-s_y$ are rotated in opposite directions, reinforcing each other. Due to spin-momentum locking, a tilt in the spin becomes a tilt in the wave vector, increasing the conductivity: spin dynamics create a feedback effect on charge transport. This feedback effect is even clearer in the fact that scattering doubles the enhancement, a feature that was demonstrated rigorously using the density matrix formalism. This doubling is valid for \textit{any} elastic scattering. Take ${\bm k} = \hat{\bm k}_1$ in Fig.~\ref{ee_SpinPol}. Scattering from ${\bm k} = \hat{\bm k}_1$ to ${\bm k} = k \hat{\bm x}$ increases the spin polarization at that point on the Fermi surface, while scattering from ${\bm k} = \hat{\bm k}_1$ to ${\bm k} = - k \hat{\bm x}$ decreases the spin polarization at that point. Because the Fermi surface is shifted by the electric field, the net effect of this mechanism is to increase the spin polarization.

This concludes the discussion of the theory of non-magnetic surface transport in topological insulators. Aside from the predictions of some fascinating effects, the central message of this section is that \textit{three} primary signatures of surface transport may be identified: (a) a non-universal minimum conductivity as in graphene, due to the formation of electron and hole puddles as the Dirac point is approached (b) a current-induced spin polarization and (c) a thickness independent conductance would be a likely indication of TI surface transport. Although not elaborated upon, the latter fact is obvious, since the number of surface carriers does not change with film thickness, wheres the number of bulk carriers does.

\section{Magnetotransport theory}

Both in theory and in experiment, considerably more attention has been devoted to magnetotransport than to non-magnetic transport. The primary reason for this focus is that specific signatures of surface transport can be extracted without having to tune through the Dirac point or measure the spin polarization of a current. In Shubnikov-de Haas oscillations and in the anomalous Hall effect, both discussed below, qualitatively different behavior is expected from the bulk and surface states. 


Broadly speaking, magnetotransport encompasses two categories: transport in magnetic fields and transport in the presence of a magnetization. A magnetic field couples both to the orbital degree of freedom, giving rise to the Lorentz force and cyclotron orbits, and to the spin degree of freedom via the Zeeman interaction. A magnetization couples only to the spin degree of freedom. The electrical response in the presence of a magnetic field and that in the presence of a magnetization do share common terms, yet they ultimately contain different contributions and are best treated separately. I will concentrate first on transport in a magnetic field, then on transport in the presence of a magnetization. Well-known aspects that have been covered extensively before, primarily the topological magnetoelectric effect, \cite{Hasan_TI_RMP10, Qi_TI_RMP_10, Qi_TFT_PRB08} will not be covered here.

\subsection{Magnetic fields}

In a magnetic field ${\bm B}$ the energy spectra of both the bulk and the surface consist of Landau levels. The cyclotron frequency $\omega_c = eB/m^*$, with $m^*$ the effective mass, is conventionally used to differentiate weak and strong magnetic fields. In a weak magnetic field $\omega_c \tau \ll 1$, while in a strong magnetic field $\omega_c \tau \gg 1$, with $\tau$ the momentum relaxation time as in the previous section. In the bulk case, with ${\bm B} \parallel \hat{\bm z}$, the diagonal conductivity $\sigma_{xx}$ contains a classical contribution as well as an oscillatory term periodic in $1/B$, referred to as Shubnikov-de Haas (SdH) oscillations. Unlike non-magnetic transport, the classical term represents \textit{scattering-assisted} transport, \cite{Vasko} which can be interpreted as the hopping of electrons between the centers of the classical cyclotron orbits. In both weak and strong magnetic fields this contribution is $\propto 1/\tau$. For details of its form in different regimes the reader is referred to the comprehensive discussion in Ref.~\onlinecite{Vasko}. The oscillatory contribution to the diagonal conductivity reflects the fact that consecutive Landau levels pass through the Fermi energy as the magnetic field is swept. The oscillatory contribution is present even at small magnetic fields, that is, in the ordinary Hall effect. The corresponding oscillations in the magnetization are termed the de Haas-van Alphen (dHvA) effect. In the bulk in a magnetic field only the ordinary Hall effect will occur, and experimentally one expects SdH oscillations for all orientations of the magnetic field.

In a strong perpendicular magnetic field in 2D systems the quantum Hall effect is observed. The diagonal conductivity $\sigma_{xx}$ consists of a series of peaks, while the Hall conductivity $\sigma_{xy}$ consists of a number of plateux. The Landau levels corresponding to the topological surface states are effectively the same as in graphene and the theory will not be developed here, primarily because it is either obvious or it has been covered elsewhere. \cite{Hasan_TI_RMP10, Qi_TI_RMP_10} Neglecting the Zeeman splitting, the Landau level energies indexed by $n$ are given by the formula
\begin{equation}
\varepsilon_n = \varepsilon_0 + {\rm sgn} (n) \, \frac{A}{\hbar}\, \sqrt{2\hbar eB|n|},
\end{equation}
with $\varepsilon_0$ the energy at the Dirac point, which is independent of the magnetic field for massless Rashba-Dirac fermions. The Landau levels show the $\sqrt{|n|B}$ dependence on the magnetic field and Landau level index characteristic of the linear Rashba-Dirac cone dispersion. In two-dimensional systems the considerations above for $\sigma_{xx}$ remain valid, while in a large perpendicular magnetic field the quantization of $\sigma_{xy}$ becomes apparent. The quantized Hall conductivity is given by 
\begin{equation}
\sigma_{xy} = \bigg(n + \frac{1}{2}\bigg) \, \frac{e^2}{h}.
\end{equation}
The half-quantization is due to the $\pi$ Berry phase, which is also present in the scattering term in non-magnetic transport. The half-quantization of the QHE can also be understood in terms of a term in the bulk Lagrangian $\propto {\bm E} \cdot {\bm B}$. Shubnikov-de Haas oscillations in $\sigma_{xx}$ are described by the same formalism as in graphene. Only the levels at the Fermi energy at one particular time contribute to the diagonal conductivity $\sigma_{xx}$, yet the Hall conductivity $\sigma_{xy}$ has contributions from all levels below the Fermi energy. 


In a magnetic field of several Tesla, if the magnetotransport signal can be ascribed to the topological surface states, one thus expects to observe the quantum Hall effect, and expects to see SdH oscillations only for a magnetic field normal to the TI surface. An in-plane magnetic field would not yield SdH oscillations. In addition, due to the different effective mass, the frequency of SdH oscillations for the bulk is considerably different from that for the surface. Specifically, the cyclotron frequency $\omega_c = eB/m^*$ is a constant for bulk transport. For surface transport one may define a cyclotron effective mass $m^* = \hbar^2 k_F/A$, which depends on the Fermi wave vector, given by $k_F = \sqrt{4\pi n}$. In the case of coexisting bulk and surface transport one therefore expects two sets of SdH oscillations with different frequencies, as well as angular dependence of the SdH oscillations, both of which should be clearly distinguishable experimentally.

Shubnikov-de Haas and de Haas-van Alphen oscillations in Bi$_2$Se$_3$ were studied theoretically in Ref.\ \onlinecite{Wang_TI_Shub_PRB10}. The authors determine the band structure from first principles, then calculate the magnetization induced by an external magnetic field. Depending on whether the zero-mode Landau level is occupied or empty, the location of center of the dHvA oscillations in the magnetization changes. Interestingly, the magnetic oscillation patterns for the cases of filled and half-filled $n=0$ Landau level are out phase.

Chu \textit{et al.} \cite{Chu_TI_HalfHall_10} studied numerically the transport properties of the topological surface states in a Zeeman field. The main argument of their work is that a chiral \textit{edge} state forms on the surface, but is split into two spatially-separated halves. Each half carries one half of the conductance quantum. They propose a four-terminal setup in which this theory can be tested. Numerical simulations reveal that the difference between the clockwise and counterclockwise transmission coefficients of two neighboring terminals is $\approx 1/2$. This suggests a half-quantized ordinary Hall conductance could be measured in a four-terminal experiment.

Zhang \textit{et al.}\cite{Zhang_TI_B_QHE_11} set out to explain why the half-quantized QHE has not been observed in experiments to date. The authors investigated a 3DTI in a uniform magnetic field using tight binding model.They point out that all surface states of a finite 3DTI are connected to each other. In a finite 3D TI, the two surfaces cannot simply be separated by the bulk as in usual 2D quantum Hall systems. As a result, surface states may move from one side of the sample to another. Hall voltage measurements for 3DTI are therefore necessarily ambiguous. The Hall current is due to 2D surface states, rather than 1D edge states. The structures of these side surface states are more complicated than the edge states in 2DEG. Quantum Hall plateaux of $(2n + 1) \, e^2/h$ can only be defined from transverse current instead of the Hall voltage. The tight-binding calculations are in agreement with earlier predictions based on an effective theory of Dirac fermion in curved 2D spaces, \cite{Lee_TI_Curved_PRL09} which showed that, based on the argument above that the current is carried by 2D rather than 1D states, the robustness of the QH conductance against impurity scattering is determined by the oddness and evenness of the Dirac cone number. 

\begin{figure}[tbp]
\bigskip
\includegraphics[width=\columnwidth]{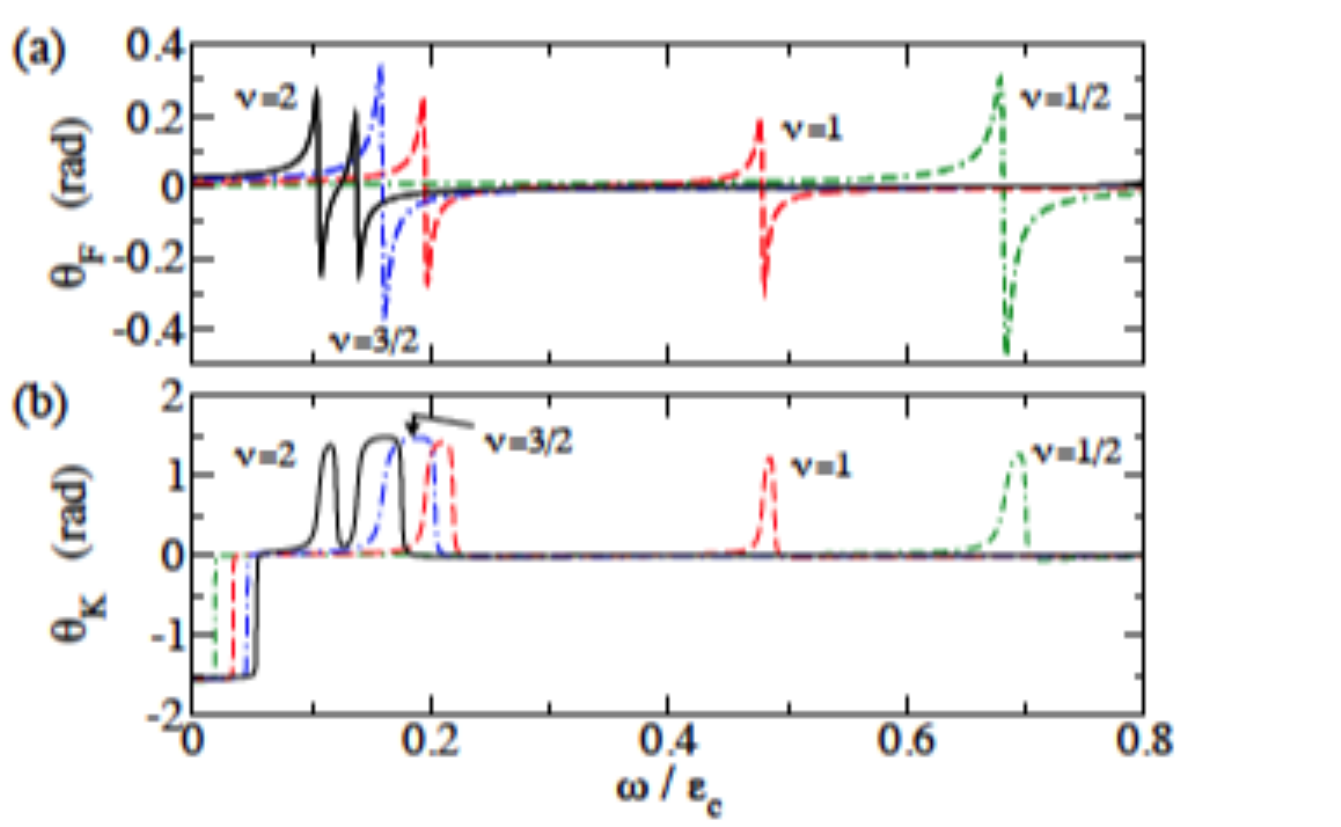}
\caption{\label{Tse_MOKE} (a) Faraday rotation angle versus frequency $\omega/(\varepsilon c)$ at equal filling factors on both surfaces for a 30nm-thick Bi$_2$Se$_3$ film. The densities on both surfaces are $5 \times 10^{11}$cm$^{-2}$. The bulk band gap is 0.35 eV, the Fermi velocity is $v_F = 5 \times 10^5$ms$^{-1}$ and the dielectric constant $ \epsilon_r = 29$. (b). Kerr rotation angle versus frequency for the same parameters. This figure is adapted from Ref.~\onlinecite{Tse_TI_QuantB_MOKE_PRB10}.}
\end{figure}

Tse and MacDonald \cite{Tse_TI_QuantB_MOKE_PRB10} developed a theory of the magneto-optical and magneto-electric response of a topological insulator thin film in a strong (quantizing) magnetic field. Their work was driven by the realization that the quantum Hall effect due to the topological surface states could be observed optically, even when bulk carriers are present in the system and contribute to conduction. Interestingly, these authors showed that the low-frequency magneto-optical properties depend only on the sum of the Dirac-cone filling factors of the top and bottom surfaces. In contrast, the low-frequency magneto-electric response is determined by the difference in these filling factors. Furthermore, the Faraday rotation angle is quantized, being given by the sum of the filling factors of the top and bottom surfaces multiplied by the fine structure constant. The Faraday rotation angle exhibits sharp cyclotron resonance peaks and changes sign near each allowed transition frequency, as shown in Fig.~\ref{Tse_MOKE}. At half-odd-integer filling factors a single dipole-allowed intraband transition exists. At other filling factors two resonances are allows at different frequencies associated with transitions into and out of the partially filled Landau level. This behavior is completely different from usual 2DEGs, in which all dipole-allowed transitions have the same energy. The Kerr rotation angle is $\pi/2$, as found by the same authors in the case of magnetically-doped TI (see below.) It is strongly enhanced at the frequencies at which cyclotron resonance occurs. Unlike the low-frequency giant Kerr effect, the resonant peaks corresponding to absorptive Landau level transitions are non-universal.

Tkachov and Hankiewicz \cite{Tkachov_Galvano_PRB11} studied the quantum Hall effect due to the topological surface states in a TI film of thickness $d$ in the situation in which bulk carriers also contribute via the ordinary Hall effect. The authors took into account the contributions of the bulk as well as the top and bottom surfaces, showing that the contribution to the conductivity due the bulk carriers can be suppressed by an external magnetic field and an ac-electric field. They work in the regime $\omega, \omega_c \ll \Delta$, where $\omega$ is the frequency of the applied electric field, and $\Delta$ is the Zeeman-energy gap in the Dirac cone. Under these conditions, the cyclotron resonance frequency is shifted by a thickness-dependent term,
\begin{equation}
\omega_c \rightarrow \omega_c + \frac{d\sigma_{xx}^{bulk}}{\tau |\sigma_{xy}^{sfc}|},
\end{equation} 
where $\sigma_{xx}^{bulk}$ is the bulk dc conductivity at zero magnetic field, and $\sigma_{xy}^{sfc}$ is the surface Hall conductivity. The nonzero surface conductivity alters the cyclotron drift in the direction of the current, and induces a linear negative magnetoresistivity. At the same time, the Hall angle depends on the magnetic field quadratically

Zyuzin \textit{et al.} \cite{Zyuzin_TI_ThinPerp_PRB11} investigated a thin TI film in a strong perpendicular magnetic field, allowing for hybridization between the top and bottom surfaces. The authors determined the Landau-level spectrum of the film as a function of the applied magnetic field and the magnitude of the hybridization matrix element, taking into account both the orbital and the Zeeman spin splitting effects of the field. Most interestingly, they identified a quantum phase transition between a state with a zero Hall conductivity and a state with a quantized Hall conductivity of $e^2/h$ as a function of the magnitude of the applied field. The transition is driven by the competition between the Zeeman and the hybridization energies.

\subsection{Magnetizations}

Doping with magnetic impurities, whose moments align beyond a certain doping concentration, makes topological insulators magnetic and opens a gap. Topological insulators in the presence of a strong out-of-plane magnetization exhibit rather different physics than their counterparts placed in magnetic fields.

The interplay of magnetism, strong spin-orbit coupling, disorder scattering and driving electric fields leads to the anomalous Hall effect in magnetic TIs. The anomalous Hall effect has been a mainstay of condensed matter physics, studied extensively over the years. \cite{Nagaosa_AHE_PRB08} A recent groundbreaking paper by Yu \textit{et al.} \cite{Yu_TI_QuantAHE_Science10} studied tetradymite semiconductors doped with Cr and Fe using first principles calculations, determining a critical temperature as high as 70K. Yu \textit{et al.} further demonstrated that in a thin TI film near a band inversion, in which the chemical potential lies in the magnetization-induced gap between the surface valence and conduction bands, edge states give rise to an integer quantized anomalous Hall effect. \cite{Yu_TI_QuantAHE_Science10} The quantized anomalous Hall effect requires a band inversion, provided by spin-orbit coupling, and ferromagnetic order, provided by doping. The $4 \times 4$ Hamiltonian describing the surface states can be broken up into two blocks, and a sufficiently large exchange field gives rise to a difference in Chern number of one between the two blocks, resulting in an anomalous Hall conductivity of $e^2/h$. Remarkably, this is true regardless of whether the system is in the topologically trivial phase or in the topologically insulating phase. The quantized anomalous Hall effect is to be distinguished from the anomalous quantum Hall effect, which has also been predicted to occur in topological insulators. \cite{Hasan_TI_RMP10}

In contrast to the 2D case, in 3D magnetic TIs with the chemical potential in the band gap the anomalous Hall conductivity due to one surface is \textit{half-quantized} $e^2/2h$, \cite{ZangNagaosa_TI_Monopole_PRB10, Nomura_3DTI_QAHE_PRL11} rather than the integer quantization seen for 2D edge states. The value of $e^2/2h$ found for $\varepsilon_F$ in gap was demonstrated to be robust against disorder. \cite{Nomura_3DTI_QAHE_PRL11} Ref.~\onlinecite{Culcer_TI_AHE_PRB11} studied the anomalous Hall effect due to the surface states of 3D topological insulators in which $\varepsilon_F$ lies in the surface conduction band, showing that the topological band structure contribution ($e^2/h$) found in previous studies, with a substantial renormalization due to disorder, yields the dominant term, overwhelming \textit{all} contributions due to skew scattering and side jump. The end result is of the order of the conductivity quantum and independent of the magnetization. These results are explained in what follows.

Theoretically, the already challenging problem of the anomalous Hall effect in 3D TI is complicated by the necessity of a quantum mechanical starting point and of a matrix formulation describing strong band structure spin-orbit coupling, spin-dependent scattering including relativistic corrections, Klein tunneling and Zitterbewegung on the same footing. The density matrix-Liouville equation formalism has a quantum mechanical foundation and accounts for the three mechanisms known to be important in the anomalous Hall effect. The interplay of band structure spin-orbit interaction, adiabatic change in ${\bm k}$ in the external electric field, and out-of-plane magnetization leads to a sideways displacement of carriers even in the absence of scattering (\textit{intrinsic}). \cite{Luttinger_AHE_PR58} The spin dependence of the impurity potentials via exchange and extrinsic spin-orbit coupling causes asymmetric scattering of up and down spins (skew scattering), \cite{Smit_SS_58} and a sideways displacement during scattering (side jump). \cite{Berger_SJ_PRB1970} 

In magnetically doped TI  the magnetization ${\bm M} \ne 0$ and the kinetic equation can be written in the form \cite{Culcer_TI_AHE_PRB11}
\begin{equation}
\label{fmag} 
\td{f_{{\bm k}}}{t} + \frac{i}{\hbar} \, [H_{m{\bm k}}, f_{{\bm k}}] + \hat{J}(f_{\bm k}) = \mathcal{D}_{\bm k}, 
\end{equation}
with $H_{m{\bm k}}$ given by Eq.\ (\ref{magHam}). All extrinsic spin-orbit terms $\propto \lambda$ must be taken into account, including the second Born approximation scattering term $\hat{J}^{3rd}(f_{\bm k})$ of Eq.\ \ref{Jss3rd}, the spin-dependent correction to the position operator $\propto \lambda$ and the corresponding modification of the interaction with the electric field. Consequently, the driving term $\mathcal{D}_{\bm k}$ contains a host of contributions not found in the non-magnetic case. It is given in detail in Ref.~\onlinecite{Culcer_TI_AHE_PRB11}. At zero temperature the splitting due to ${\bm M}$ is resolved, and the Fermi energy $\varepsilon_F \gg |{\bm M}|$ as usual lies in the \textit{surface} conduction band. Three orthogonal directions are identified in reciprocal space, 
\begin{equation}
\arraycolsep 0.3ex
\begin{array}{rl}
\displaystyle \hat{\bm \Omega}_{\bm k} = & \displaystyle - a_k \, \hat{\bm \theta} + b_k \, \hat{\bm z} \\ [3ex] 
\displaystyle \hat{\bm k}_{eff} = & \displaystyle \hat{\bm k} \\ [3ex] 
\displaystyle \hat{\bm z}_{eff} = & \displaystyle a_k \, \hat{\bm z} + b_k \, \hat{\bm \theta},
\end{array}
\end{equation}
where $a_k = 2Ak/\hbar\Omega_k$ and $b_k = 2M/\hbar\Omega_k$, with ${\bm \Omega}_{\bm k}$ defined in Eq.\ (\ref{magHam}), and $a_F$ and $b_F$ at $k = k_F$, with $b_F \ll 1$ at usual transport densities. Considering the negligible size of $b_k$ at $k = k_F$, the condition yielding the suppression of backscattering is effectively unmodified from the non-magnetic case. The vector ${\bm \sigma}$ of Pauli matrices is projected onto the three directions in reciprocal space in an analogous fashion to the non-magnetic case, and the details will not be reproduced here. \cite{Culcer_TI_AHE_PRB11} Likewise $S_{\bm k}$ is also projected onto these three directions. 
 
The kinetic equation is solved perturbatively in $\hbar/(\varepsilon_F\tau)$, characterizing the disorder strength, as well as in $\lambda$, which quantifies the strength of the extrinsic spin-orbit coupling. When a magnetization is present a number of terms independent of $\tau$ appear in the conductivity, representing transvers (Hall) transport. The dominant term in the anomalous Hall conductivity is \cite{Culcer_TI_AHE_PRB11} 
\begin{equation}
\arraycolsep 0.3ex
\begin{array}{rl}
\displaystyle \sigma_{yx} = & \displaystyle - \frac{e^2}{2h} \, (1 - \alpha),
\end{array}
\end{equation}
where $\alpha$ is a disorder renormalization, equivalent to a vertex correction in the diagrammatic formalism. \cite{Culcer_TI_AHE_PRB11} The bare contribution to $\sigma_{yx}$ is $ \displaystyle - \frac{e^2}{2h}$, which can also be expressed in terms of the Berry (geometrical) curvature \cite{Culcer_AHE_PRB03}, and is thus a topological quantity. It is similar to the result of Ref.~\onlinecite{ZangNagaosa_TI_Monopole_PRB10} in the vicinity of a ferromagnetic layer, and can be identified with a monopole in reciprocal space. It arises from the integral
\begin{equation}
\int_0^{k_F} \frac{AkM}{(A^2 k^2 + M^2)^{3/2}} = \frac{1}{A} \, \bigg( 1 - \frac{M}{\sqrt{A^2 k_F^2 + M^2}} \bigg).
\end{equation}
If we ignore terms of order $b_F^2$ it can be approximated as simply $1/A$, corresponding to the limit $k_F \rightarrow \infty$. In this regime the integral is therefore a constant, and the bare conductivity $\sigma_{yx} = \displaystyle - \frac{e^2}{2h}$, independent of the magnetization, number density, or Rashba spin-orbit coupling strength $A$. Contributions $\propto \lambda$ turn out to be negligible in comparison with this term. \cite{Culcer_TI_AHE_PRB11} 

Ref.~\onlinecite{Culcer_TI_AHE_PRB11} showed that there are no terms from extrinsic spin-orbit scattering that commute with the band Hamiltonian, and thus, following the argument above, no terms in the density matrix $\propto \tau$ due to extrinsic spin-orbit mechanisms. Neither extrinsic spin-orbit coupling nor magnetic impurity scattering give a driving term in the kinetic equation \textit{parallel} to $H_{0{\bm k}}$ (that is, $\parallel {\bm \Omega}_{\bm k}$), thus the anomalous Hall response does not contain a term of order $[\hbar/(\varepsilon_F\tau)]^{-1}$ due to extrinsic spin-orbit scattering or magnetic impurity scattering. Such a term would have overwhelmed the renormalized topological term $\sigma_{yx} = \displaystyle - \frac{e^2}{2h} \, (1 - \alpha)$ in the ballistic limit.

In TI the spin and charge degrees of freedom are inherently coupled and the anomalous Hall current can also be viewed a steady-state in-plane \textit{spin} polarization in a direction \textit{parallel} to the electric field. The (bare) topological term has two equal contributions, which are part of the correction to the density matrix \textit{orthogonal} to the effective Zeeman field, therefore they represent an electric-field induced displacement of the spin in a direction transverse to its original direction. They are also obtained in the Heisenberg equation of motion if one takes into account the fact that ${\bm k}$ is changing adiabatically. Physically, the $\hat{\bm x}$-component of the effective Zeeman field ${\bm \Omega}_{\bm k}$ is changing adiabatically, and the out-of-plane spin component undergoes a small rotation about this new effective field. Consequently, each spin acquires a steady state component parallel to ${\bm E}$, which in turn causes ${\bm k}$ to acquire a small component in the direction perpendicular to ${\bm E}$. Elastic, pure momentum scattering [contained in $\hat{J}_0 (f_{\bm k})$] reduces this spin polarization because, in scattering from one point on the Fermi surface to another, the spin has to line up with a different effective field ${\bm \Omega}_{\bm k}$. The extra spin component of each electron is $\propto M$, however the final result is independent of $M$, as the integrand contains a monopole located at the origin in ${\bm k}$-space. \cite{Culcer_AHE_PRB03, ZangNagaosa_TI_Monopole_PRB10} As $M\rightarrow 0$ the effect disappears, since the correction to the orthogonal part of the density matrix vanishes. Finally, though the form of this term would hint that it is observable for infinitesimally small $M$, it was tacitly assumed that $M$ exceeds the disorder and thermal broadening. 

The appearance of terms to order zero in $\hbar/(\varepsilon_F\tau)$, not related to weak localization, is standard in spin-dependent transport. The expansion of the density matrix starts at order $\hbar/(\varepsilon_F\tau)^{(-1)}$, i.e. the leading-order term is $\propto \tau$, so the fact that the next term is independent of $\hbar/(\varepsilon_F\tau)^{(-1)}$ is to be expected. Of course this term is \textit{not} independent of the angular characteristics of the scattering potential ($\alpha$ above depends on the scattering potential), which is why it is not enough to consider only the band structure contribution, but also its disorder renormalization. Disorder renormalizations are crucial in spin-related transport, where they can go as far as to cancel band structure contributions, as they do in the spin-Hall effect in spin-orbit coupled systems described by the Rashba Hamiltonian. \cite{Inoue_RashbaSHE_Vertex_PRB04} 

Experimentally, for charged impurity scattering the figures depend on the Wigner-Seitz radius $r_s$, representing the ratio of the Coulomb interaction energy and the kinetic energy, with $\epsilon_r$ the relative permittivity. For Bi$_2$Se$_3$ with $r_s = 0.14$ \cite{Culcer_TI_Kineq_PRB10}, the dominant term by far is $\sigma_{yx} \approx - 0.53 \, (e^2/2h) \approx -e^2/4h$. We can consider also the (artificial) limit $r_s \rightarrow 0$, which implies $\varepsilon_r \rightarrow \infty$, artificially turning off the Coulomb interaction. As $r_s \rightarrow 0$, the prefactor of $-e^2/2h$ tends to $0.61$, while at $r_s = 4$, the limit of RPA in this case, it is $\approx 0.12$. Interestingly, for short-range scattering, the anomalous Hall current changes sign, with $\sigma_{yx} \approx 0.18 \, (e^2/2h)$. The expression above represents the contribution from the conduction band, and in principle an extra $e^2/2h$ needs to be added to the total result to obtain the signal expected in experiment. Since the conduction band is expected to contribute $\approx -e^2/4h$, this does not make a difference in absolute terms. However, if one bears in mind that the bare contribution from the conduction band is $-e^2/2h$, the inescapable conclusion is that the remaining, observable part of the conductivity, is exactly the disorder renormalization $\alpha$. The same issue as in the quantum Hall effect is present with regard to the detection of the anomalous Hall effect, since there is always more than one surface and the surfaces are connected. Unlike the quantum Hall effect, the surfaces give opposite contributions to the half-quantized anomalous Hall effect. \cite{Nomura_3DTI_QAHE_PRL11} But since the disorder renormalizations should be different on the two surfaces a finite total signal should exist. 

In a related context, Tse and MacDonald \cite{Tse_TI_GiantMOKE_PRL10} studied a TI thin film weakly coupled to a ferromagnet, but focused on the magneto-optical Kerr and Faraday effects, since these are less likely to be affected by unintentional bulk carriers. Using linear-response theory, these scientists discovered that at low frequencies the Faraday rotation angle has a universal value, which is determined by the vacuum fine structure constant, when $\varepsilon_F$ lies in the Dirac gap for both surfaces. The Kerr rotation angle in this regime, is again universal and equal to $\pi/2$, hence the nomenclature of a giant Kerr rotation. The physics overlaps significantly with the anomalous Hall effect described above leading to the half-quantized anomalous Hall effect, except in this case the effect explicitly involves both surfaces. The effect reflects the interplay between the chiral nature of the topological surface states and interference between waves reflected off the top surface and waves scattered off the bottom surface. The surface Hall response creates a splitting between the reflected left-handed and right-handed circularly polarized fields along the transverse direction. The presence of the bottom topological surface is important since the first-order partial wave reflected from the top surface undergoes only a small rotation. The higher-order partial waves from subsequent scattering with the bottom surface yield a contribution that strongly suppresses the first-order partial wave along the incident polarization plane. As a result, the left-handed and right-handed circularly polarized fields each acquire a phase of approximately $\pi/2$ in opposite directions, leading to a $\pi/2$ Kerr rotation.

This concludes the discussion of the anomalous Hall effect. The topic of spin transfer in magnetic topological insulators has also been studied recently. Nomura and Nagaosa \cite{Nomura_TI_ElCrgMgnTxt_PRB10} investigated magnetic textures in a ferromagnetic thin film deposited on a 3D TI, showing that they can be electrically charged thanks to the proximity effect with the Dirac surface states. They consider an insulating magnet on the TI surface, with a magnetization ${\bm M}$ pointing along. Due to the interplay of the chiral surface states with the proximity-induced magnetization, electric charge and current densities are induced in the TI,
\begin{equation}
\arraycolsep 0.3ex
\begin{array}{rl}
\displaystyle n = & \displaystyle - \bigg(\frac{M \sigma_H}{e v_F}\bigg) \, {\bm \nabla}\cdot\hat{\bm M} \\ [3ex] 
\displaystyle {\bm j} = & \displaystyle \bigg(\frac{M \sigma_H}{e v_F}\bigg) \, \pd{\hat{\bm M}}{t},
\end{array}
\end{equation}
where $\sigma_H$ is the quantized Hall conductivity. The electrical current, which we have established corresponds to a steady-state spin density, is the analog of the spin transfer torques. Spin transfer in systems with strong spin-orbit interactions was discussed in the related context of ferromagnetic semiconductors in Ref.~\onlinecite{Culcer_Fmg_SpinTransfer_PRB09}. In topological insulators, precisely because of the equivalence of charge currents and spin densities, this electric charging of magnetic textures can be used for electrical manipulation of domain walls and vortices. In the case of vortices, an electric field breaks the vortex-antivortex symmetry on the TI surface. In the case of domain walls, the authors show that an electric field can displace a domain wall in a topological insulator, much like a magnetic field can accomplish this in an ordinary magnet.

The action of an electrical current on a magnetization at the interface between a topological insulator and a ferromagnet was also studied by Garate and Franz. \cite{Garate_TI/FM_InverseSpinGalv_PRL10} While their results were similar, these authors focused on the magnetization dynamics induced by the steady-state spin polarization, inlcuding the possibility of current-induced magnetization reversal. In their language, the Hall current changes the effective anisotropy field, which in turn affects the magnetization dynamics in the ferromagnet. Interestingly, their work shows that the magnetization may be flipped by the Hall charge current in tandem with a small magnetic field of approximately $20$mT. 

The same authors subsequently focused on the magnetoelectric response of a disorder-free time-reversal invariant TI. \cite{Garate_HelMtl_MgnEl_PRB10} By determining the spin-charge response function, these researchers confirmed the observation that a uniform static in-plane magnetic field does not produce a spin polarization in topological insulators. The equilibrium RKKY interaction between localized spins separated by $R$ decays as $1/R^3$ for $k_F = 0$, while for $k_F R \gg 1$ it oscillates and decays as $1/R^2$. Unlike ordinary two-dimensional electron systems, the RKKY coefficients depend on the carrier number density. The equilibrium RKKY interaction can be strongly modified by moderate electric fields, in particular when the localized impurities are separated by distances in excess of 5nm. The authors also found that the helical metal mediates a strong Dzyaloshinskii-Moriya coupling at finite doping, which favors noncollinear magnetization configurations. Lastly, thanks to the inherent spin-momentum locking of topological insulators, an impurity magnetic moment that precesses in an external Zeeman field gives rise to an alternating charge current. 


Other researchers have brought to light different facets of the exciting new physics of magnetic topological insulator. A recent work by Burkov and Hawthorn \cite{Burkov_TI_SpinCharge_PRL10} focused on spin-charge coupling due to the topological surface states of a \textit{diffusive} topological insulator. The authors derived a set of diffusion equations describing spin-charge coupled transport of the helical metal surface. The most interesting finding of Ref.~\onlinecite{Burkov_TI_SpinCharge_PRL10} is that the continuity equation for the carrier density $n$ and particle current density ${\bm j}$ a new, spin-dependent term appears,
\begin{equation}
\arraycolsep 0.3ex
\begin{array}{rl}
\displaystyle \pd{n}{t} = & \displaystyle - {\bm \nabla} \cdot {\bm j} \\ [3ex] 
\displaystyle {\bm j} = & \displaystyle - D{\bm \nabla}n - v_F \, {\bm S} \times \hat{\bm z},
\end{array}
\end{equation}
where $D$ is the usual diffusion coefficient and ${\bm S}$ represents the spin density. The second, unconventional term in the current, related to the chirality (helicity) of the topological surface states, leads to a new magnetoresistance effect. The authors envisage a TI surface on which current flows between a ferromagnetic electrode and a non-magnetic electrode. The ferromagnetic electrode injects a spin-polarized current. Solving the diffusion equation, it emerges that the voltage drop consists of two contributions, the first being the usual Ohmic resistance, while the second is given by
\begin{equation}
V = \frac{4\pi I \eta}{e^2 k_F},
\end{equation}
with $I$ the current and the dimensionless $\eta$ parameterizing the spin polarization of the current. This contribution is proportional to the cross product of the polarization of the ferromagnetic electrode and the charge current, and disappears when the spin polarization of the ferromagnetic electrode is along the transport direction. The effect is independent of disorder and of the separation between the electrodes and only depends on the spin polarization of the injected current and the Fermi momentum. However, the degree of the injected spin polarization depends on the (non-universal) properties of the interface between the ferromagnetic electrode and the topological insulator. The authors further show that this effect is tunable by a gate voltage, enabling operation of the device as a transistor. 

Ref.\ \onlinecite{Mondal_TI_Magneto_PRB10} analyzed the tunneling conductance of a series of ballistic structures involving topological insulators and quantizing magnetic fields or magnetic materials. For a TI in crossed electric and strong (quantizing) magnetic fields, with the magnetic field perpendicular to the surface, the orbital and Zeeman effects due to an applied magnetic field can be modulated by the electric field. The Landau levels have the form
\begin{equation}
\arraycolsep 0.3ex
\begin{array}{rl}
\displaystyle E_n (k_y) = & \displaystyle \pm \frac{\hbar v_F}{l_B \gamma^{3/2}} \, \sqrt{2|n| + \frac{g^2\mu_B^2 \gamma B \, \sin\theta}{e \hbar v_F^2}} - \beta\hbar v_Fk_y, \\ [3ex]
& \displaystyle ({\rm for} \,\,\, n \ne 0) \\ [1ex]
 & \displaystyle - \frac{1}{\gamma} \, |g \mu_B B \sin\theta| - \beta\hbar v_Fk_y, ({\rm for} \,\,\, n = 0).
\end{array}
\end{equation}
Here the magnetic length $l_B = \sqrt{\hbar/(eB\sin\theta)}$, the boost $\beta = E/(v_FB\sin\theta) \le 1$, and $\gamma = (1 - \beta^2)^{-1/2}$. This equation shows that the Landau levels can be collapsed by varying either the strength of the electric field ${\bm E}$ or the out-of-plane tilt $\theta$ of the magnetic field. The tunneling conductance depends on $\theta$, a fact easily understood by recalling that the in-plane component of the magnetic field can be removed from the Dirac-cone by a gauge transformation, and is therefore irrelevant. The Zeeman term yields an exponential suppression of the tunneling conductance, which is explained by the fact that the Zeeman term in TI is equivalent to a mass term. In normal metal-magnetic film-normal metal (NMN) junctions transport can be controlled using the proximity-induced exchange field due to a ferromagnetic film. At a critical value of this exchange field, the tunneling conductance of the junction changes from oscillatory to a monotonically decreasing function of the junction width $d$. In principle such a setup could realize a magnetic switch. Normal-metal-barrier-magnetic film (NBM) junctions are characterized by the height $V_0$ of the barrier. In the absence of a magnetization, such a junction displays maxima in the tunneling conductance at integer quantized values $n$ of the quantity $eV_0d/\hbar v_F$. In the magnetic case, beyond a critical value of the exchange field, trasnmission resonances occur at values $n + 1/2$ of this quantity. At even higher exchange fields the junction conductance can be fully suppressed. For normal-metal-magnetic film-superconductor (NMS) junctions, the position of the peaks of the zero-bias tunneling conductance can be tuned using the magnetization of the ferromagnetic film. 

\begin{figure}[tbp]
\bigskip
\includegraphics[width=\columnwidth]{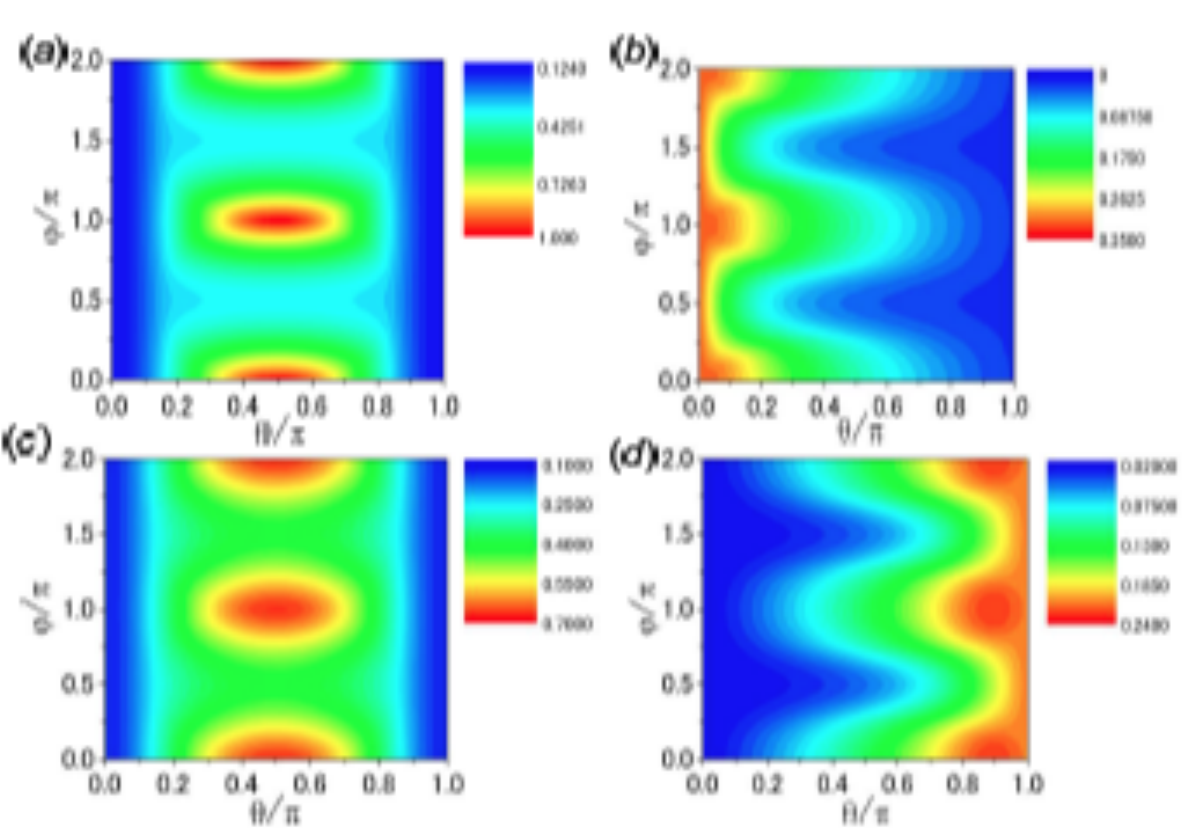}
\caption{\label{Nagaosa_AMR}
Anomalous magnetoresistance for a ferrromagnet-ferromagnet interface on top of a topological insulator. The figure shows the  tunneling conductance for the case when the barrier potential between the two regions is zero. On the incident side the magnetization is defined as $m_1 (\sin\theta \cos\phi, \sin\theta \sin\phi, \cos\theta)$ while on the other side of the interface it is $(0, 0, m_2)$. In (a) and (c) below $m_2 = 0$, while in (b) and (d) $m_2$ is finite. In (a) and (b) the results are plotted for an $n-n$ junction, while in (c) and (d) they are plotted for a $p-n$ junction. The figure is adapted from Ref.~\onlinecite{Nagaosa_AMR_TopSfc_PRB10}.}
\end{figure}

Yokoyama \textit{et al.} \cite{Nagaosa_AMR_TopSfc_PRB10} investigated conduction across a ferromagnet/feromagnet interface on the surface of a topological insulator. The device considered is the analog of the more usual spin-valve devices, except the substrate in this case is a topological insulator (the current itself flows in the TI). The tunneling conductance across the interface has a strong dependence on the directions of the magnetizations of the two ferromagnets, related to the manner in which the wavefunctions on the two sides of the interface connect. A misfit of the in-plane momentum between the two sides gives rise to a strong dependence of the tunneling conductance on the \textit{in-plane} rotation angle $\phi$. Unlike conventional spin-valve devices, the conductance in the antiparallel configuration can greatly exceed that in the parallel configuration. A representative part of their results is reproduced in Fig.~\ref{Nagaosa_AMR}. 

Magnetic impurity scattering has been considered in TI transport, \cite{Nomura_3DTI_QAHE_PRL11, Culcer_TI_AHE_PRB11}  although work to date solely focused on straightforward potential scattering, without considering in detail spin-flip terms that lead to the Kondo effect. All studies above require the magnetization to be large enough so that it is resolved, implying that the local moments are all lined up, thus neglecting Kondo physics in the anomalous Hall effect seems sensible. Transport theory can be modified to include spin-flip scattering and Kondo physics, provided that scattering terms $\propto \sigma_x, \sigma_y$ are allowed, and provided that one works explicitly in the grand canonical ensemble in order to account fully for the spin-flip factors. In the context of the Kondo effect, local moment formation and Friedel oscillations associated with magnetic impurities in TI have been studied in Ref.~\onlinecite{Tran_TI_Kondo_PRB10}. In addition, Biswas and Balatsky \cite{Biswas_TI_ImpurStt_PRB10} calculated the $t$-matrix of TI, determining the modiÞcation of the local electronic structure caused by a single local impurity on the surface of a 3DTI. They found that the local density of states around the Dirac point is disrupted near the impurity by the creation of low-energy resonance states, yet this does not destroy the Dirac point locally. The RKKY coupling between spin textures created near the magnetic impurities was found to be anisotropic. Aside from these papers, the study of the Kondo effect in TI is in its infancy, and there is no reliable indication of how high the Kondo temperature is expected to be, and the nontrivial effect of spin-momentum locking on Kondo singlet formation has not been determined. 

\section{Weak localization}

On a fundamental level, weak localization is an interference phenomenon between two time-reversed paths in a scattering event, which gives a correction to the conductivity of a solid. Weak localization refers to interference between the incident and backscattered paths, and is often referred to as coherent backscattering. To obtain the weak localization correction in the linear response Kubo formalism, one would have to return to Eq.\ \ref{Kubo} and determine the impurity average of the product of two Green's functions in the next order in the small parameter $\hbar/(\varepsilon_F\tau)$. Logistically,  this is tantamount to calculating the Cooperon in the Kubo formula, that is, the sum of maximally crossed diagrams, which can be mapped onto a sum of twisted ladder diagrams. In the classical interpretation of this phenomenon, the maximally crossed diagrams represent the probability that a particle will return to its starting position after a scattering process. Although the Liouville-density matrix formalism has not been extended to cover weak localization, an analogous and widely used kinetic equation formalism can be devised starting from the Keldysh Green's function approach. \cite{Vasko} This can be straightforwardly extended to incorporate weak localization, and the weak localization correction to the conductivity in this language is also expressed in terms of the Cooperon.

Being an interference phenomenon, weak localization reflects the existence of phase coherence between two time-reversed paths, incident and backscattered. This is quantified by defining a phase relaxation time $\tau_\phi$, which represents the time over which the electron wave function retains its phase. Phase relaxation is typically caused by dynamic perturbations such as electron-electron scattering and phonon scattering, although in practice the phase relaxation time is frequently taken as a phenomenological quantity. In an ordinary 2DEG the weak localization correction to the conductivity takes the form 
\begin{equation}
\delta \sigma_{xx} = - \frac{e^2}{\pi h} \, \ln\frac{\tau_\phi}{\tau} 
\end{equation}

Phase coherence between time-reversed paths is also destroyed by magnetic interactions, thus weak localization is correspondingly suppressed. For example, there is no weak localization correction to the ordinary Hall effect. Magnetic impurities also reduce weak localization, although they do not eliminate it. The disappearance of weak localization in a magnetic field makes this phenomenon a useful experimental characterization tool. Weak localization gives a negative contribution to the conductivity, therefore application of a magnetic field yields a positive magnetoconductance (negative magnetoresistance.) In 2D weak localization has a logarithmic temperature-dependence, and electron-electron interactions also lead to a conductivity correction logarithmic in the temperature. 

If spin-orbit coupling is present in the impurity potential or in the band structure, the correction to the conductivity is positive, and is termed weak antilocalization. The same is true for the case of a pseudospin dependent Hamiltonian, such as that of graphene monolayers and bilayers, \cite{Kechedzhi_Gfn_MonoBil_WL_EPJ07} and is expected to apply to topological insulators. The occurrence of weak antilocalization in topological insulators can be easily understood by recalling that backscattering is absent. Simply put, there cannot be coherent backscattering when there is no backscattering at all. Weak antilocalization manifests itself in experiment as a negative magnetoconductance (positive magnetoresistance.) The full magnetic field dependence in an ordinary 2DEG is encapsulated in the Hikami-Larkin-Nagaoka\cite{Hikami_WAL_PTP80} formula for $\delta \sigma_{xx}^{mag} = \delta \sigma_{xx} (B) - \delta \sigma_{xx} (0)$
\begin{equation}\label{dsmag}
\delta \sigma_{xx}^{mag} = \frac{e^2}{2 \pi h} \, \bigg[\ln \frac{\hbar}{4eBD\tau_\phi} - \Psi \bigg(\frac{1}{2} + \frac{\hbar}{4eB D \tau_\phi}\bigg) \bigg],
\end{equation}
where $\Psi$ represents the digamma function, which is the logarithmic derivative of the Euler gamma function, and $D = v_F\tau^2/2$ the diffusion coefficient.

At sufficiently high impurity densities, coherent backscattering leads to Anderson localization, and the material undergoes a transition to an insulating state which bears the name of the Anderson metal-insulator transition. Although in topological insulators the absence of backscattering is believed to suppress Anderson localization, one remark is in order. This problem has been studied with the aid of renormalization group techniques, and it was determined that the lower critical dimension for the localization problem is $dim = 2$. This seems to imply that all states are localized in 2D. Yet the bare textbook argument that no extended states exist in 2D is not borne out by experiment, since the localization length can easily exceed the size of the sample. Taken literally, this would imply that the 2D electron gas, MOSFETs and other semiconductor devices do not exist, whereas quasi-2D systems display the highest known mobilities in the world (tens of millions cm$^2$/Vs in GaAs), as a result of modulation doping. 
 
Tkachov and Hankiewicz \cite{Tkachov_HgTe_WAL_PRB11} studied weak antilocalization in HgTe quantum wells, which have a finite mass, and topological surface states, which are massless. In HgTe quantum wells, the weak localization correction is shown to be
\begin{equation}
\delta \sigma_{xx} = - \frac{2e^2\alpha}{\pi h} \, \ln \frac{1/\tau_e}{1/\tau_m + 1/\tau_\phi},
\end{equation}
where $\alpha$ is a constant, $\tau_e$ is the elastic scattering lifetime, while $\tau_m$ is given by
\begin{equation}
\frac{1}{\tau_m} = \frac{2}{\tau_e} \, \frac{(m + Bk_F^2)^2}{A^2 k_F^2 + (m + Bk_F^2)^2},
\end{equation}
and the mass parameter in the Dirac equation is $m + Bk_F^2$, with $m$ and $B$ material specific constants, and $A$ as defined in the Rashba-Dirac Hamiltonian. The finite mass suppresses weak antilocalization in HgTe quantum wells at times larger than $\tau_m$, which corresponds to the opening of a relaxation gap $\hbar/\tau_m$ in the Cooperon diffusion mode. Since in HgTe quantum wells $m$ can be either positive or negative depending on the well width, $\tau_m$ has a non-monotonic dependence on both $m$ and $n$, and vanishes in the case when
\begin{equation}
m + 2\pi n B = 0.
\end{equation}
Along this line the weak antilocalization term in the conductivity reaches its maximum value, given by the usual expression $\delta \sigma_{xx} = \displaystyle \frac{e^2}{\pi h} \, \ln \frac{\tau_\phi}{\tau_e}$. In the case of topological insulators, the authors find that the weak antilocalization correction takes the form
\begin{equation}
\delta \sigma_{xx} = \displaystyle \frac{e^2}{2\pi h} \, \ln \frac{\tau_\phi}{\tau_e},
\end{equation}
the same as a 2DEG with spin-orbit interaction. However, in a parallel magnetic field, due to the exponential decay of the surface states into the bulk, topological insulators exhibit a weak-antilocalization magnetoconductivity which is a function of the penetration length into the bulk. The time scale for the suppression of the quantum interference by the magnetic flux is determined by the thickness of the surface state, which is expected to be much smaller than the magnetic length.

Concomitantly, Lu \textit{et al.} \cite{Lu_TI_WL/WAL_PRL11} identified a competition between weak localization and anti-localization in topological surface states. They find that, as expected, the gapless topological surface states indeed give rise to weak antilocalization. At the same time, however, when the gap is opened (by magnetic doping, for example), a new weak localization contribution emerges. The gap and magnetic scattering can change the magnetic field dependence from weak antilocalization-like to a parabolic dependence, and increasing the size of the gap with respect to $\varepsilon_F$ may drive the system into the weak localization regime. In this regime, however, the weak localization correction is half that found in the usual 2DEG. The transition from weak antilocalization to weak localization is understood as a change in Berry phase as the magnetization gap increases from zero to $2\varepsilon_F$.

In a subsequent paper, Lu and Shen \cite{LuShen_TI_WL/WAL_11} concentrated on the bulk channel contribution to weak localization. The authors argue that the recently observed weak antilocalization in TI  appears to have contributions from the bulk states. Therefore they devised a unified model that allowed them to study bulk and surface contributions to weak localization and antilocalization and magnetoconductivity. Within the framework of ${\bm k} \cdot {\bm p}$ perturbation theory the authors derived a modified Dirac Hamiltonian 
\begin{equation}
H_{3D} = C + D(k^2 + k_z^2) + A \, {\bm \alpha} \cdot {\bm k} + [m - B(k^2 + k_z^2)] \, \beta,
\end{equation}
where ${\bm k} = (k_x, k_y)$, ${\bm \alpha}$ and $\beta$ are the usual matrices with the same name found in the Dirac equation, and $A$, $B$, $C$, $D$ and $m$ are model parameters. From this Hamiltonian the topological surface bands and 2D bulk subbands are easily determined. Following a linear response calculation along the lines of their previous work cited above, the authors derived the magnetoconductivity for both the 2D bulk subbands and surface bands. They find that the bulk states contributes a weak localization term in the magnetoconductivity, which, in certain parameter regimes, may compensate the weak antilocalization terms arising from the surface states. When the bulk weak localization channels outnumber the surface weak antilocalization channels a crossover will occur at a finite magnetic field. The authors suggest that the competition between the two mechanisms may explain the observed experimental deviation of $\delta \sigma_{xx}$ from the Hikami-Larkin-Nagaoka formula.

Finally, an interesting paper by Ostrovsky \textit{et al.} \cite{Ostrovsky_TI_IntCrit_PRL10} focused on electron-electron interactions in disordered topological insulators. The authors used the perturbative renormalization group to calculate the scaling function $\beta(\tilde{\sigma}_{xx}) = \td{\tilde{\sigma}_{xx}}{\ln L}$, where the dimensionless $\tilde{\sigma}_{xx}$ is the conductivity measured in units of $e^2/h$ and $L$ the system size, in the presence of Coulomb interactions for the case when $\tilde{\sigma}_{xx} \gg 1$. The one-loop renormalization leads to the following equation
\begin{equation}
\beta(\tilde{\sigma}_{xx}) = \frac{N}{2} - 1 + (N^2 - 1) \, \mathcal{F},
\end{equation}
where $N$ represents the degeneracy, which is 4 in graphene and 1 in TI. The term $N/2$ represents weak antilocalization, while the second term arises from the Coulomb interaction in the singlet channel and evidently suppresses the conductivity. It is independent of $N$ since all the flavors are involved in screening. The last term is due to the Coulomb interaction in the multiplet channel, with $\mathcal{F}$ an interaction parameter. In TI the last term vanishes and the scaling function turns out to be negative in the presence of electron-electron interactions, implying a tendency towards localization. Since at small $\tilde{\sigma}_{xx}$ the scaling function is positive due to topological protection, it must cross zero at some critical point. The authors thus identify an interaction-induced critical point, concluding that interactions can fully localize topological surface states in 3D TI. 

\section{Transport experiments}

Experimental work on transport in topological insulators has begun to advance at a brisk pace, and is without doubt entering its heyday, in the way ARPES and STM work did two years ago (for a recent experimental review see Ref.~\onlinecite{Yongqing_TI_Transp_FOP11}.) It will help to summarize in the following what one expects to see in a sample in which bulk carriers have been eliminated and transport is due to the chiral surface states. When bulk carriers are still present one expects variable range hopping characterized by an activation energy. The semi-logarithmic plot of resistivity versus temperature in this case displays a clear plateau. Observation of a graphene-like minimum conductivity requires tuning through the Dirac point. Measuring the spin polarization of a current is not an easy experiment. In contrast, weak localization and antilocalization both disappear in a weak magnetic field. A magnetic field also allows observations of Shubnikov-de Haas oscillations, which may be grouped into oscillations due to bulk states and oscillations due to surface states. The surface SdH oscillations are expected to vary as $\Delta(1/B) = e/(hn)$, where we recall that $n$ is the 2D carrier density, and to show angular dependence: for surface transport one anticipates that SdH oscillations will only be present for ${\bm B}$ perpendicular to the surface. On the other hand, for bulk transport SdH oscillations are expected for all orientations of the magnetic field. The cyclotron mass, discussed above under magnetotransport theory, is typically extracted from cyclotron resonance experiments. It is given by $\varepsilon_F/v_F^2$, where $v_F$ is the (constant) Fermi velocity given by $A/\hbar$. These latter considerations show that certain signatures of chiral surface states may be easier to observe using magnetotransport. Consequently, thus far significantly more experimental work has focused on magnetotransport than on non-magnetic transport. 

In the initial experimental efforts, it appeared impossible to identify any signatures whatsoever of the elusive surface states. One early avenue towards the removal of bulk carriers involved compensation by doping with Ca. Checkelsky \textit{et al.} \cite{Checkelsky_Bi2Se3_QmIntfr_PRL09} performed transport experiments on Ca-doped samples noting an increase in the resistivity, but concluded that surface state conduction alone could not be responsible for the smallness of the resistivities observed at low temperature. A subsequent paper by the same group \cite{Checkelsky_Bi2Se3_SfcSttCond_PRL11} claims to have observed surface conduction, inferred from the sign change in the Hall coefficient. Such a sign change would indicates a change in the carrier type from electrons to holes, and therefore the fact that the Dirac point has been crossed. The results are interpreted based on a model containing parallel bulk and surface conduction channels. The chemical potential is assumed to lie in the bulk gap, implying that bulk carriers are present due to thermal excitation across the gap. Curiously, the bulk gap is observed to be narrowed, greatly increasing the number of thermally excited bulk carriers. At the same time, the resistivity displays a power law variation as a function of temperature, whereas the activated temperature dependence typical of variable range hopping is not observed. 

Analytis \textit{et al.} \cite{Analytis_Bi2Se3_ARPES_SdH_3DFS_PRB10} used SdH oscillations and ARPES to probe the Fermi surface of single crystals of Bi$_2$Se$_3$. The Bohr radius and Thomas-Fermi screening length place this material near the boundary of the metal-insulator transition. The SdH and ARPES probes agreed quantitatively on measurements of the effective mass and bulk band dispersion, and, in samples with high electron density ($10^{19}$cm$^{-3}$), they agreed on the position of $\varepsilon_F$. Yet discrepancies emerge regarding the location of $\varepsilon_F$ in samples with reduced carrier densities. For samples with carrier densities down to $10^{17}$cm$^{-3}$ ARPES identifies the Dirac dispersion of the surface states, whereas SdH oscillations reveal a 3D Fermi surface, pinning the Fermi level in the bulk conduction band. None of the data displayed any thickness dependence. 

Eto \textit{et al.} \cite{Eto_Bi2Se3_SdH_3DFS_PRB10} reported angular-dependent Shubnikov-de Haas (SdH) oscillations due only to the residual 3D Fermi surface in Bi$_2$Se$_3$ with an electron density of $5 \times 10^{18}$cm$^{-3}$. A significant finding by these authors is the fact that the bulk SdH oscillations themselves are anisotropic. A general guideline emerges that the observation of anisotropic SdH oscillations by itself does not constitute a demonstration of surface transport. 

Steinberg \textit{et al.} \cite{Steinberg_Bi2Se3_Ambipolar_NL10} reported signatures of ambipolar transport analogous to graphene in Bi$_2$Se$_3$ samples with thicknesses of up to 100 nm, using a top-gate with a high-dielectric constant insulator. Ambipolar transport would be consistent with tuning between electron and hole carriers. Transport results were explained in terms of parallel surface and bulk contributions, and the dominant bulk contribution was subtracted so as to obtain the surface contribution. Surprisingly, the surface effect (i.e. an apparent minimum conductivity that would indicate ambipolar transport) was observed only for top-gate voltage sweeps and was absent during back-gate sweeps, a fact that to date remains unexplained. Further studies by the same group \cite{Steinberg_Bi2Se3_Film_Sfc2Blk_11} on weak antilocalization in Bi$_2$Se$_3$ thin films revealed that the weak antilocalization contribution to the conductivity reflects the presence of various conduction channels -- bulk as well as the top and bottom surfaces.

Experimental studies of tetradymite materials have to date reported Landau levels but no quantum Hall effect, no conclusive measurement exists of the $\pi$-Berry phase in Shubnikov-de Haas oscillations. Cheng \textit{et al.}\cite{Cheng_TI_STM_LL_PRL10} claim to have observed Landau levels analogous to graphene in 100 nm-thick Bi$_2$Se$_3$ films, with Landau level energies $\propto{\sqrt{nB}}$, where $n$ is the Landau level index. The fitted data indicate an expected spin and valley degeneracy of one, yet puzzling features such as the absence of Landau levels \textit{below} the Fermi energy and enhanced oscillations near the conduction band edge have not been resolved, and some arbitrariness is involved in extracting the Landau level filling factor $\nu$. Hanaguri \textit{et al.}\cite{Hanaguri_TI_STM_LL_PRB10} report similar findings in the same material, with analogous features in the Landau level spectrum.

Butch \textit{et al.} \cite{Butch_Bi2Se3_Search_PRB10} carried out SdH, ordinary Hall effect and reflectivity measurements on Bi$_2$Se$_3$ and concluded that the observed signals could be ascribed solely to bulk states, even at 50mK and low Landau level filling factor. Nominally undoped single crystals of Bi$_2$Se$_3$ with carrier densities of $\approx 10^{16}$cm$^{-3}$ and mobilities in excess of 2 $m^2$/(Vs) were studied. Oscillations due to the surface states were expected at a very different frequency from those due to bulk states. The absence of a significant surface contribution to bulk conduction demonstrates that even in very clean samples, the surface mobility is lower than that of the bulk, despite its topological protection. A similar conclusion was reached by Sushkov \textit{et al.} \cite{Sushkov_Bi2Se3_Cyclo_Faraday_PRB10} based on far-infrared cyclotron resonance and the Faraday effect in Bi$_2$Se$_3$ in magnetic fields up to 8 T, as well as by Jenkins \textit{et al.} \cite{Jenkins_TI_Kerr_Reflect_PRB10} based on the Kerr effect in the same material.

Taskin \textit{et al.} \cite{Taskin_TI_eh_PRL11} studied bulk and surface transport in the material Bi$_{1.5}$Sb$_{0.5}$Te$_{1.7}$Se$_{1.3}$, calculating that 70$\%$ of the total conductance arises from the surface. Data fitting yielded a surface $g$-factor of approximately 20. Many intriguing unexplained findings were reported, including the observation that the surface states \textit{change} as a function of time. The scientists report quantum oscillations from Dirac holes \textit{and} Dirac electrons in the same sample at different points in time over one month. This seems to be supported by the fact that the low-temperature Hall coefficient changes sign with time in thin samples, while the high temperature Hall coefficient is essentially unchanged as a function of time. Furthermore, the SdH maxima and minima were found to depend only on the perpendicular magnetic field.

\begin{figure}[tbp]
\bigskip
\includegraphics[width=\columnwidth]{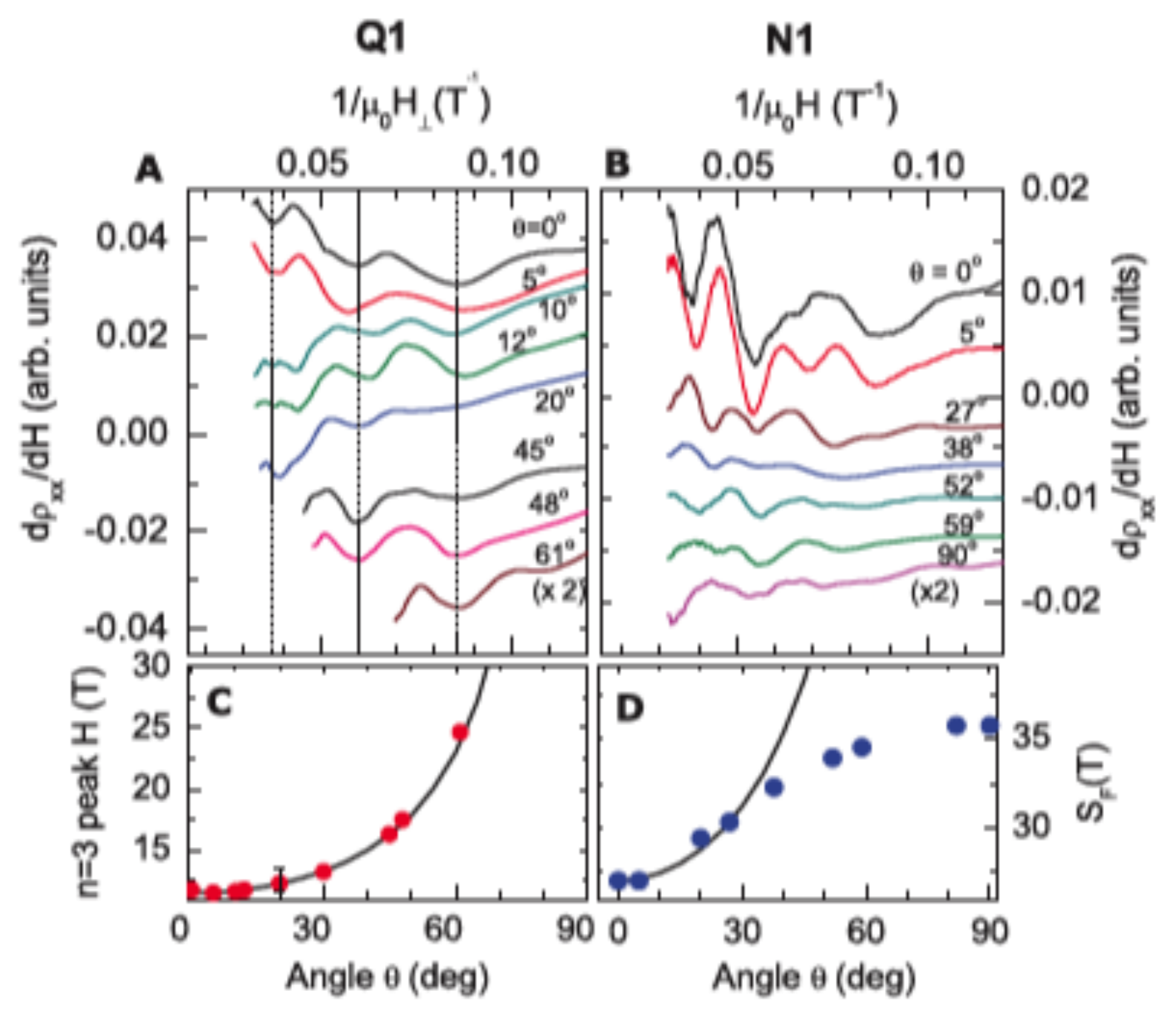}
\caption{\label{Qu_SdH}
Shubnikov-de Haas oscillations seen by Qu \textit{et al}. \cite{Qu_Bi2Te3_QmOsc_Sci10} in two samples (labeled Q1 and N1 in the article.) Figures (A) and (B) show the derivative $d\rho_{xx}/dH$ of the diagonal resistivity with respect to the magnetic field $H$ versus the perpendicular component $H_\perp$ of the magnetic field at selected tilt angles $q$. (A) Sample Q1. (B) Sample N1. In (A), the minima lie on the vertical dashed lines consistent with a 2D Fermi surface, whereas in (B), the minima shift systematically with $q$. (C) The field position of the $n = 3$ Landau level peak for sample Q1 (red circles) varies with $q$ as $1/\cos q$ (black curve), consistent with a 2D Fermi surface. (D) The period for sample N1 (blue circles) increases by $35\deg$ as $q \rightarrow 90\deg$, deviating strongly from $1/\cos q$ (black curve), consistent with a 3D Fermi surface. This figure has been adapted from Ref.~\onlinecite{Qu_Bi2Te3_QmOsc_Sci10}.}
\end{figure}

Qu \textit{et al.} \cite{Qu_Bi2Te3_QmOsc_Sci10} studied quantum oscillations in Bi$_2$Te$_3$, claiming to resolve a surface electron current consistent with a Fermi velocity $v_F \approx 4 \times 10^5$ms$^{-1}$, in agreement with ARPES measurements. The authors analyzed a variety of samples, of which some were metallic and some had a high bulk resistivity. Selective cleaving resulted in non-metallic crystals. Though in Bi$_2$Te$_3$ $\varepsilon_F$ lies in the bulk valence band, the authors were able to manufacture both $n$- and $p$-type samples, which suggests a degree of competition and compensation between the bulk valence and conduction bands. Data from the weak-magnetic field Hall effect yield a surface mobility of 9000-10,000 and substantially higher than in the bulk. SdH oscillation data is shown in Fig.~\ref{Qu_SdH}. In non-metallic samples, the SdH oscillation period depends only on the perpendicular magnetic field, with oscillations not resolved for magnetic field tilt angles $q$ in the range $65\deg < q < 90\deg$, whereas oscillations in the metallic sample survive up to $90\deg$. The Hall signal changes sign abruptly as $\varepsilon_F$ approaches the Dirac point. However, the SdH oscillations appear to be weak, and were found by subtracting two large numbers. Even though the surface mobility is calculated to be very large and to exceed the bulk, even in the most resistive sample, the bulk \textit{conductance} (i.e. total signal) exceeds the surface conductance by a factor of $\approx$ 300. The authors explain that the bulk hole density and mobility are not easily determined in the nonmetallic crystals. They also point out certain unconventional findings, for example, the transverse magnetoresistance of several non-metallic samples increases linearly with the magnetic field, a feature not seen in metallic samples.

Coexistence of bulk and 2D Fermi surfaces was also observed by Taskin \textit{et al.} \cite{Taskin_BiSb_MagRes_PRB10} in their study of angular-dependent magnetoresistance and SdH oscillations in Bi$_{0.91}$Sb$_{0.09}$. SdH oscillations fall into two categories. Oscillations observed at low magnetic fields are believed to originate from the surface states. A different pattern of SdH oscillations becomes prominent at higher magnetic fields, which at present remains unexplained. 

\begin{figure*}[tbp]
\bigskip
\includegraphics[width=2\columnwidth]{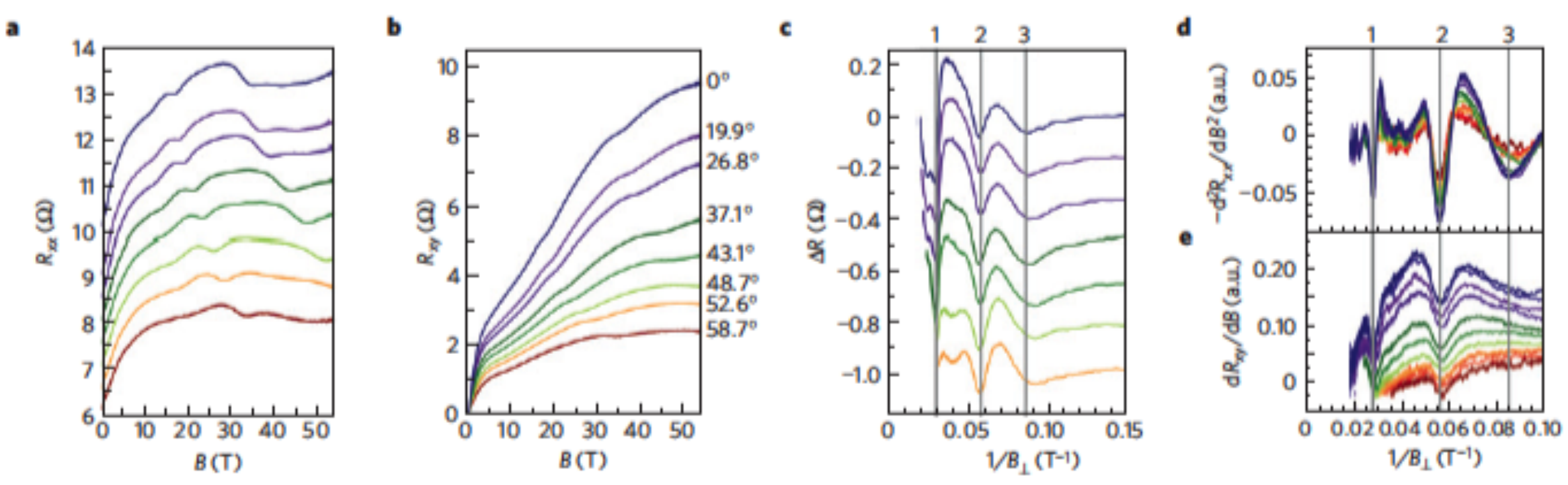}
\caption{\label{Analytis_SdH}
Shubnikov-de Haas oscillations seen by Analytis \textit{et al.} \cite{Analytis_Bi2Se3_QL_SfcStt_NP10} Longitudinal resistance $R_{xx}$ (a) and Hall resistance $R_{xy}$ (b) data traces as a function of magnetic field for indicated angles for a single crystal with carrier density $n \approx 4 \times 10^{16}$ cm$^{-3}$ (labeled sample $\Sigma 1$ in the paper.) The $R_{xx}$ traces are offset for clarity. The deviation from linearity of the low-field $R_{xy}$ indicates the 3D quantum limit $(B \approx 4 T)$, which does not vary with angle. Beyond this limit, additional features in both $R_{xy}$ and $R_{xx}$ move smoothly up in field as the tilt angle $\theta$ is increased. $\theta = 0$ is defined to be when the field is perpendicular to the surface (parallel to the trigonal $c$-axis). (c) A smooth second-third order polynomial can be fitted as background and subtracted from the raw $R_{xx}$, revealing oscillations that grow with field. (d) $-d^2R_{xx}/dB^2$ and (e) $dR_{xy}/dB$ (e) as a function of $1/B_\perp$, where $B_\perp = B\cos\theta$ aligns all features associated with the 2D surface state. The vertical lines in c-e indicate the first three Landau levels $n = 1,2,3$ of the 2D state. This figure has been adapted from Ref.~\onlinecite{Analytis_Bi2Se3_QL_SfcStt_NP10}.}
\end{figure*}

In a beautiful experiment, Analytis \textit{et al.} \cite{Analytis_Bi2Se3_QL_SfcStt_NP10} recently detected the topological surface states of Bi$_2$Se$_3$ in which Sb was partially substituted for Bi to reduce the bulk carrier density to $10^{16}$cm$^{-3}$. Results from this paper are shown in Fig.~\ref{Analytis_SdH}. The bulk carrier density was systematically reduced by Sb doping until the \textit{quantum limit} was reached, that is, the point where a magnetic field can collapse the bulk carriers to their lowest Landau level. Beyond this field surface state could be seen clearly. The carrier density determined by the low field Hall effect matches the Fermi surface parameters measured via SdH oscillations. When the field is rotated away from the crystal trigonal axis, the SdH frequency changes according to the morphology of the Fermi surface. SdH oscillations depend only on the perpendicular magnetic field and the oscillatory features grow with increasing field. Dips that are periodic in the inverse of the magnetic field in both the Hall derivative and the magnetoresistance align at all angles, leading the authors to conclude that the plateau-like features in the Hall and minima in the SdH oscillatins originate from the topological surface states. On the other hand, curiously, the quantization of the conductance is not evident. The 2D Fermi surface in the samples considered is small enough that the 2D quantum limit could be reached at pulsed fields of up to 60T. In this limit an altered phase of the oscillations was observed, as well as quantum oscillations corresponding to fractions of the Landau integers, suggesting that correlation effects play an important role in the physics. These latter observations have not been fully explained.

Xiu \textit{et al.} \cite{Xiu_Bi2Te3_Ribbon_NN11} report experimental evidence for the modulation of topological surface states by using a gate voltage to control quantum oscillations in back-gate Bi$_2$Te$_3$ nanoribbons. The bulk resistance increases below 50K as bulk carriers are frozen out, but little change in the resistance is measured as the temperature is reduced below 20 K, and a clear plateau of resistance versus temperature was not seen. Surface conduction was inferred from a small variation in the resistance below 4 K. Under zero gate bias at 1.4 K SdH oscillations are not seen. The magnetoresistance is linear with superimposed universal conductance fluctuations. Although $\varepsilon_F$ moves, $v_F$ extracted from SdH oscillations remains constant, consistent with 2D transport. Up to half of the conductance is attributed to the surface states. The authors find that a gate voltage yields a significant enhancement of surface conduction, and the surface mobility may reach 5800 cm$^2$/(Vs), much higher than previously recorded. At the same time Aharonov-Bohm oscillations with a period of $h/2e$ are reported. 

Chen \textit{et al.} \cite{Chen_Bi2Se3_Tunable_PRB11} studied the low-magnetic field weak antilocalization magnetoconductivity of Bi$_2$Se$_3$ thin films, in which carrier densities can be tuned over a wide range with a back gate. Weak localization magnetoconductivity was fitted to the well-known Hikami-Larkin-Nagaoka formula. For the case in which the carriers can be divided into an electron layer at the top surface and a hole layer at the bottom, the magnetoconductivity deviates strongly from the single-component Hikami-Larkin-Nagaoka equation. The authors found that, as the electron density is lowered, the magnetotransport behavior deviates from the single-component description, and strong evidence exists for independent conducting channels associated with the top and bottom surfaces. This description is found to be valid for surface electron densities in the range $0.8-8.6 \times 10^{13}$ cm$^{-2}$ in samples in which the Fermi energy lies in the conduction band. 

\begin{figure}[tbp]
\bigskip
\includegraphics[width=\columnwidth]{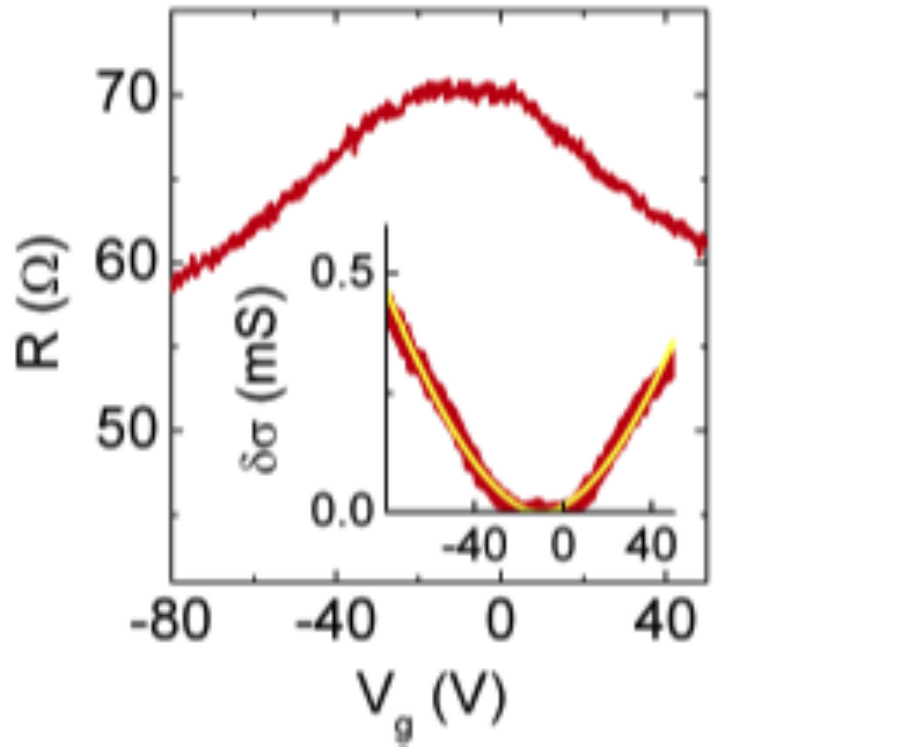}
\caption{\label{Sacepe_MaxRes}
Maximum resistivity, a signature of ambipolar transport. The figure shows the dependence of the normal state resistance on the gate voltage $V_g$. The maximum at $V_g \approx -10 V$, provides a first indication of ambipolar transport. The $V_g$
dependence originates from the modulation of the conductivity of the surface close to the gate, $\delta\sigma (V_g )$. The inset shows the calculated $\delta\sigma (V_g )$. This figure has been adapted from Ref.~\onlinecite{Sacepe_TI_NS_11}.}
\end{figure}

\begin{figure}[tbp]
\bigskip
\includegraphics[width=\columnwidth]{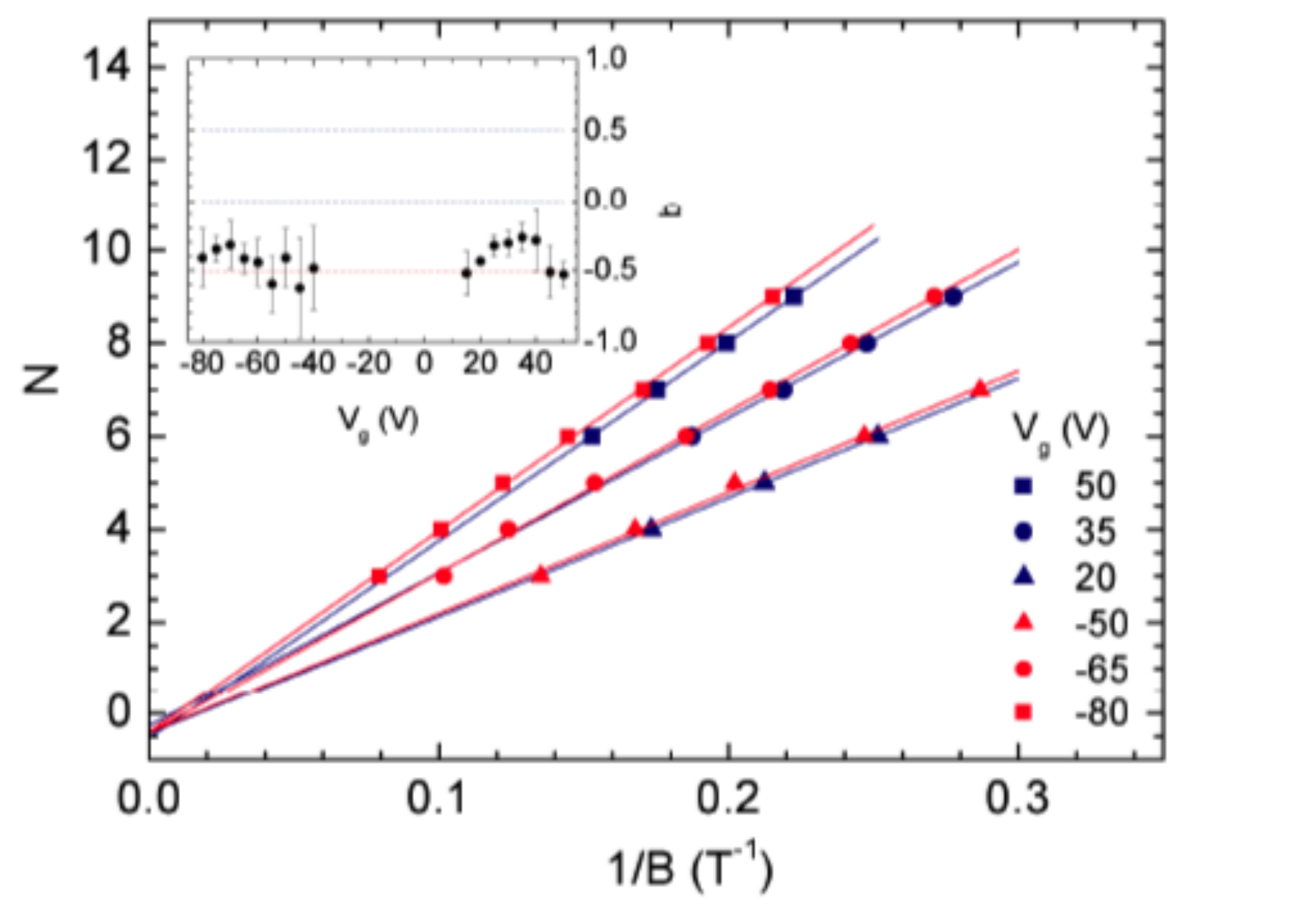}
\caption{\label{Sacepe_LL}
2D Landau levels for massless Rashba-Dirac fermions. The main panel shows the Landau level index $N$ versus $1/B$ plotted for selected values of gate voltage, corresponding to different concentration of holes (red symbols) and electrons
(blue symbols). According to the relation characteristic for massless Dirac fermions, $N = \frac{nh}{eB} - \frac{1}{2}$ , where $n$ is the density of charge carriers (electrons or holes), whereas for normal fermions $N = \frac{nh}{eB}$. Examining the way $N$
extrapolates for $1/B \rightarrow 0$ allows one to discriminate between massless Rashab-Dirac and normal fermions. The continuous lines are linear least-square fits to the data $(N =a/B + b)$. They show that as $1/B \rightarrow 0$, $N$ extrapolates to close to $-1/2$ as expected for massless Rashba-Dirac fermions (and not for normal fermions). The inset shows the extrapolation value (b) obtained by least square fitting for approximately 20 different values of gate voltage (in the region with $V_g$ between $-40$ and $+15 V$, the oscillations are not sufficiently pronounced to perform the analysis.) The data show that b fluctuates close to $-1/2$ and not around $0$, and, consistently, the average of $b$ over gate voltage is $\bkt{b} = - 0 . 4 \pm 0.1$, compatible only with massless Rashba-Dirac fermions rather than normal fermions. This figure has been adapted from Ref.~\onlinecite{Sacepe_TI_NS_11}.}
\end{figure}

A more recent experimental breakthrough by Sacepe \textit{et al.} \cite{Sacepe_TI_NS_11} investigated surface transport in thin films of Bi$_2$Se$_3$ of thickness $\approx$ 10nm. Results from this work are shown in Figs.~\ref{Sacepe_MaxRes} and \ref{Sacepe_LL}. The authors fabricated superconducting junctions on thin Bi$_2$Se$_3$ single crystals. A gate electrode allowed them to vary the electronic density. In the presence of a sufficiently large perpendicular magnetic field, sweeping the gate voltage enabled observation of the filling of Landau levels of the Dirac fermions. The Landau levels evolve continuously from electron-like to hole-like. At zero magnetic field and with the electrodes are in the superconducting state, a supercurrent appears, whose magnitude can be gate tuned, and is minimum in correspondence of the charge neutrality point determined from the Landau level filling. While these results offer grounds for hope, they have not been reproduced by other experimental groups to date.

Kim \textit{et al.} \cite{KimOh_Bi2Se3_WAL_11} investigated transport and weak antilocalization in Bi$_2$Se$_3$, finding a strong thickness-dependences over the thickness range 3 nm -- 170 $\mu$m. The mobility was found to increase linearly with thickness in the thin film regime and saturated in the thick limit, indicating dominant bulk transport. The weak localization measurements were consistent with surface transport over two decades of thickness, yet the phase coherence length increased monotonically with thickness. This latter observation hints at strong bulk-surface hybridization. The same group later reported giant, thickness-independent surface transport in Bi$_2$Se$_3$ thin films, \cite{Bansal_Bi2Se3_Giant_11} with a surface-to-bulk conductance ratio believe to be $\approx$ 470$\%$ and surface-to-bulk conductivity ratio of 11,000$\%$. The films ranged in thickness from 2 to 2750 quintuple layers. The Hall coefficient was measured to be nonlinear in the magnetic field, in agreement with a two carrier model with different mobilities. Over the entire thickness range the inferred surface density was unaltered. Interestingly, this surface density persists down to two quintuple layers, which, as the scientists themselves remark, is the lowest thickness at which top and bottom surfaces can still be defined. 

Quantum oscillations were also recently seen in Bi$_2$Te$_2$Se. \cite{Ren_Bi2Te2Se_PRB10} In this material a high resistivity was recorded, in excess of 1 $\Omega$cm as well as variable-range hopping features, accompanied by SdH oscillations attributed to the topological surface states. The authors inferred a peculiar modified Berry phase from the SdH data, which is not fully understood. The surface accounts for 6 $\%$ of the total conductivity, implying a significant number of bulk carriers, and the calculated carrier density from SdH oscillations does not match the low field Hall carrier density. These findings are ascribed to an impurity band, yet the presence of an impurity band is puzzling in well-ordered chalcogenide layers. For deep levels the carrier density needed to form an impurity band greatly exceeds the calculated bulk density. A different group working on the same material \cite{Xiong_Bi2Te2Se_QmOsc_LrgBlkR_11} reported SdH oscillations in a sample with a bulk resistivity of 6 $\Omega$cm, and a surface mobility of 2800 cm$^2$/(Vs), two orders of magnitude larger than the bulk mobility.

\begin{figure}[tbp]
\bigskip
\includegraphics[width=\columnwidth]{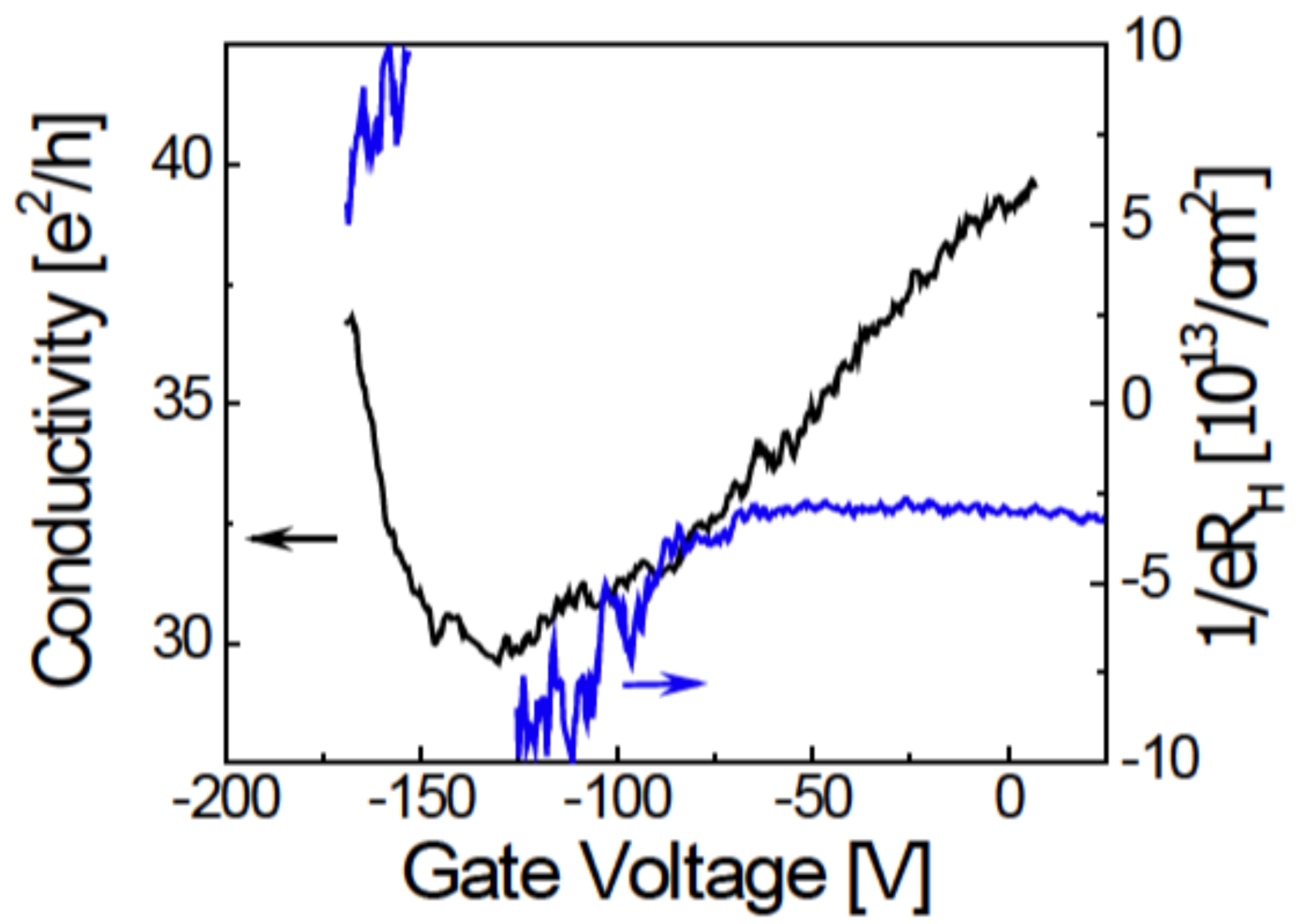}
\caption{\label{Kim_MinCond}
Minimum conductivity of a Bi$_2$Se$_3$ thin film. The figure shows the conductivity $\sigma_{xx}$ and Hall carrier density measured in an $Al_2O_3$ capped device at $T=4.2K$ as a function of back gate voltage. This figure has been adapted from Ref.~\onlinecite{Kim_TI_Gate_MinCond_11}.}
\end{figure}

Cho \textit{et al.} \cite{Cho_Bi2Se3_Insul_NL11} grew ultrathin (approximately three quintuple layer) $n$-type field-effect transistors of Bi$_2$Se$_3$. The devices have a clear OFF state and display activated temperature-dependent conductance and energy barriers up to 250 meV, identifying them as conventional insulators. In a more recent breakthrough, the same group \cite{Kim_TI_Gate_MinCond_11} showed that Bi$_2$Se$_3$ surfaces in samples with thicknesses of $<10$nm, with carrier densities of $<10^{17}$cm$^{-3}$ are strongly electrostatically coupled. A gate electrode entirely removes bulk carriers and takes both surfaces through the Dirac point simultaneously. Ambipolar transport was observed with with well-defined $p$ and $n$ regions, together with a minimum conductivity of $3-5 \, e^2/h$ per surface, reflecting the presence of electron and hole puddles. Their results are shown in Fig.~\ref{Kim_MinCond}. This experiment also measured the weak localization magnetoresistance consistent with topological surface transport.

\begin{figure}[tbp]
\bigskip
\includegraphics[width=\columnwidth]{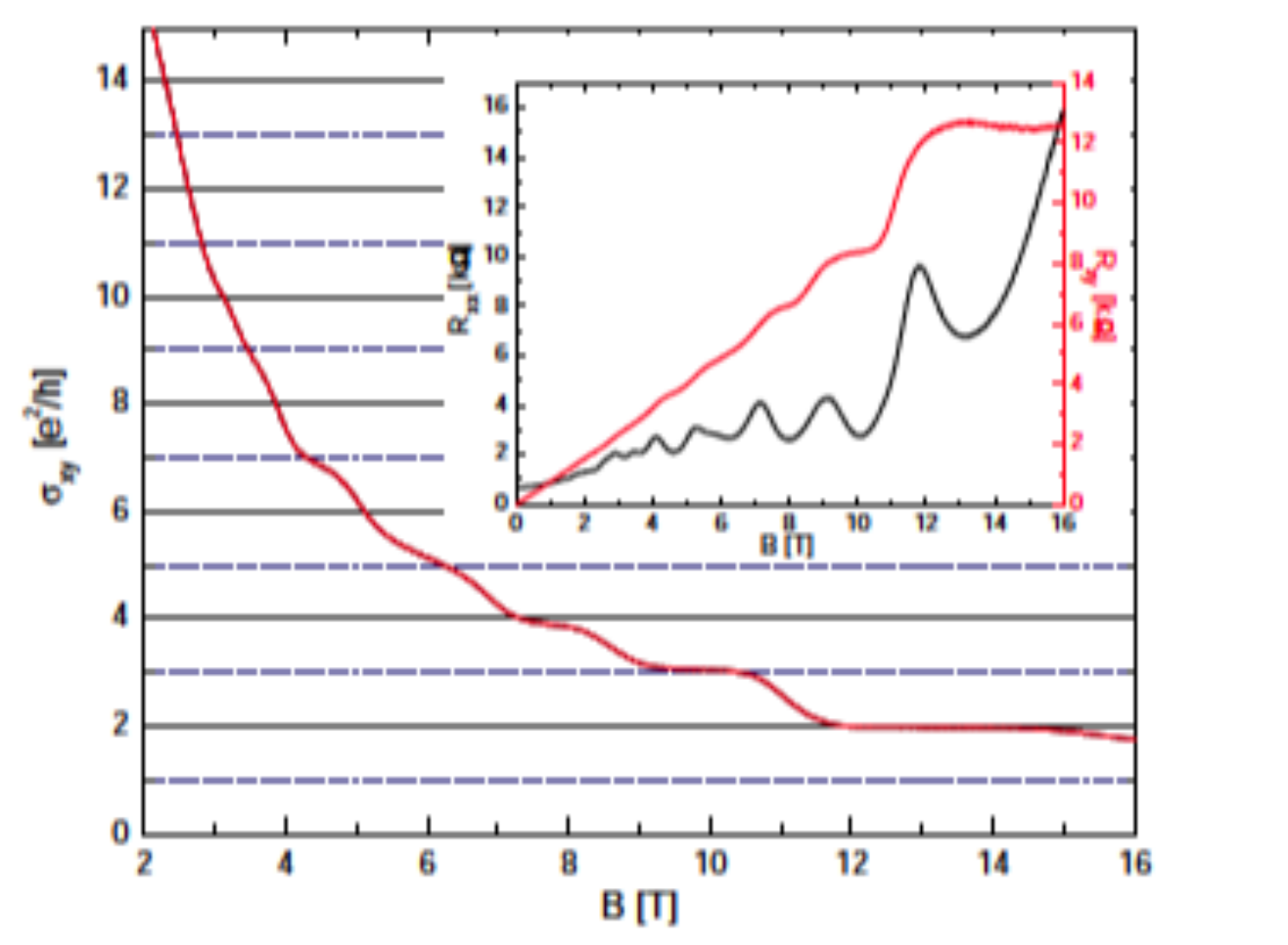}
\caption{\label{Bruene_QHE}
2D Landau levels in strained HgTe. The figure shows transport data on a strain-gapped 70-nm-thick HgTe sample. The Hall conductivity measured at 50 mK shows plateaus at the quantized values. The inset shows the Hall resistance $R_{xy}$, together with the longitudinal resistance $R_{xx}$. This figure has been adapted from Ref.~\onlinecite{Bruene_StrainedHgTe_QHE_PRL11}.}
\end{figure}

Two recent papers from the W\"urzburg group have reported exciting developments in HgTe, a semimetal. The first was a study of zero-gap HgTe quantum wells, \cite{Buettner_HgTe_ZeroGapQW_NP11} which realizes a two-dimensional electron system with a single spin-degenerate Dirac valley. The quantum well was grown at a critical thickness where the bandgap vanishes. Application of a perpendicular magnetic field resulted in the observation of the quantum Hall effect in this novel material. The quantum Hall plateaux displayed the anomalous sequence characteristic of a (single-valley) Dirac systems. The second paper concerned the observation of the quantum Hall effect due to the topological surface states of a strained 70 nm-thick HgTe layer. \cite{Bruene_StrainedHgTe_QHE_PRL11} This marks the first time that a well-defined Hall conductance from the surface states is observed experimentally. The HgTe layer was strained by epitaxial growth on a CdTe substrate, which opens a gap in the bulk material. Strained bulk HgTe thus becomes a 3D TI. Contributions from residual bulk carriers to the transport properties of the gapped HgTe layer are negligible at mK temperatures. As a result, the sample exhibits a quantized Hall effect that results from the 2D single cone Dirac-like topological surface states, shown in Fig.~\ref{Bruene_QHE}. 

In a related paper from the same group, Tkachov \textit{et al.} \cite{Tkachov_HgTe_Bxct_PRL11} presented a theoretical and experimental study of the mobility of 5-12 nm-wide HgTe quantum wells with a finite gap. Whereas semiconductor heterostructures exhibit an increase in mobility with carrier density, a mobility maximum is seen in HgTe quantum wells. This is attributed to fluctuations in the well width, which lead to fluctuations in the mass of the carriers, a phenomenon that allows backscattering between states of opposite momenta. 

\begin{figure}[tbp]
\bigskip
\includegraphics[width=\columnwidth]{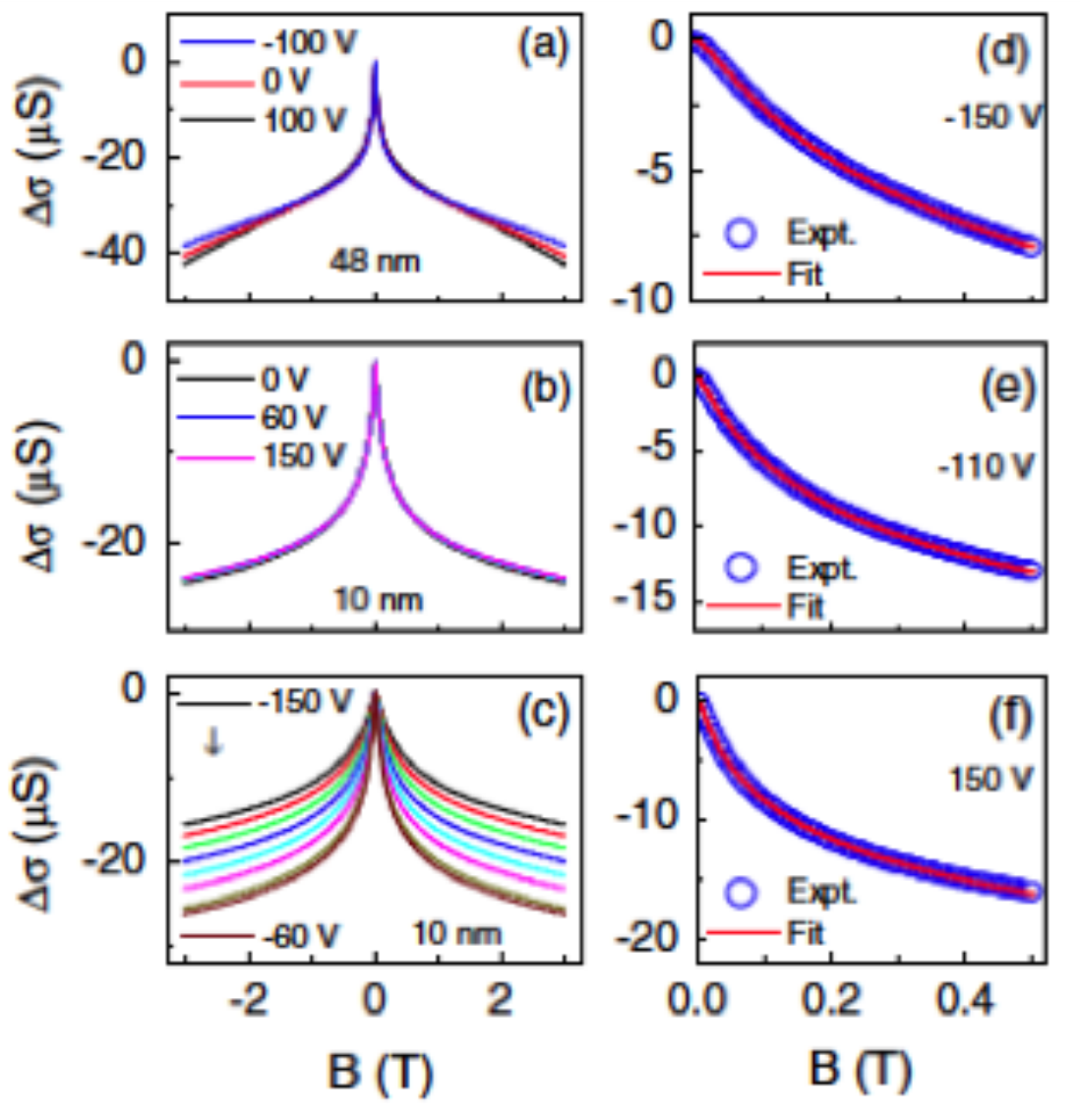}
\caption{\label{Chen_WAL}
Weak antilocalization correction to the conductivity. The figure shows the magnetoconductivity $\delta\sigma^{mag}_{xx}$ of Eq.\ (\ref{dsmag}) as a function of magnetic field measured in various samples. Figure (a) is for the sample labeled A in the article, and (b) and (c) are for the sample labeled B at different sets of gate voltages. Figures (d)-(f) are fitted curves of the magnetoconductivity of sample B at different gate voltages using the Hikami-Larkin-Nagaosa formula, given in the same equation. This figure has been adapted from Ref.~\onlinecite{Chen_Bi2Se3_GateControlEFWAL_PRL10}.}
\end{figure}

Chen \textit{et al.} \cite{Chen_Bi2Se3_GateControlEFWAL_PRL10} investigated weak antilocalization in Bi$_2$Se$_3$ thin films epitaxially grown on SrTiO3 substrates. The dielectric constant of this substrate is extremely large (10000), enabling significant tunability of the carrier density with a back gate. With the exception of very low electron densities, the low field weak antilocalization magnetoconductivity, shown in Fig.~\ref{Chen_WAL}, shows a very weak dependence on the gate voltage. These results indicate a heavily suppressed bulk conductivity at large negative gate voltages and a possible role of surface states in weak antilocalization. Even though a feature similar to the expected minimum conductivity was observed, a residual bulk conductivity cannot be ruled out based on present data.

\begin{figure}[tbp]
\bigskip
\includegraphics[width=\columnwidth]{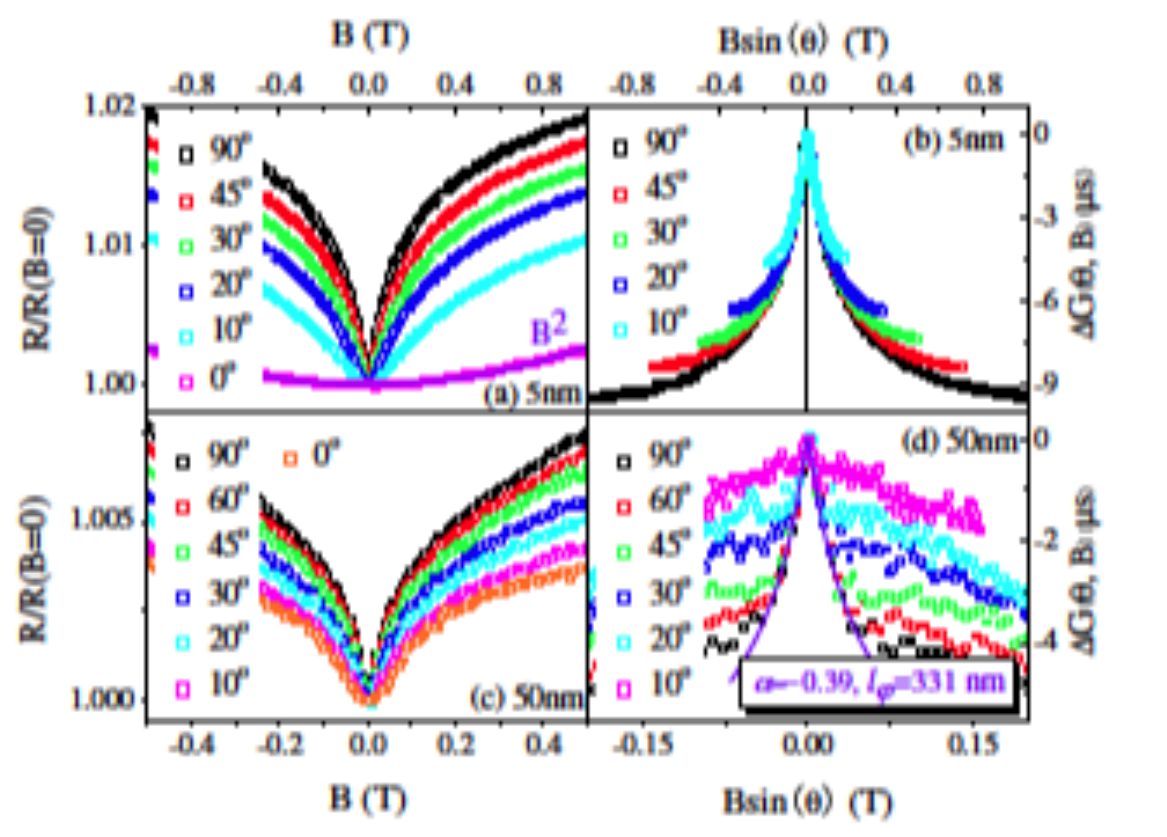}
\caption{\label{He_WAL}
Weak antilocalization correction to the conductivity. The figure shows the normalized magnetoresistance of a 5 nm Bi$_2$Te$_3$ film (a) and a 50 nm Bi$_2$Te$_3$  film (c) measured in tilted magnetic fields at $T = 2K$. The solid curve in (a) is a parabolic fit to the magnetoresistance data measured at zero tilt angle. The magnetoconductance as a function of the normal magnetic field component with the magnetoconductance at zero tilt subtracted for the 5 nm sample (b) and for the 50 nm sample (d). This figure has been adapted from Ref.~\onlinecite{He_Bi2Te3_Film_WAL_ImpEff_PRL11}.}
\end{figure}

Weak antilocalization was also studied in thin films of  Bi$_2$Te$_3$. \cite{He_Bi2Te3_Film_WAL_ImpEff_PRL11} The observed signal is believed to arise from the topological surface states due to the anisotropy of its dependence on the magnetic field. The weak antilocalization signal is robust against deposition of nonmagnetic Au impurities on the surface of the thin films, but it is quenched by the deposition of magnetic Fe impurities which destroy the $\pi$ Berry phase of the topological surface states. Their results are shown in Fig.~\ref{He_WAL}.

Liu \textit{et al.} \cite{Liu_TI_ee_PRB10} studied ultrathin Bi$_2$Se$_3$ films with thickness from one quintuple layer to six quintuple layers grown on sapphire. At low temperatures, the film resistance increases logarithmically with decreasing temperature, which is believed to represent an insulating ground state. The resistance increase as a function of temperature becomes more pronounced in thinner films. The sharp increase of resistance with magnetic field is interpreted as weak antilocalization due to the topological surface states. The authors claim the insulating ground states is induced by electron-electron interactions, yet it is puzzling that electron-electron interactions can be observed when the dielectric permittivity of Bi$_2$Se$_3$ is approximately 100. 

Wang \textit{et al.} \cite{Wang_Bi2Se3_Film_ee_PRB11} studied thin films of Bi$_2$Se$_3$ with and without Pb doping. ARPES clearly identifies the topological surface states in the Pb doped films, with $\varepsilon_F$ in the bulk gap. The thickness of the films is smaller than the inelastic mean free path. Transport data show weak localization behavior, but the authors claim the temperature and magnetic field dependences of resistance cannot be explained without appealing to electron-electron interactions. On the other hand, both the bulk and the surface participate in transport, and a clear separation is not possible in the samples investigated. 

\section{Experimental challenges}

The biggest obstacle to the observation of $Z_2$ topological order in transport has been a factor that may be termed the bad metal syndrome. The existence of chiral surface states is well established experimentally through angle-resolved photoemission (ARPES) and scanning-tunneling microscopy (STM) studies, yet in transport their contribution is overwhelmed by the bulk, which, due to unintentional defect-induced doping, is a low-density metal. To date no experimental group has produced a true topological insulator. Consequently, all existing topological insulator systems are in practice bulk-doped materials.

One key aim of current TI efforts is the identification of a definitive signature of surface transport. Several recent experimental works point to a possible way forward, at least in the detection of surface transport, if not a permanent solution to the bulk doping problem. In recent gated thin film measurements,\cite{Kim_TI_Gate_MinCond_11} the two notable features are gate voltage tuning well past the Dirac point, and the resemblance of the conduction to that of dirty graphene dominated by charge puddles, signifying the removal of bulk electrons. Cho \textit{et al.} also recently reported completely insulating Bi2Se3 devices. \cite{Cho_Bi2Se3_Insul_NL11} These researchers used mechanical exfoliations to fabricate ultrathin field-effect transistors of approximately three quintuple layers of Bi2Se3 on a Si/SiO$_2$ substrate. The devices are $n$-type and have a clear off state at negative gate voltage. The conductance shows an activated temperature dependence with a large activation energy up to 250 meV.
  
Kong \textit{et al.} \cite{Kong_Bi2Se3_Bi2Te3_Nanoplates_NL10} reported the synthesis and characterizations of ultrathin Bi2Te3 and Bi2Se3 few-layer tunable nanoplates with thickness down to 3 quintuple layers, using a catalyst-free vapor-solid (VS) growth mechanism. Optical images reveal thickness-dependent color and contrast for nanoplates grown on Si/SiO$_2$. Ultrathin TI nanoplates have an extremely large surface-to-volume ratio and can be electrically gated more effectively than the bulk form, potentially enhancing surface state effects in transport measurements. Low-temperature transport measurements of a single nanoplate device, with a high-k dielectric top gate, show decrease in carrier concentration by several times and large tuning of chemical potential.

Teweldebrhan \textit{et al.} \cite{Teweldebrhan_Bi2Te3_Exfol_NL10} obtained atomically thin crystalline films of Bi$_2$Te$_3$ on Si/SiO$_2$ by exfoliation, which were characterized by atomic force microscopy and micro-Raman spectroscopy. The crystals have low thermal conductivity and high electrical conductivity. Remarkably, the films are non-stoichiometric: altering the thickness and sequence of atomic planes enables one to change the composition and doping of the films. 

The above presentation demonstrates that, thus far, experimentalists have only succeeded in eliminating bulk carriers \textit{temporarily} using a gate. The undesirable residual carrier density can be reduced by doping, since the presence of dopants drains the excess carriers and brings the chemical potential down into the bulk gap. One potential way forward was opened by Hor \textit{et al.}, \cite{Hor_Bi2Se3_Ca_p_PRB09} who demonstrated that Ca doping ($\approx 1 \%$ substituting for Bi) brings the Fermi energy into the valence band, followed by the demonstration by Hsieh \textit{et al.} \cite{Hsieh_BiSb_QmSpinTxtr_Science09} that Ca doping can be used to tune the Fermi energy so that it lies in the gap. However, the Ca-doped material displays irregular magnetotransport features that have not been explained yet. Furthermore, a larger dopant concentration translates into a larger impurity concentration, and in the current TI materials this may cause the formation of impurity bands. A large impurity concentration may also lead to Anderson localization in the bulk, which is dangerous for TI. The bulk conductivity vanishes at zero temperature for an Anderson insulator, but bulk hopping contributes to transport at finite temperatures and may still dominate over the contribution from the surface states. Mechanically exfoliated thin TI films on a substrate have turned out to have large carrier densities, hinting that further doping is introduced during the cleaving process or via interaction with the substrate.

An additional experimental problem is the observation that the atmosphere leads to $n$-type doping of the surface. \cite{Analytis_Bi2Se3_BlkSfc_PRB10} It is known that Bi$_2$Se$_3$ acquires additional $n$-type bulk doping after exposure to atmosphere, thereby reducing the relative contribution of surface states to the total observed conductivity. Kong \textit{et al.} demonstrated this \cite{Kong_Bi2Se3_SfcOx_Degrad_ACS11} via transport measurements on Bi$_2$Se$_3$ nanoribbons. Systematic surface composition analyses by X-ray photoelectron spectroscopy reveal fast formation and continuous growth of native oxide on Bi$_2$Se$_3$ under ambient conditions. In addition to $n$-type doping at the surface, such surface oxidation is likely the material origin of the degradation of topological surface states. Appropriate surface passivation or encapsulation may be required to probe topological surface states of Bi2Se3 by transport measurements. In this context, I will note that ARPES results obtained by different groups on the same material tend to show differences in the measured dispersion relations, with slightly different values of $v_F$, and a significant time-dependence.

Considerable work has focused on the detection of surface states using quantum oscillations. One recurring problem in current experimental reports of coexisting bulk and surface magnetotransport it the fact that interpretations typically rely on the angle dependence of the quantum oscillations, which in many cases is in fact very weak, and, until convincing proof is devised, may still reflect the anisotropy of 3D bulk transport. Moreover, substantial disagreements exist between the Hall coefficient measured in different ways, and in many experimental studies, although signatures of 2D electrons were reported, they were frequently accompanied by a host of unconventional extraneous findings, such as time-dependent Hall coefficients. Despite the reported observation of SdH oscillations by many experimental groups, one salient issue, pointed out above in the discussion of the Hall effect is that the TI surfaces are all connected. Therefore in principle one does not expect the period of SdH oscillations to diverge. A full understanding of the surface SdH oscillations without doubt requires further investigation.

These obstacles are exacerbated by the fact that, whenever an indication of surface mobility can be obtained, it is characteristically very low, often below 1000 cm$^2$/Vs.  Below some threshold, when exfoliating bulk crystals, both the effective 3D carrier density and scattering go up, which is the primary reason why a substantial effort has been devoted to the gating process. A working model that has gained popularity nowadays is that the cleaving process mechanically introduces various kinds of defects into the samples, an observation borne out by the fact that MBE sample growers have experienced great difficulty in obtaining carrier densities lower than $10^{18}$ cm$^{-3}$. The band structure may be distorted near the surface due to space-charge accumulation. For a host of practical reasons, the procedure of removing the bulk electrons by reducing the sample thickness appears to bring in unwanted problems of its own. On a more fundamental level, surface electrons experience the random potential due to impurities, inhomogeneities, and roughness found at the interfaces. Topological insulator surfaces remain poorly understood, and experimentally the next big challenge may well turn out to be the task of engineering cleaner and more tractable surfaces.

From the point of view of data interpretation, bearing in mind that the bulk conductance tends to overwhelm the surface conductance, subtracting the bulk from the total in order to obtain the surface contribution is tricky and may be unreliable. A small number that is obtained by subtracting two large numbers can often lead to erroneous conclusions. In addition, many interpretations have hinged on a transport model that considers two types of carriers, which are taken to be surface and bulk carriers, although the model itself does not distinguish 2D from 3D transport. 

One final factor to bear in mind, though not a problem in itself, is that the Rashba-Dirac cone can be buried in the bulk conduction or valence bands. Its location within the bulk gap is not topologically protected.

\section{Conclusions and Outlook}

A feature of this review that should be plainly obvious is that the work covered spans a period of only three years -- in fact, the bulk of it was reported within the last two years alone. The fact that a transport review is necessary after such a short period of time illustrates both the overwhelming fundamental interest in this field and its phenomenal pace of growth. Three-dimensional topological insulators have grown from non-existence to a vastly developed mature field involving hundreds of researchers practically overnight. Within this time span, chiral surface states started out as a mere theoretical concept, were predicted to exist in several materials, were subsequently imaged and have finally been seen in transport. The rise of topological insulators is following a close parallel to the rise of graphene a short time ago. Credit for this unprecedented development must be attributed to the tireless efforts of many experimental groups and to the state-of-the-art technology available in the present day.

As the previous section made abundantly clear, several stumbling blocks remain in transport, particularly in experiment. The growth of new materials is a nontrivial issue, and obtaining high-quality samples where only the surface electrons can be accessed in transport has proved to be a difficult task. It is especially important to bear in mind that future work may initially be hampered by factors such as roughness and dirt inherent in solid-state interfaces. The theoretical results described in this review will become useful in discerning transport by the topological surface states only after a real topological insulator material, that is, a system which is a bulk insulator, becomes available. It remains true that all current TI materials are effectively bulk \textit{metals} because of their large unintentional doping -- at present, bulk carriers are only removed temporarily by gating. Such systems are, by definition, unsuitable for studying surface transport properties since bulk transport overwhelms any signatures of surface transport. Discussing TI surface transport in such bulk-doped TI materials retains some ambiguity, since it necessarily involves complex data fitting and a series of assumptions required by the necessity of distinguishing bulk versus surface transport contributions. Although such analyses are being carried out in bulk doped TI materials, in order to understand the transport behavior, real progress is expected when surface TI transport can be carried out unambiguously, without any complications arising from the (more dominant) bulk transport channel. 

Coexisting bulk and surface transport remains a relevant problem to TI experiments at present, and will be until scientists can produce a material that does not require gating in order to shift the chemical potential into the bulk gap. It could also be useful in the case of very small gap TIs that have large numbers of thermally excited bulk carriers. Analysis of current experimental data often uses a simple one carrier model, rather than relying on a systematic two-carrier analysis as a function of thickness. Also, it is not clear whether minimalistic heuristic models based on \textit{parallel} surface and bulk conduction are valid in systems in which these two types of carriers are present. 

Undoubtedly, the next major tasks facing experimentalists are getting the chemical potential in the gap without the aid of a gate, further experimental studies confirming the existence of a minimum conductivity, the measurement of a spin-polarized current, the search for the anomalous Hall effect (whether half-quantized or not quantized) and the associated giant Kerr effect, and potentially measurement of the giant spin-Hall effect. Theoretically, areas that have received little attention include a comprehensive determination of the role of electron-electron interactions, the Kondo effect and the associated resistance minimum, the possibility of large extrinsic spin-orbit interactions, a clear resolution of the weak localization problem and its extension to cover universal conductance fluctuations.

The observations above should not be considered a pessimistic assessment, since topological insulators are truly new materials, and materials science must explore ways to handle novel problems, such as accessing the surface states. That is necessarily a challenging undertaking. Although fine control over carrier densities remains difficult with existing bulk-doped material, considering that the field has only been active for three years, the progress registered is astounding, and tempering an occasional impatience for instant success can only be fruiftul in the long run.

I am extremely indebted to N.~P.~Butch, S.~Das Sarma, Yongqing Li, Changgan Zeng, Zhong Fang, Junren Shi, Di Xiao, and Shun-Qing Shen. This work was in part supported by the Chinese Academy of Sciences.


\end{document}